\documentclass[fleqn,usenatbib]{mnras}
 
\usepackage[T1]{fontenc}

\DeclareRobustCommand{\VAN}[3]{#2}
\let\VANthebibliography\thebibliography
\def\thebibliography{\DeclareRobustCommand{\VAN}[3]{##3}\VANthebibliography}

\usepackage{graphicx}	
\usepackage{amsmath}	
\usepackage{amssymb}	

\usepackage{lipsum}

\AtBeginDocument{}

\usepackage{newtxtext,newtxmath}

\title[Halo baryon content in COLIBRE]{The influence of feedback on the baryonic content of haloes in the COLIBRE simulations}

\author[J. J. Davies et al.]{Jonathan J. Davies,$^{1}$\thanks{E-mail: j.j.davies@ljmu.ac.uk},
Joop Schaye,$^{2}$
Filip Hu\v{s}ko,$^{2}$
Ruby J. Wright,$^{3}$
Evgenii Chaikin,$^{4,2}$
Matthieu Schaller,$^{2}$
\newauthor
Robert J. McGibbon,$^{2}$
Alejandro Ben\'{i}tez-Llambay,$^{5}$
Sylvia Ploeckinger$^{6}$
and Alexander J. Richings$^{7,8}$
\\
$^{1}$Astrophysics Research Institute, Liverpool John Moores University, 146 Brownlow Hill, Liverpool L3 5RF, UK \\
$^{2}$Leiden Observatory, Leiden University, PO Box 9513, 2300 RA Leiden, the Netherlands \\
$^{3}$International Centre for Radio Astronomy Research, University of Western Australia, 7 Fairway, Crawley, WA 6009, Australia \\
$^{4}$Institute for Computational Cosmology, Department of Physics, University of Durham, South Road, Durham, DH1 3LE, UK \\
$^{5}$Dipartimento di Fisica ``Giuseppe Occhialini'', Universit\`{a} degli Studi di Milano-Bicocca \\
$^{6}$Department of Astrophysics, University of Vienna, T\"{u}rkenschanzstrasse 17, A-1180 Vienna, Austria \\
$^{7}$Centre for Data Science, Artificial Intelligence and Modelling, University of Hull, Cottingham Road, Hull, HU6 7RX, UK \\
$^{8}$E. A. Milne Centre for Astrophysics, University of Hull, Cottingham Road, Hull, HU6 7RX, UK
}

\date{Accepted XXX. Received YYY; in original form ZZZ}

\pubyear{2025}

\begin{document}
\label{firstpage}
\pagerange{\pageref{firstpage}--\pageref{lastpage}}
\maketitle
\begin{abstract}
We present predictions for the relationship between halo gas mass fraction and halo mass, $f_{\rm gas}^{200}-M_{200}$, from the COLIBRE cosmological simulations of galaxy formation, and explore how the gas content of haloes is influenced by feedback from supernovae and active galactic nuclei (AGN) over time. The $f_{\rm gas}^{200}-M_{200}$ relation in COLIBRE is non-monotonic, with a peak at $M_{200}\sim 10^{11.5-12}$~M$_\odot$. Below this mass, supernova feedback efficiently expels gas from the haloes of dwarf galaxies; above it, AGN feedback efficiently depletes the haloes of galaxy groups. The relation rises for $M_{200}\gtrsim 10^{13}$~M$_\odot$, as AGN-driven expulsion becomes less efficient. The fiducial COLIBRE model yields group and cluster gas fractions that agree with Chandra and XMM-Newton X-ray constraints. While these gas fractions are lower than those in simulations such as EAGLE and IllustrisTNG, they are high relative to constraints from some recent eROSITA stacking analyses, and to gas fractions in hydrodynamical and baryonification models that reproduce both the eROSITA stacks and kinetic Sunyaev-Zel'dovich (kSZ) effect measurements. COLIBRE's hybrid AGN feedback model, which combines thermal and jet-driven feedback, produces lower gas fractions in better agreement with eROSITA stacks and kSZ constraints (but worse agreement with Chandra and XMM-Newton). We investigate why COLIBRE exhibits better agreement with these observational constraints in comparison to EAGLE and other contemporary simulations, and explore how simulation resolution affects the impact of feedback. Our results demonstrate that halo gas fractions are a sensitive probe of feedback physics, and can differ significantly between simulations that otherwise produce very similar galaxy populations.

\end{abstract}




\section{Introduction}
\label{sec:intro}

The dark matter haloes of galaxies, galaxy groups and galaxy clusters contain reservoirs of diffuse, multiphase gas spanning a wide range of temperatures and densities, from the cold, dense clouds present in the circumgalactic medium (CGM) of individual galaxies to the hot, shock-heated gas that is increasingly prevalent in higher-mass haloes \citep[for a review, see][]{tumlinson17}. Gaseous haloes are of crucial importance to galaxy formation and evolution, as inflows from (and through) them are essential to sustain the formation of stars in central galaxies \citep[e.g.][]{keres05,vandevoort11a,sanchezalmeida14}. In turn, the properties of gas in haloes are strongly influenced by outflows driven by supernovae (SNe) and active galactic nuclei (AGN) that are associated with these processes, which can recycle gas between the galaxy and halo, or expel gas beyond the virial radius \citep[][]{oppenheimer10,vandevoort11b,fauchergiguere11,heckmanbest14,harrison18,veilleux20,mitchell20a,mitchell20b,mitchell22a}. The total mass of gas in haloes is typically expressed as a fraction of the total mass, i.e. the gas fraction, $f_{\rm gas}$; this quantity is markedly sensitive to the integrated history of feedback across cosmic time. The relationship between the gas fraction and halo mass therefore represents a powerful diagnostic for galaxy formation models that is complementary to the observable properties of galaxies \citep[][for a review see \citealt{crainvandevoort23}]{mccarthy10,davies20,mitchell22b}.

Observational constraints on the halo gas fraction have historically been obtained through converting X-ray surface brightness profiles into gas density profiles, and combining these with measurements of the total halo mass, either from the X-ray data under the assumption of hydrostatic equilibrium, or from weak lensing data. Measurements made with observatories such as Chandra and XMM-Newton provide a broadly consistent picture of the relation between $f_{\rm gas}$ and halo mass for galaxy groups and clusters \citep[e.g.][]{sun09,lovisari15, eckert16, akino22}. More recently, however, this consensus has become less clear. Studies of eROSITA X-ray survey data for stacks of galaxy groups \citep[both optically-selected, e.g.][and X-ray selected, e.g. \citealt{siegel26b}]{popesso26, zhang26} have indicated that gas fractions may typically be lower than previously measured at this mass scale, suggesting a preference for stronger and more expulsive AGN feedback. Conversely, other recent analyses of eROSITA and XMM-Newton data \citep{eckert26,seppi26,khalil26} and of the X-ray--cosmic-shear cross-correlation \citep{mcdonald26} do not suggest lower gas fractions, and instead indicate gas fractions consistent with those inferred from pre-eROSITA X-ray data. These discrepancies between different analyses primarily arise because the conversion from X-ray observable to gas fraction requires many assumptions about the temperature and metallicity profiles of the gas, the form of the density profile, and the accuracy of the halo mass estimate \citep{zhang26,eckert26,seppi26,bucko26}. These systematic uncertainties are typically larger than the statistical uncertainty on any individual measurement \citep[see e.g.][]{zhang26}, and particularly severe for lower-mass group and $\sim L^\star$ haloes, where the X-ray surface brightness is low and the gas properties are poorly constrained. 

Complementary constraints have recently become available from alternative tracers that are less biased towards the hot, dense gas at the centres of haloes that dominates the X-ray surface brightness. Two such tracers are the thermal and kinetic Sunyaev-Zel'dovich (tSZ \& kSZ) effects, observable through the combination of full-sky data from {\it Planck} with higher-resolution and higher-sensitivity data from the Atacama Cosmology Telescope (ACT) or the South Pole Telescope (SPT). The tSZ signal quantifies the total thermal energy of the halo gas (when integrated over an area), whereas the kSZ signal quantifies the gas mass (or more specifically, the gas momentum). Evidence for the expulsion of baryons from haloes by feedback is given by the suppression of the tSZ power spectrum on scales corresponding to haloes \citep[e.g.][]{mccarthy14,mccarthy23,louis25,efstathiou25,raghunathan26} and cross-correlations of the tSZ signal with galaxy positions or cosmic shear \citep[e.g.][]{schaan21,gatti22,troster22,mccarthy23,liu25,pandey25,dalal26}. Likewise, stacked kSZ measurements can now be used to constrain the suppression of the matter power spectrum, and when combined with `baryonification' frameworks, can yield constraints on $f_{\rm gas}$ \citep[e.g.][]{schneider22,bigwood24,bigwood25kSZ,mccarthy25,hadzhiyska25a,kovac25,siegel26a}. Recent results from these studies collectively suggest that feedback has a stronger and more expulsive influence on the gas content of galaxy groups than was indicated by pre-eROSITA X-ray data. A third alternative tracer of halo gas content is becoming available through the dispersion measures of fast radio bursts (FRBs), which probe the column density of free electrons along the line of sight. Recently, \citet{sharma26} combined dispersion measures from 109 FRBs with a halo model framework, and inferred higher gas fractions that are in line with pre-eROSITA studies, and discrepant with constraints inferred from the eROSITA analyses of \citet{popesso26} and \citet{siegel26b}. 

The challenges involved in observationally characterising the gas content of haloes motivate the use of cosmological, hydrodynamical simulations, which self-consistently follow the co-evolution of galaxies and their gaseous haloes across cosmic time. These simulations have revealed how gas in haloes of different masses is influenced by feedback processes; in lower-mass haloes, feedback from supernovae (SNe) efficiently drives outflows and gas recycling \citep[e.g.][]{nelson19,mitchell20b,wright24}, but this feedback ceases to be effective in the hot, shock-heated media of more massive haloes, causing their central supermassive black holes to grow rapidly and become the dominant influence on halo properties through AGN feedback \citep[e.g.][]{bower17,mcalpine18}.

Cosmological simulations must employ `subgrid' prescriptions for the physical processes driving SN and AGN feedback that occur on unresolved scales, and there are several strategies for designing and calibrating such models. Feedback prescriptions may be calibrated with the aim of reproducing the observed properties of galaxies, such as the galaxy stellar mass function and the sizes of galaxies, with the properties of gas within haloes effectively left as a prediction of the model. Contemporary examples of this approach include EAGLE \citep{schaye15}, SIMBA \citep{dave19}, and COLIBRE \citep{schaye26,chaikin_calibration}, the last of which is the subject of this study. In another approach, the properties of halo gas, such as $f_{\rm gas}$, are explicitly considered in the calibration, typically in simulations that focus on the impact of feedback on large-scale structure and cosmology. Examples of this approach include BAHAMAS \citep{mccarthy17}, (X)FABLE \citep{henden18,bigwood25xfable}, and FLAMINGO \citep{schaye23,kugel23}. Other models that are not specifically focused on large-scale structure, such as IllustrisTNG \citep{pillepich18, nelson19dr}, also attempt to calibrate feedback to both the properties of galaxies and halo gas.

In the current generation of cosmological simulations, there is not a ``unique solution'' for the impact of feedback on gas in haloes. Models employing different subgrid prescriptions for feedback processes can produce galaxy populations with similar properties, but can differ strongly in their relationships between $f_{\rm gas}$ and halo mass \citep[e.g.][]{davies20,oppenheimer21,crainvandevoort23,popesso26}, their ``baryon cycles'' of inflow and outflow \citep{mitchell22a,mitchell22b,wright24,oren26}, and the length scales over which baryons are redistributed, as quantified by the ``closure radius'' \citep{ayromlou23}, or the suppression of the matter power spectrum \citep{vandaalen20,mccarthy25}. Within the context of a single model, the impact of feedback is sensitive to the chosen values of model parameters \citep[e.g.][]{mitchell22b,gebhardt24,medlock25,grayson26}, and the relationship between $f_{\rm gas}$ and halo mass is typically highly sensitive to the method of coupling feedback energy to the halo gas \citep[e.g.][]{davies20,sorini22,khrykin24,husko24}.

Whilst the shape and normalisation of the relation between $f_{\rm gas}$ and halo mass are highly sensitive to the adopted models for stellar and AGN feedback, a near-universal prediction of these simulations is that the scatter in the relation at a given halo mass correlates strongly with the integrated amount of efficient AGN feedback injected into haloes over time. The mass of the central supermassive black hole (SMBH), $M_{\rm BH}$, is an excellent proxy for this, and in many simulations, haloes that are gas-poor for their mass host overmassive SMBHs and vice-versa \citep{davies19,davies20,dave19,terrazas20,robson23,marini25b}. In turn, scatter in $M_{\rm BH}$ at a given halo mass is driven by differences in the halo assembly time, which influences the binding energy that AGN feedback must work against \citep{boothschaye10,davies19}, and by differences in merger history, as galaxy interactions can disrupt the ordered co-rotation of gas in galaxies and fuel rapid SMBH growth \citep{davies22,davies24,roberts26}. There are exceptions to this general picture; no such correlations exist in the ROMULUS25 simulation, which features AGN feedback that is less expulsive on halo scales \citep{sanchez24}, and while this picture holds for the haloes of galaxies in the FLAMINGO simulations, the correlation between $f_{\rm gas}$ and $M_{\rm BH}$ reverses and becomes positive for group and cluster haloes, in which overmassive SMBHs are indicative of earlier expulsion and subsequent replenishment of halo gas \citep{costello25}.

The scatter in $f_{\rm gas}$ at a fixed halo mass is of crucial importance to the diversity of galaxy properties produced by cosmological simulations. The expulsion of a large fraction of a halo's baryons reduces its characteristic density, giving rise to longer radiative cooling times and preventing replenishment of the star-forming gas reservoir in the central galaxy. This transformation of the baryon cycle can quench galaxies over long (several Gyr) timescales, or maintain quiescence in galaxies that have already depleted their gas reservoirs. Quenched galaxies therefore tend to reside in gas-poor haloes, while star-forming galaxies typically reside in gas-rich haloes, at a given halo mass \citep[e.g.][]{davies20,davies22,zinger20,appleby21,voit24,ni25}. Some tentative evidence for this connection is emerging from recent eROSITA X-ray \citep{chadayammuri22,zhang25sf} and ACT tSZ \citep{das25} data, though such analyses remain challenging, particularly due to difficulties in deriving halo mass estimates, and in the case of X-ray observations, handling contamination from unresolved AGN and X-ray binaries.

The studies listed thus far have demonstrated that an understanding of the whole galaxy-halo ecosystem is vital for a holistic and complete model of galaxy formation and evolution. Motivated by this, herein we present predictions for the relation between $f_{\rm gas}$ and halo mass from the COLIBRE suite of cosmological hydrodynamical simulations. COLIBRE incorporates a number of significant improvements over its predecessor, EAGLE \citep{schaye15,crain15}, including a new chemical network, coupled to a dust grain evolution model, which allows modelling of the cold, multiphase interstellar medium (ISM) in large cosmological volumes. This significant change to the structure of the ISM motivated the development of new, more advanced models for feedback from SNe and AGN, and our primary goal for this study is to explore how these new models shape the $f_{\rm gas}-M_{\rm halo}$ relation. In addition to simulations performed with the fiducial COLIBRE model, the suite also includes simulations performed with an alternative `hybrid' AGN model, which features jets, and in which the SMBH spin and accretion disc state are tracked and influence the efficiency and nature of the feedback. 

The COLIBRE model has been calibrated to reproduce the observed present-day galaxy stellar mass function and size-mass relation at multiple mass resolution levels and with both AGN feedback models, as detailed in the reference articles \citep{schaye26,chaikin_calibration,husko26}. The simulated populations produced by the model show excellent agreement with a range of other observables, such as the evolution of the galaxy stellar mass function \citep{chaikin_gsmf} and the sizes of galaxies \citep{ludlow26}, the present-day galaxy luminosity function from far-UV to near-IR and from far-IR to sub-mm wavelengths \citep{lu26a}, the metallicities and H\textsc{i} and H$_{2}$ masses of galaxies as a function of stellar mass \citep{schaye26}, the evolution of the stellar mass - gas metallicity relation \citep{sharda26}, the evolution of the relations between spatially-resolved star formation rates, H\textsc{i} and H$_{2}$ masses and metallicity \citep{lagos26}, and the halo X-ray luminosity as a function of halo mass \citep{schaye26}.

We begin by describing the COLIBRE simulations and our analysis methods in \S\ref{sec:methods}, and then present the relation between $f_{\rm gas}$ and halo mass (\S\ref{sec:results:flagship}), show the breakdown of halo gas temperatures as a function of $M_{\rm halo}$ (\S\ref{sec:results:temperature}), and discuss convergence with resolution (\S\ref{sec:results:convergence}). We compare the predictions of the fiducial COLIBRE model with current observational constraints on the $f_{\rm gas}$ in \S\ref{sec:results:obs}, and further show this comparison for the alternative hybrid AGN model in \S\ref{sec:results:hybrid}. We compare COLIBRE's predictions with those of other contemporary simulations in \S\ref{sec:results:simcompare}, with a particular focus on how the new subgrid physics in COLIBRE has changed the gas fractions of haloes relative to EAGLE. Finally, we explore how halo gas fractions are shaped by feedback over time in \S\ref{sec:results:histories}, before summarising and discussing our results in \S\ref{sec:summary}.

\section{Methods}
\label{sec:methods}

The COLIBRE simulations are described in detail in \citet{schaye26}, and the strategy for their calibration in \citet{chaikin_calibration}. For brevity, we will only briefly cover the key information about the simulation suite here, and refer the reader to these articles for an in-depth description of the model. We will, however, include a more complete description of COLIBRE's prescriptions for stellar and AGN feedback (\S\ref{sec:methods:sf} \& \S\ref{sec:methods:bh}), as they are crucial for setting the gas fractions presented in this work.

All simulations in the COLIBRE suite were performed using the {\sc Swift} simulation code \citep{schaller24}, employing the {\sc Sphenix} smoothed-particle hydrodynamics (SPH) scheme \citep{borrow22}. The initial conditions for COLIBRE were generated using {\sc monofonIC} \citep{hahn20,michaux21}, with a starting redshift $z=63$. They adopt the Dark Energy Survey year three `3x2pt + all external
constraints' $\Lambda$CDM cosmology \citep{abbott22}: $\Omega_{{\rm M},0} = 0.306$,
$\Omega_{{\rm b},0} = 0.0486$, $\sigma_8 = 0.807$, $h= 0.681$, $n_{\rm s} = 0.967$, and a
single massive neutrino species with mass 0.06 eV.

The COLIBRE model has been calibrated at three resolution levels to match key observational diagnostics. These levels are, in terms of gas particle mass ($m_{\rm gas}$) and dark matter (DM) particle mass ($m_{\rm DM}$): m7 ($m_{\rm gas}=1.47\times10^7$~M$_\odot$, $m_{\rm DM}=1.94\times10^7$~M$_\odot$), m6 ($m_{\rm gas}=1.8\times10^6$~M$_\odot$, $m_{\rm DM}=2.4\times10^6$~M$_\odot$) and m5 ($m_{\rm gas}=2.3\times10^5$~M$_\odot$, $m_{\rm DM}=3.0\times10^5$~M$_\odot$). A notable feature of COLIBRE is the use of similar baryonic and DM particle masses, achieved by using 4$\times$ more DM particles than baryonic particles, to mitigate the spurious numerical heating of the stellar components of galaxies that occurs when using more massive DM particles \citep[e.g.][]{ludlow19,ludlow21,ludlow23,wilkinson23}. The suite also includes two different implementations of AGN feedback (\S\ref{sec:methods:bh}), and calibrated versions of the model have been run at all resolutions with each implementation. In this study, we present results at all resolution levels and with both AGN prescriptions, in each case using the largest available simulation volume\footnote{We have verified that our results are very well converged with respect to simulation volume; for the relations presented in this study, smaller-volume simulations performed at the same resolution and with the same model yield near-identical median relations to their larger counterparts.} at $z=0$ to achieve the best sampling of high-mass (and hence rare) haloes. The largest simulations run with the fiducial model have side length 400 cMpc at m7 resolution (denoted L400m7), 200 cMpc at m6 resolution (L200m6) and 25 cMpc at m5 resolution (L025m5). The largest simulations run with the hybrid AGN model have side length 200 cMpc at m7 resolution (L200m7h) and 100 cMpc at m6 resolution (L100m6h).

\subsection{Subgrid physics}
\label{sec:methods:subgrid}

COLIBRE includes modelling of the cold gas phase, with radiative cooling allowed down to $\approx 10$~K. As detailed by \citet{ploeckinger25}, new physical prescriptions have been implemented to allow for this, including the formation of molecules (in gas and on the surface of dust grains), photo-ionisation, photo-dissociation and photoheating by cosmic rays and the metagalactic and intergalactic radiation fields, self-shielding from this radiation, and photo-electric heating by dust. Radiative cooling, heating and chemical evolution are calculated using the \textsc{chimes} network \citep{richings14a,richings14b}, with hydrogen and helium treated in full non-equilibrium. Cooling due to other elements assumes equilibrium species fractions but is adjusted according to the non-equilibrium free electron density provided by the H and He reaction network. COLIBRE also incorporates a prescription for the formation, growth and destruction of dust (which is coupled to the cooling and chemical network as previously noted), using a model incorporating three grain species and two grain sizes \citep{trayford26}. The abundances of the s-process elements Ba and Sr, plus the r-process element Eu are tracked, and turbulent diffusion of the mass fractions of all elements and dust grains are modelled \citep{correa26}. As a result of these advancements, COLIBRE galaxies exhibit a multiphase ISM that includes much higher densities and lower temperatures than are possible in models that impose an effective equation of state for dense gas. 

Star formation is implemented using a Schmidt law, subject to a gravitational instability criterion, following \citet{nobels24}. Mass loss and enrichment from asymptotic giant branch (AGB) stars, core-collapse supernovae (CCSNe) and Type Ia supernovae (SNIa) are modelled using up-to-date nucleosynthetic stellar yields \citep{correa26}. Star particles inject energy into their environments through feedback processes, starting with stellar winds, radiation pressure, and the creation of H\textsc{ii} regions \citep{benitezllambay26}, followed by feedback from CCSNe and SNIa.

\subsection{Supernova feedback}
\label{sec:methods:sf}

Injection of feedback from CCSNe in COLIBRE is implemented using the method of \citet{chaikin23}. The majority of the CCSN energy is injected thermally and stochastically following \citet{dallavecchia12}, with the key parameters being the energy injected per supernova, $f_{\rm E}$ (in units of $10^{51}$ erg), and the temperature increment by which gas particles are heated, $\Delta T_{\rm SN}$. Ten percent of the CCSN energy injected in the form of low-velocity kicks, which are implemented by kicking pairs of particles in random, opposite directions with a target velocity of 50 km s$^{-1}$. This kinetic feedback stabilises the ISM against star formation by driving turbulence, whereas the thermal component of the feedback drives outflows.

The energy injected per supernova, $f_{\rm E}$, is a function of the thermal gas pressure at the birth of a given star particle, $P_{\rm birth}$:
\begin{equation}
\label{eq:fE}
    f_{\rm E} = f_{\rm E,min}\,+\,\frac{f_{\rm E,max}-f_{\rm E,min}}{1+\exp\left(-\frac{\log_{10}P_{\rm birth}-\log_{10}P_{\rm E,pivot}}{\sigma_{\rm P}}\right)},
\end{equation}
where $f_{\rm E,min}$ and $f_{\rm E,max}$ are the minimum and maximum energies (in units of $10^{51}$ erg) that can be injected by a single CCSN, the parameter $\sigma_{\rm P}$ governs the width of the transition from $f_{\rm E,min}$ to $f_{\rm E,max}$, and $P_{\rm E,pivot}$ is the birth pressure for which $f_{\rm E} = (f_{\rm E,min}+f_{\rm E,max})/2$. In the simulations presented here, $f_{\rm E,max}=4$, $\sigma_{\rm P}=0.3$, $f_{\rm E,min}$ is a calibrated parameter that increases with resolution, with values ranging from $0.1-0.8$, and $P_{\rm E,pivot}$ is also a calibrated parameter that varies with resolution in the range $8.0\times 10^3\,{\rm K\, cm}^{-3}<P_{\rm E,pivot}/k_{\rm B}<1.5\times 10^4\,{\rm K\, cm}^{-3}$, where $k_{\rm B}$ is the Boltzmann constant.

In EAGLE, which injected all CCSN energy following the method of \citet{dallavecchia12}, the value of $\Delta T_{\rm SN}$ was fixed at $10^{7.5}$~K to ensure that each star particle heated at least one gas particle in its lifetime. In COLIBRE, $\Delta T_{\rm SN}$ instead varies as a function of the gas density, $n_{\rm H}$:
\begin{equation}
\label{eq:deltaTSN}
    \Delta T_{\rm SN}(n_{\rm H})=10^{6.5}\,{\rm K}\,\left(\frac{n_{\rm H}}{n_{\rm H,pivot}}\right)^{2/3},
\end{equation}
where $n_{\rm H,pivot}$ is a free parameter that is calibrated for each COLIBRE model variant and resolution, with values ranging between $0.5-1.5$ cm$^{-3}$, and $\Delta T_{\rm SN}$ is limited to calibrated minimal and maximal values \citep[for details see \S 3.10 of][]{schaye26}. This variable temperature increment improves the sampling of feedback in low-density gas and ensures that overcooling is avoided in very dense regions; we will show that this is a highly impactful change for the halo gas content of lower-mass galaxies. A further improvement relative to EAGLE is that feedback is not injected in a weighted sense within the SPH kernels of star particles, but injected isotropically using the method of \citet{chaikin22}. This change reduces radiative cooling losses, as particle weighting (and hence mass weighting) biases energy injection into the densest regions. In Appendix \ref{sec:app:feedback} we show how the use of a variable $\Delta T_{\rm SN}$ and other features of COLIBRE's stellar feedback model influence our results.

\subsection{Black holes and AGN feedback}
\label{sec:methods:bh}

Supermassive black holes (BHs) are seeded in haloes exceeding a mass threshold, $M_{\rm FoF,seed}$, as identified by a friends-of-friends (FoF) halo finder which is periodically run on-the-fly during the simulation. In haloes exceeding $M_{\rm FoF,seed}=5\times 10^{10}$~M$_\odot$ (m7 resolution) or $1\times 10^{10}$~M$_\odot$ (m6 and m5 resolution), the densest gas particle in the halo is converted to a BH particle. Following \citet{dimatteo05}, BHs are seeded with a subgrid mass, $m_{\rm BH,seed}$, which is initially less than the true dynamical mass of the particle, and represents the BH mass for the purposes of subgrid accretion and feedback physics. The value of $m_{\rm BH,seed}$ is a calibrated parameter, varying between $2\times 10^{4}$~M$_\odot$ and $5\times 10^{5}$~M$_\odot$, and has some influence on the onset of rapid BH growth through accretion.

BHs can merge in the simulations if three conditions are met: the particles must be separated by less than 3 gravitational softening lengths, the less massive BH must be within the SPH smoothing length of the more massive BH, and the relative velocity between the particles must satisfy $\Delta v<\left(2G(m_{\rm BH,1}+m_{\rm BH,2})/r\right)^{1/2}$, where $m_{\rm BH,i}$ are the BH particle masses and $r$ is their spatial separation. The energy lost to gravitational waves during the merging process is subtracted from the rest mass energy of the descendant BH. To account for unresolved dynamical friction, BH particles are repositioned using the method of \citet{bahe22}; at each timestep, BH particles are moved to the location of the SPH neighbour within three gravitational softening lengths that has the lowest potential (excluding the potential of the BH itself). This repositioning is crucial for keeping BH particles at the centres of haloes; \citet{bahe22} showed that without it, BH growth is strongly suppressed, rendering AGN feedback less effective.

\subsubsection{The fiducial (thermal) model}
\label{sec:methods:bh:fiducial}

In the fiducial COLIBRE model, BHs grow via accretion of gas at the Bondi-Hoyle-Lyttleton (BHL) rate \citep{bondi52,hoylelyttleton39}, modified by corrections for the turbulence and vorticity of the surrounding gas from \citet{krumholz06}. This prescription is similar to EAGLE, which also used a BHL prescription, but differs in its accounting for the relative motion between BHs and their surrounding gas \citep[see][]{rosasguevara15}. BH accretion also differs from EAGLE in that super-Eddington rates are allowed, at up to 100 times the Eddington rate\footnote{This limit is imposed for numerical reasons and is very rarely reached in practice.}. Gas accretion causes the subgrid BH mass to increase at the rate
\begin{equation}
\label{eq:bhgrowthrate}
    \dot{m}_{\rm BH} = (1-\epsilon_{\rm r})\dot{m}_{\rm accr},
\end{equation}
where $\dot{m}_{\rm accr}$ is the accretion rate and $\epsilon_{\rm r}=0.1$ is the assumed radiative efficiency of the subgrid accretion disc. Accounting for the energy radiated away causes the particle (dynamical) mass to decrease at a rate given by
\begin{equation}
    \dot{m}_{\rm BH,part} = -\epsilon_{\rm r}\dot{m}_{\rm accr}.
\end{equation}
If the subgrid BH mass is greater than the particle mass following accretion, mass and momentum are transferred to the BH from the surrounding gas particles, weighted by the contribution of those particles to the SPH density at the BH's location. Gas neighbours with less than half of their initial mass are excluded from this process.

AGN feedback in the fiducial model is implemented purely thermally, based on the method of \citet{boothschaye09}. To prevent numerical overcooling, energy is not injected continuously into the gas surrounding the BH, but is instead accumulated until there is sufficient energy to heat a single gas particle by a pre-defined temperature increment $\Delta T_{\rm AGN}$. Energy is accumulated at a rate given by
\begin{equation}
    \dot{E}_{\rm AGN} = \epsilon_{\rm f}\epsilon_{\rm r}\dot{m}_{\rm accr}c^2,
\end{equation}
where $c$ is the speed of light and $\epsilon_{\rm f}$ is the fraction of the energy that couples to the surrounding gas. The parameter $\epsilon_{\rm f}$ affects the mass scale at which BHs are able to self-regulate their growth through feedback, and was chosen to produce realistic BH masses in present-day high mass galaxies; $\epsilon_{\rm f}=0.1$ is used at m7 resolution, and $\epsilon_{\rm f}=0.05$ is used at m6 and m5 resolutions. The energy per feedback event is
\begin{equation}
\label{eq:deltaE_thermal}
    \Delta E_{\rm AGN, thermal} = \frac{m_{\rm gas}k_{\rm B}\Delta T_{\rm AGN}}{(\gamma-1)\mu m_{\rm p}},
\end{equation}
where $m_{\rm gas}$ is the gas particle mass, $\gamma=5/3$ is the ratio of specific heats for a monatomic gas and $\mu=0.6$ is the mean molecular weight of ionized, primordial gas. This energy is injected into the SPH neighbour nearest the BH; multiple particles may be heated if enough energy has accumulated within a single timestep.

The value of $\Delta E_{\rm AGN, thermal}$ is determined by $\Delta T_{\rm AGN}$; in the fiducial EAGLE model, a fixed increment of $\Delta T_{\rm AGN}=10^{8.5}$~K was used, whereas in COLIBRE $\Delta T_{\rm AGN}$ scales with the BH mass,
\begin{equation}
\label{eq:deltaTAGN}
    \Delta T_{\rm AGN} = 10^9\,{\rm K}\,\left(\frac{m_{\rm BH}}{10^8\,{\rm M}_\odot}\right).
\end{equation}
This scaling is limited to $10^6$~K $\leq \Delta T_{\rm AGN}\leq T_{\rm AGN,max}$, where $T_{\rm AGN,max}=10^{9.5}$~K at m7 resolution and $10^{10}$~K at m6 and m5 resolution. This scaling greatly improves the sampling of AGN feedback events for low-mass BHs, which would otherwise need to accrue energy over long timescales to inject energy in a fixed-value implementation. The efficiency or `strength' of AGN feedback, in terms of its influence on the gas within galaxies and their haloes, is primarily determined by the value of $\Delta T_{\rm AGN}$, and hence this scaling also increases the efficiency of AGN feedback in the most massive haloes. The value of $\Delta T_{\rm AGN}$ and its scaling with $m_{\rm BH}$ are of crucial importance to our results; we show how this prescription and the choice of the variable $\epsilon_{\rm f}$ influence our results in Appendix \ref{sec:app:feedback}. 

\subsubsection{The hybrid (thermal/jet) model}
\label{sec:methods:bh:hybrid}

The COLIBRE suite also includes simulations performed with an alternative `hybrid' model for AGN feedback, which implements a combination of jets driven by kinetic feedback, and winds represented by the thermal feedback prescription detailed above. The balance of energy injection between these modes, and the efficiency in each mode, is set by a subgrid accretion disc model which tracks the BH spin and accretion disc state. The model assumes a magnetically-arrested disc \citep{narayan03} with jets powered by the \citet{blandfordznajek77} mechanism. This model is detailed in \citet{husko26}, and we provide only a summary of its key features here.

In the hybrid model, the BH growth rate (equation \ref{eq:bhgrowthrate}) is instead given by
\begin{equation}
    \dot{m}_{\rm BH} = (1-\epsilon_{\rm r}-\epsilon_{\rm jet}-\epsilon_{\rm wind})\epsilon_{\rm accr}\dot{m}_{\rm accr},
\end{equation}
where $\epsilon_{\rm jet}$ and $\epsilon_{\rm wind}$ are the efficiencies of jets and accretion disc winds respectively, and $\epsilon_{\rm accr}$ is the fraction of the mass accretion rate onto the disc that is actually accreted onto the BH. These efficiencies are functions of the BH spin, and (except for $\epsilon_{\rm r}$) the magnetisation of the disc, in a manner that varies between three accretion disc states. The active state depends on the Eddington ratio, $f_{\rm Edd}$: BHs are in the ``thick disc'' state for $f_{\rm Edd}<0.01$, the ``thin disc'' state for $0.01<f_{\rm Edd}<1$, and the ``slim disc'' state for $f_{\rm Edd}>1$. In general, kinetic jets are dominant in the thick disc state or when $f_{\rm Edd}>0.1$, otherwise thermal feedback dominates.

In all states, winds are driven by the thermal energy injection method detailed in \S\ref{sec:methods:bh:fiducial}. Jets are driven by kicking two particles in opposite, random directions, limited to within 7.5 degrees of the BH spin axis. Analogously to the scaling of $\Delta T_{\rm AGN}$ for thermal feedback, the kick velocity is scaled with the BH mass. Equations \ref{eq:deltaE_thermal} and \ref{eq:deltaTAGN} are therefore replaced respectively by:
\begin{equation}
    \Delta E_{\rm AGN,jet} = 2 \times \frac{1}{2}m_{\rm gas}v_{\rm jet}^2,
\end{equation}
and
\begin{equation}
\label{eq:vjet}
    v_{\rm jet} = v_{\rm jet,0}\left(\frac{m_{\rm BH}}{10^9\,{\rm M}_\odot}\right)^{1/2},
\end{equation}
where $v_{\rm jet}$ is the jet velocity, which is limited to $10^{2.5}$ km s$^{-1} \leq v_{\rm jet} \leq v_{\rm jet,0}$, with $v_{\rm jet,0}=10^{4.5}$ km s$^{-1}$.

\subsection{Identifying and characterising haloes}
\label{sec:methods:haloes}

Haloes and subhaloes are identified within the simulations using HBT-HERONS \citep{forouhar25}, an updated version of the Hierarchical Bound Tracing structure finder \citep[HBT+,][]{han18}. Self-bound haloes are identified as they emerge in the simulations by performing an iterative unbinding algorithm on particles in friends-of-friends (FOF) groups. The particles bound to a halo are then used to identify the descendant halo in the next simulation snapshot, and in each case, haloes are checked to determine whether they remain bound (as a central halo or a satellite subhalo within a more massive system), have merged with another halo, or have been disrupted. This ``history-based'' approach to structure finding gives more reliable identification of satellite haloes, and produces a complete merger tree with minimal ambiguity over the progenitor and descendant of any given halo.

Once haloes are identified, we use the Spherical Overdensity and Aperture Processor \citep[SOAP,][]{mcgibbon25} to produce a catalogue of halo and galaxy properties for each simulation. In this study, quantities relating to haloes, such as the total mass and gas mass, are calculated within spherical apertures of radius $r_\Delta$ that enclose a mean density equal to $\Delta\rho_{\rm crit}$, where $\rho_{\rm crit}$ is the redshift-dependent critical density of the Universe. Throughout this study, we use $\Delta=200$, except when comparing to observational constraints, where we use $\Delta=500$ to match the definition typically adopted for X-ray observations. We define the halo gas fraction, $f_{\rm gas}^{\Delta} \equiv M_{\rm gas}^{\Delta}/M_{\Delta}$, where $M_{\rm gas}^{\Delta}$ is the total gas mass within $r_\Delta$, and $M_{\Delta}$ is the total mass within this radius. We also define the stellar fraction, $f_{\star}^{\Delta} \equiv M_{\star}^{\Delta}/M_{\Delta}$, and total baryon fraction $f_{\rm b}^{\Delta} \equiv (M_{\rm gas}^{\Delta}+M_{\star}^{\Delta})/M_{\Delta}$, where $M_\star^\Delta$ is the total stellar mass within $r_\Delta$. We normalise each mass fraction by the cosmic baryon fraction, $f_{\rm b}^{\rm cosmic}=\Omega_{\rm b}/\Omega_{\rm M}$, where $\Omega_{\rm b}$ and $\Omega_{\rm M}$ are the density parameters for baryons and matter respectively. Quantities relating to galaxies, such as the stellar mass, are calculated from all bound particles within a 50 pkpc spherical aperture, consistent with previous analyses of COLIBRE \citep[e.g][]{schaye26,chaikin_calibration}. In all cases, apertures are centred on the most-bound particle in the halo. We only show results for central galaxies/haloes, and do not examine satellites in this study, as their halo gas fractions are not well defined.

\section{Results}
\label{sec:results}

In this section we present our results as follows. In \S\ref{sec:results:flagship} we present COLIBRE's predictions for the $z=0$ halo gas mass fraction as a function of halo mass, and examine how this relation changes with simulation resolution. In \S\ref{sec:results:obs} we compare COLIBRE's predictions with gas fractions inferred from X-ray surveys, and from baryonification models constrained jointly by kSZ and X-ray measurements. In \S\ref{sec:results:hybrid} we compare COLIBRE's fiducial thermal AGN feedback model to its `hybrid' thermal+jet model, and in \S\ref{sec:results:simcompare} we compare COLIBRE's gas fraction-halo mass relations with those produced by other galaxy formation models. Finally, in \S\ref{sec:results:histories} we examine how the gas fraction evolves with time in haloes of different mass.

\subsection{The baryon content of haloes in COLIBRE}
\label{sec:results:flagship}

\begin{figure}
\includegraphics[width=\columnwidth]{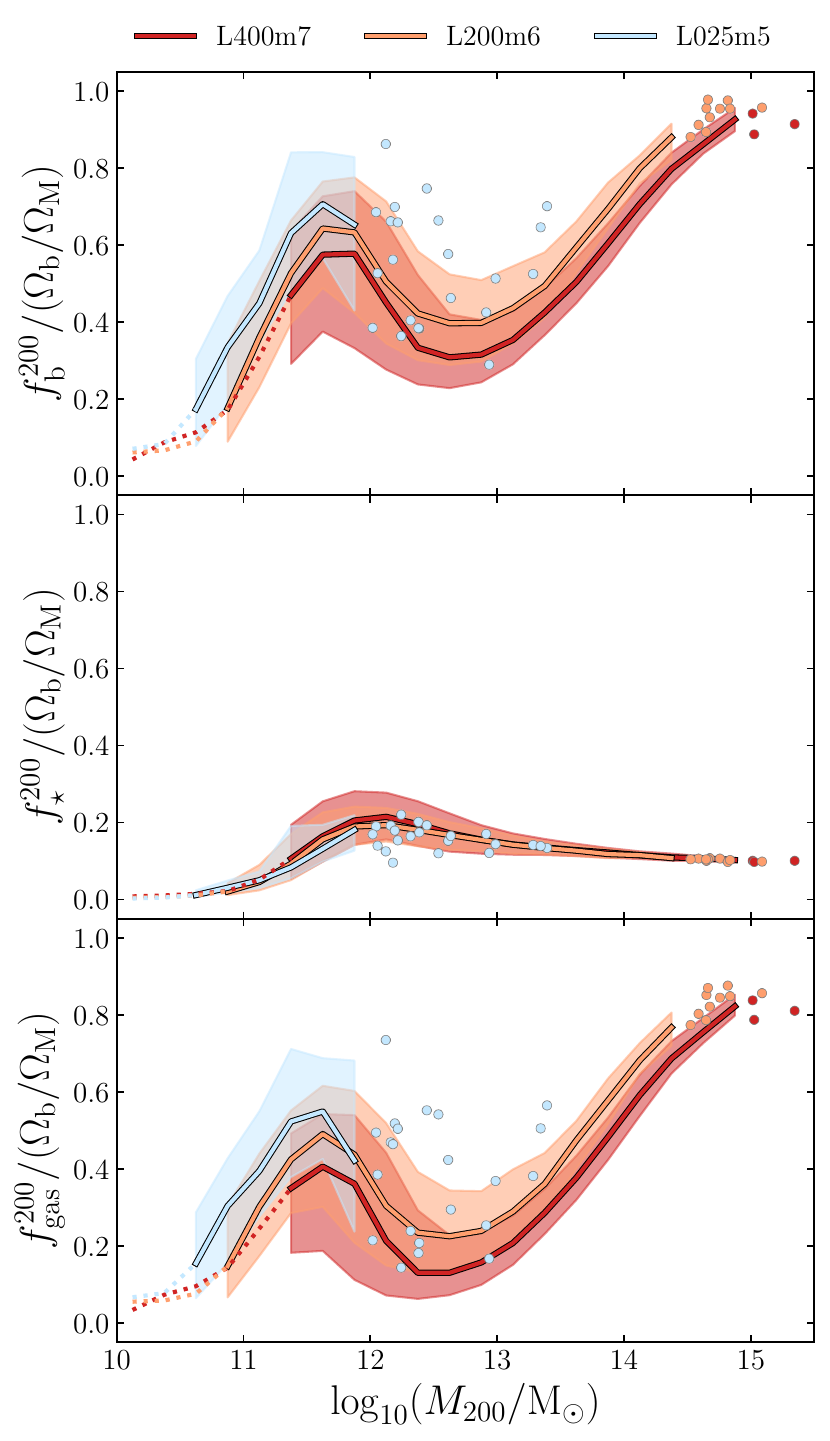}
\vspace{-4mm}
\caption{Halo baryon fractions ($f_{\rm b}^{200}$, upper panel), stellar fractions ($f_\star^{200}$, middle panel) and gas mass fractions ($f_{\rm gas}^{200}$, lower panel), within radius $r_{200}$, as a fraction of halo mass $M_{200}$, for the L400m7, L200m6 and L025m5 COLIBRE simulations at $z=0$. All fractions are normalised to the cosmic baryon fraction, $f_{\rm b}^{\rm cosmic}=\Omega_{\rm b}/\Omega_{\rm M}$. Solid lines show the median relation in bins of halo mass, and shading shows the 16$^{\rm th}$-84$^{\rm th}$ percentile scatter in each bin. We use dotted lines where the stellar mass of the central galaxy is below 100 times the initial baryonic particle mass, indicating that the galaxy is poorly resolved, and show individual data points where there are fewer than 10 haloes per mass bin. The relations have similar shapes in each simulation, though the baryon and gas fractions are typically higher at a given $M_{200}$ as resolution increases.}
\label{fig:flagship}
\end{figure}

We begin by exploring the relationship between halo baryon content and mass, $M_{200}$, at the present day. In \autoref{fig:flagship} we show the baryon fractions ($f_{\rm b}^{200}$, upper panel), stellar fractions ($f_\star^{200}$, middle panel) and gas mass fractions ($f_{\rm gas}^{200}$, lower panel) of haloes as a function of $M_{200}$ in the L400m7, L200m6 and L025m5 COLIBRE simulations, all run with the fiducial model (i.e. with purely thermally-driven AGN feedback). Solid lines indicate the median value in 0.25 dex wide bins across the mass range $10<\log_{10}(M_{200}/{\rm M}_\odot)<15.5$, and shading indicates the 16$^{\rm th}$-84$^{\rm th}$ percentile scatter. Following the convention used in earlier analyses of COLIBRE \citep[e.g.][]{schaye26}, at low masses we show only the median with dotted lines, for bins where the median stellar mass corresponds to less than 100$\times$ the initial baryonic particle mass, indicating that the galaxy is poorly resolved. In sparsely-sampled high-mass bins, we show individual data points for each halo where the number of galaxies in the bin is less than 10.

The $f_{\rm b}^{200}-M_{200}$ and $f_{\rm gas}^{200}-M_{200}$ relations closely resemble one another, as gas dominates the halo baryon content at all masses. Stars constitute a maximum of $\approx 20$ percent of $f_{\rm b}^{\rm cosmic}$ in haloes hosting Milky Way-like galaxies ($M_{200}\sim 10^{12}$~M$_\odot$), where galaxy formation is most efficient; here $f_{\rm gas}^{200}$ lies significantly below $f_{\rm b}$. In the haloes of dwarf galaxies ($M_{200}\lesssim 10^{11}$~M$_\odot$) $f_{\rm gas}^{200}$ is very similar to $f_{\rm b}^{200}$, whilst in galaxy groups and clusters ($M_{200}\gtrsim 10^{13}$~M$_\odot$) $f_{\rm gas}^{200}$ lies below $f_{\rm b}^{200}$ with a steadily decreasing offset as halo mass increases. Given that the baryon fractions and gas mass fractions exhibit such similar relationships with mass, we focus solely on the $f_{\rm gas}^{200}-M_{200}$ relation for the remainder of this study.

The relations shown in \autoref{fig:flagship} are non-monotonic, reflecting the changing physical processes that shape the gas fractions of haloes as their masses increase. In low-mass haloes ($M_{200}\lesssim 10^{11}$~M$_\odot$), feedback from SNe is readily able to both expel gas beyond the virial radius and prevent gas inflow at the halo scale. As a result, both the stellar and gas fractions (and, consequently, the total baryon fractions) of haloes are low in this mass range. Gas fractions increase as halo mass increases, reaching a peak in the $M_{200}\sim 10^{11.5-12}$~M$_\odot$ mass range, as feedback from SNe becomes less efficient. This increase likely occurs because the potential well becomes deeper, and because a gravitationally shock-heated hot halo develops, in which gas at the characteristic entropy of SN-driven outflows becomes less buoyant with increasing virial temperature \citep[see e.g.][]{bower17,mcalpine18}. This decreased efficiency of SN feedback also corresponds to an increase in $f_\star^{200}$ at this mass scale.

\begin{figure*}
\includegraphics[width=\textwidth]{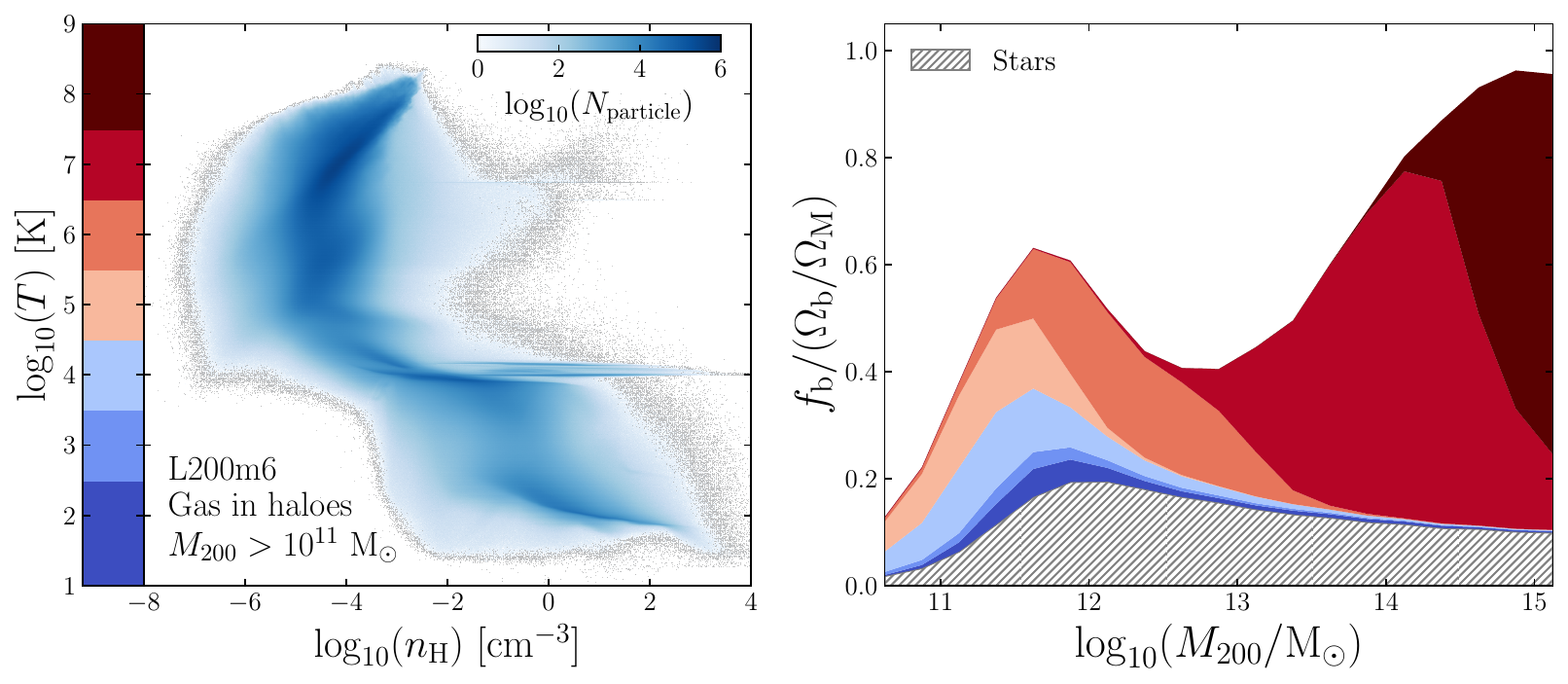}
\vspace{-4mm}
\caption{Left panel: the distribution of temperatures, $T$, and hydrogen number densities, $n_{\rm H}$, for gas particles bound to haloes with $M_{200}>10^{11}$~M$_\odot$ in the L200m6 simulation at $z=0$. The coloured regions highlighted along the temperature axis indicate the bins used in the right-hand panel, which shows the contribution of gas within these temperature bins to the mean $f_{\rm b}^{200}-M_{200}$ relation, in addition to the contribution from stars (shown with grey hatching). Results in the right panel are for all gas within $r_{200}$, including the ISM. Gas in the haloes of galaxies ($M_{200}\lesssim 10^{12.5}$~M$_\odot$) is multiphase, with gas at a variety of temperatures from 10 K to $10^{6.5}$~K contributing significantly to the total mass. Group and cluster haloes are dominated by gas at the virial temperature.}
\label{fig:temperature}
\end{figure*}

As SN-driven outflows become less efficient and gas is increasingly confined within the galaxy, the central SMBH can grow more readily, leading to AGN feedback that is more energetic and expulsive than SN feedback. At the onset of AGN feedback, $f_{\rm gas}^{200}$ starts to decline with increasing $M_{200}$, as the central SMBH grows with the halo and drives outflows capable of expelling gas beyond $r_{200}$. In Milky Way-mass haloes and up to the mass scale of galaxy groups ($M_{200}\sim 10^{12-13}$~M$_\odot$), halo gas fractions are at a minimum, as their SMBHs have typically begun self-regulating their growth \citep[e.g.][]{boothschaye10}. Above this mass scale, $f_{\rm gas}^{200}$ increases once more with $M_{200}$, as AGN-driven outflows become less efficient for the same reasons as SN-driven outflows do at lower mass \citep[e.g.][]{pontzen26}, and previously-evacuated haloes can either accrete fresh gas or re-accrete previously expelled material (see \S\ref{sec:results:histories}).

The $f_{\rm gas}^{200}-M_{200}$ relation exhibits significant scatter at fixed halo mass, particularly for haloes of Milky Way-mass and below. Previous work exploring the origin of scatter in simulations such as EAGLE and IllustrisTNG has demonstrated that for $\sim L^\star$ haloes, this scatter is principally driven by differences in the integrated amount of efficient AGN feedback injected, which is equivalent to (in EAGLE), or correlates with (in IllustrisTNG), the BH mass \citep{davies19,davies20,terrazas20}. In turn, scatter in the BH mass is driven by differences in halo binding energy and concentration \citep{boothschaye10,davies19,davies20}, and the galaxy's merger history \citep{davies22,davies24,roberts26}. Preliminary examination of the scatter in COLIBRE indicates that these factors also drive the scatter at $\sim L^\star$ in the COLIBRE model.

\subsubsection{Multiphase gas within galaxies and their haloes}
\label{sec:results:temperature}

In \autoref{fig:temperature} we show the contribution of gas at different temperatures to the baryon content of haloes in the L200m6 simulation, to illustrate the multiphase nature of the CGM in COLIBRE. The left-hand panel shows the distribution of temperatures, $T$, and hydrogen number densities, $n_{\rm H}$, for gas particles determined by HBT-HERONS to be bound to haloes with $M_{200}>10^{11}$~M$_\odot$. Several key gas phases are visible in this space:
\begin{itemize}
    \item The densest, coldest ($T<10^{2.5}$~K) gas, corresponding to the cold phase of the ISM. Gas in this phase has a high molecular hydrogen fraction and dust content. Modelling of this phase in a simulation of this volume is unique to COLIBRE.
    \item Gas approximately following the $T=10^4$~K isotherm at densities $10^{-3} \lesssim n_{\rm H}/$cm$^{-3} \lesssim 1$, corresponding to gas in thermal equilibrium for which photoheating is balanced by radiative cooling. This phase corresponds to cooler gas in the CGM, and has both low metallicity and a high atomic hydrogen fraction \citep[see fig. 9 of][]{schaye26}.
    \item Gas slightly above $T=10^4$~K and at densities $n_{\rm H}\gtrsim 10^{-2}$ cm$^{-3}$, corresponding to H{\sc ii} regions within the ISM.
    \item Gas at high temperatures ($T\gg 10^4$~K), and densities ($n_{\rm H}\gtrsim 10^{-2}$) corresponding to gas recently heated by feedback.
    \item Gas at intermediate temperatures ($10^5\lesssim T/{\rm K} \lesssim 10^7$) and low densities ($10^{-6} \lesssim n_{\rm H}/$cm$^{-3} \lesssim 10^{-4}$), corresponding to the shock-heated CGM, i.e. gas in the haloes of galaxies.
    \item Gas at very high temperatures ($T\gtrsim 10^7$~K), corresponding to the intragroup and intracluster media.
\end{itemize}
For a full description of these phases, see the discussion of fig. 8 in \citet{schaye26}. To illustrate the contribution of each of these phases to the gas content of galaxies and their haloes, we define five bins in $\log_{10} T$ that are 1 dex wide from $10^{2.5}$~K to $10^{7.5}$~K, plus two further bins to capture temperatures above and below these limits. These bins are illustrated with colours along the temperature axis of the left panel of \autoref{fig:temperature} to facilitate identification of the corresponding gas phase.

In the right panel of \autoref{fig:temperature} we show the contribution of gas in each of these temperature bins to the $f_{\rm b}^{200}-M_{200}$ relation in the L200m6 simulation, by colouring the area under the relation corresponding to each bin. We include the contribution from stars with grey hatching. The upper limit of the coloured regions is similar, but not equal to, the median curve for the L200m6 simulation in the upper panel of \autoref{fig:flagship}; here we find the gas mass (within $r_{200}$) corresponding to each temperature bin, sum over all haloes in each mass bin, divide by the total halo mass within the mass bin, and normalise to the cosmic baryon fraction $\Omega_{\rm b}/\Omega_{\rm M}$. The upper limit here therefore corresponds to the mean $f_{\rm b}^{200}-M_{200}$ relation once the stellar component is included. 

Gas in the haloes\footnote{This definition includes gas in the ISM.} of galaxies ($M_{200}\lesssim 10^{12.5}$~M$_\odot$) is clearly multiphase, with gas in each temperature bin contributing significantly to the overall gas content. In particular, cold gas ($T\lesssim10^{4.5}$~K) constitutes a large fraction of the total gas mass (a maximum of 50\% at $M_{200}\approx 10^{11.4}$~M$_\odot$). Whereas EAGLE (and other contemporary large-volume simulations) impose methods to prevent dense gas from cooling below $\sim 10^{3.5}$~K, in COLIBRE we can examine the contribution of gas down to $T=10$~K. The coldest gas at $T<10^{2.5}$~K, corresponding to the cold ISM, contributes a maximum of 11\% of the halo gas content at $M_{200}\approx 10^{11.6}$~M$_\odot$. The remaining hotter, shock-heated gas is present at progressively higher temperatures as $M_{200}$ (and hence the virial temperature) increases. The haloes of galaxy groups and clusters contain a very small fraction of cold gas, and are overwhelmingly dominated by hot gas at approximately the virial temperature.

In Appendix \ref{sec:app:multiphase} we show how this breakdown of gas temperatures varies between simulations of different resolution, and with different AGN feedback models. The relative balance of gas temperatures is broadly consistent across the different resolutions and models.

\subsubsection{Convergence with resolution}
\label{sec:results:convergence}

Unlike models such as BAHAMAS \citep{mccarthy17}, FLAMINGO \citep{schaye23,kugel23} and (X-)FABLE \citep{henden18,bigwood25xfable}, the subgrid prescriptions in COLIBRE and the parameter values used in those prescriptions were not calibrated at any resolution to reproduce observed constraints on halo gas fractions. As such, it is not surprising that the $f_{\rm gas}^{200}-M_{200}$ relation is not perfectly converged with resolution. The three curves in \autoref{fig:flagship} show relations produced at COLIBRE's three resolution levels, with baryonic (and dark matter) particle masses of $\sim 10^7$~M$_\odot$ (m7), $\sim 10^6$~M$_\odot$ (m6), and $\sim 10^5$~M$_\odot$ (m5); in general, as the mass resolution increases, $f_{\rm gas}^{200}$ also increases at a given $M_{200}$. Haloes in the L200m6 simulation are somewhat more gas rich than those in the L400m7 simulation, and while the small volume of the L025m5 simulation yields few haloes of Milky Way-mass and above, low-mass haloes in this simulation are systematically more gas rich than in the L200m6 simulation. There is however large overlap in the scatter between all resolution levels. \autoref{fig:flagship} shows that the stellar fractions are well converged with resolution, and so the lower gas fractions seen at lower resolution are not the result of the conversion of more halo gas into stars; the halo baryon fractions exhibit a similar trend with resolution to the gas mass fractions. A small exception to this occurs at $M_{200}\sim 10^{12}$~M$_\odot$, where $f_\star^{200}$ increases slightly with resolution (see also Appendix \ref{sec:app:multiphase}), and so $f_{\rm gas}^{200}$ decreases more with resolution than $f_{\rm b}^{200}$ does, as more stars are formed.

The differences in the ability of feedback to entrain and expel gas from haloes (and prevent inflows) at different resolution levels can be understood by considering how energy injection and cooling losses depend on resolution in COLIBRE. The two channels through which the fiducial COLIBRE model primarily drives halo-scale outflows are CCSNe in low-mass haloes, and AGN in high-mass haloes; in both cases outflows are principally launched through stochastic heating of nearby particles by a temperature increment, $\Delta T$. 

We first consider how the effects of AGN feedback are influenced by resolution. As resolution increases, heating a single particle by a given temperature increment $\Delta T$ requires less energy as a smaller mass is being heated; heating events are thus less individually energetic, but more frequent. The densities of gas particles heated by AGN are higher as resolution increases \citep[see fig. 11 of][]{schaye26}, leading to greater radiative losses since the radiative cooling rate scales with the square of the density. As a result of these two effects, feedback events become less energetic, and more prone to cooling losses, as resolution increases. Conversely, feedback events are more energetic and less frequent at lower resolution, leading to stronger outflows that both expel halo gas and prevent inflows more effectively, yielding the lower gas fractions at a given mass shown in \autoref{fig:flagship}.

To fully understand the influence of resolution, we must also consider how the parameters of the AGN feedback model are recalibrated between resolutions. The temperature increment $\Delta T$ scales with the BH mass (equation \ref{eq:deltaTAGN}), but crucially this scaling is not changed with resolution; only the maximum possible $\Delta T$ is adjusted between resolutions (to higher values at higher resolution). Thus, for all but the most massive BHs, $\Delta T$ is unchanged between resolution levels. The higher maximum $\Delta T$ used in the L200m6 simulation does allow BHs in group-mass haloes and above to attain higher heating temperatures (by up to 0.5 dex) than in the L400m7 simulation at late times, as we show in Appendix \ref{sec:app:dTAGN}. However, this does not appear to counteract the difference in feedback efficiency between resolutions, as the gas fractions are higher in this mass range in the L200m6 simulation. The AGN feedback coupling efficiency $\epsilon_{\rm f}$ is higher at lower resolution, however this difference does not result in lower gas fractions. We show in Appendix \ref{sec:app:feedback} that at fixed resolution, this parameter has little effect on the gas fraction at higher masses, and at lower masses a higher $\epsilon_{\rm f}$ in fact yields higher gas fractions. This occurs because a higher $\epsilon_{\rm f}$ causes BHs to self-regulate at lower masses, and hence lower $\Delta T$, making feedback less effective. The seed mass for BHs is also higher at lower resolution, and as a result lower-mass galaxies (with stellar mass $M_\star<10^{10}$ M$_\odot$) host more massive BHs \citep[see fig. 16 of][]{chaikin_calibration}. This difference contributes to more effective AGN feedback and lower gas fractions for low-mass haloes in the L400m7 simulation, but is not a significant effect at higher masses.

Differences in the influence of CCSN feedback on $f_{\rm gas}^{200}$ between resolution levels are less straightforward to ascertain; for CCSN feedback, $\Delta T$ scales with the local gas density (to the power $2/3$, equation \ref{eq:deltaTSN}), but with a calibrated pivot density that varies between resolutions. The pivot pressure in the energy injected per CCSN ($f_{\rm E}$, equation \ref{eq:fE}) is also a calibrated parameter that varies between resolutions. The densities of gas heated by CCSN feedback are, however, better converged with resolution than for AGN feedback \citep[see fig. 11 of][]{schaye26}. In lower-mass ($M_{200}\lesssim 10^{11.5}$~M$_\odot$) haloes for which CCSN feedback is more important than AGN feedback, convergence in $f_{\rm gas}^{200}$ between m7 and m6 resolution levels is better than at higher masses, though $f_{\rm gas}^{200}$ is higher at m5 resolution at fixed $M_{200}$.

The calibrated subgrid parameters at each resolution level have proven effective in injecting an amount of energy (after radiative losses) that results in a self-regulating balance of inflow and outflow onto galaxies that leads to a realistic GSMF, size-mass relation, and BH mass-stellar mass relation at all resolution levels \citep{schaye26,chaikin_calibration}. However, the precise amount of the gas that is expelled beyond $r_{200}$ over the course of achieving this self-regulating balance appears to vary. The fact that the same galaxy properties can be reached with significant differences in the halo gas fractions is not surprising, given that many simulation models produce realistic galaxies with very different $f_{\rm gas}^{200}-M_{200}$ relations \citep[\S\ref{sec:results:simcompare}, see also][]{crainvandevoort23}. 

\subsection{Comparison with observational constraints}
\label{sec:results:obs}

In \autoref{fig:flagship_xray_ksz}, we assess how well the fiducial COLIBRE model reproduces the best available constraints on the $f_{\rm gas}^{500}-M_{500}$ relation, as inferred from X-ray observations and from baryonification models fitted to kSZ and X-ray measurements. Note that in this section, we now show quantities measured within $r_{500}$ to facilitate these comparisons. The curves and shading are shown in the same fashion as \autoref{fig:flagship}, though we do not show the L025m5 simulation as it contains very few haloes in the mass range probed by observational data. Since X-ray and kSZ measurements trace the properties of hot gas, we also show median $f_{\rm gas}^{500}-M_{500}$ relations for hot ($T>10^6$~K) gas with dashed lines\footnote{For clarity, we do not show the hot gas fraction separately for the individually plotted clusters; the difference from the total gas fraction is small.}. We adopt a temperature cut of $T>10^6$~K as it is appropriate for comparisons with eROSITA observations, which are sensitive down to energies of $\approx 0.2$~keV. In the relevant mass range ($M_{200}>10^{12.5}$~M$_\odot$) there is little difference between the curves, as most of the gas is hot and at approximately the virial temperature (see \autoref{fig:temperature}). In Appendix \ref{sec:app:xraytemp}, we show how the $f_{\rm gas}^{500}-M_{500}$ relation for hot gas varies for different choices of temperature cut.

\subsubsection{X-ray observations pre- and post-eROSITA}
\label{sec:results:obs:xray}

\begin{figure}
\includegraphics[width=\columnwidth]{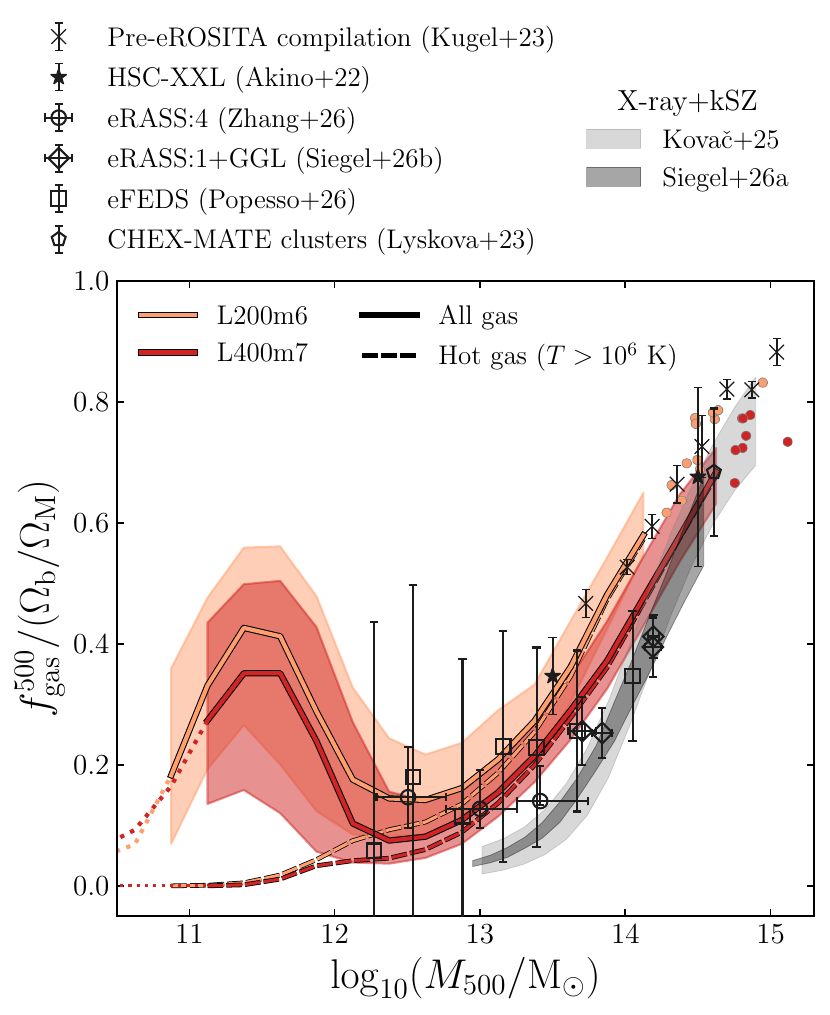}
\vspace{-4mm}
\caption{The present-day $f_{\rm gas}^{500}-M_{500}$ relation in the L400m7 and L200m6 simulations, compared with observational constraints. Data points show pre-eROSITA X-ray data compiled by \citet{kugel23}, including their two data points derived from \citet{akino22}, and the results of eROSITA stacking analyses by \citet{zhang26}, \citet{popesso26} and \citet{lyskova23}, plus data from low-redshift X-ray detected groups in eRASS:1 with halo masses obtained through galaxy-galaxy lensing \citep{siegel26b}. Uncertainties are as published except for the \citet{zhang26} data, for which we show the range of values they provide upon making different assumptions in their analysis. Grey shading shows 16$^{\rm th}$-84$^{\rm th}$ percentile confidence intervals from baryonification models jointly fitted to X-ray and kinetic Sunyaev-Zel'dovich (kSZ) effect measurements \citep{kovac25,siegel26a}. Simulation results are shown in the same fashion as in previous figures; since the observations trace the hot gas phase, we also show the median hot gas fractions ($T>10^6$~K) with dashed lines. The gas fractions of groups and clusters in the L200m6 simulation match those derived from pre-eROSITA X-ray data well, but are high relative to constraints from eROSITA X-ray studies and baryonification models. The lower gas fractions in the L400m7 simulation match the CHEX-MATE cluster data well, but are still high for galaxy groups relative to eROSITA and baryonification constraints. Both simulations are broadly consistent with constraints for $M_{500}\lesssim 10^{13}$~M$_\odot$.}
\label{fig:flagship_xray_ksz}
\end{figure}

The total mass of hot gas within haloes can be inferred from X-ray observations by converting X-ray surface brightness profiles into density profiles, after making assumptions about the temperature and metallicity of the gas, and integrating the density profile out to (in our case) $r_{500}$. If the halo mass is known, the result can then be converted into a gas fraction. This can be done for individual haloes, as is the case for the data from \citet{kugel23}, \citet{akino22} and \citet{siegel26b} shown in \autoref{fig:flagship_xray_ksz}, or for stacked X-ray surface brightness profiles of many haloes within a mass window, as is the case for the other datasets. The methods used for selection/stacking, and the assumptions made in obtaining a density profile, vary significantly between datasets.

The data from \citet{kugel23} are a collection of pre-eROSITA observations of 310 unique groups and clusters at $z<0.25$, compiled for the purpose of calibrating the FLAMINGO model on $f_{\rm gas}^{500}$. Gas masses and total masses were obtained by fitting density and temperature profiles to the X-ray data under the assumption of hydrostatic equilibrium, applying a constant correction factor to the total mass to account for hydrostatic mass bias. Data points show the median values in each mass bin, and the uncertainties are 16$^{\rm th}$-84$^{\rm th}$ percentiles obtained by bootstrap resampling the median. The uncertainties are purely statistical, and observational uncertainty is not propagated into the error bars. We supplement these data with the two data points derived by \citet{kugel23} from \citet{akino22}, who combined XMM-Newton observations with weak lensing-derived mass measurements for 136 groups and clusters. At m6 resolution, the $f_{\rm gas}^{500}-M_{500}$ relation produced by COLIBRE is in excellent agreement with these data. The COLIBRE model was not explicitly calibrated to achieve this, so this agreement was not guaranteed, and indeed the lower-resolution m7 simulations produce lower gas fractions that fall below the data.

The remainder of our observational constraints from X-ray data are derived from the eROSITA All-Sky Survey \citep{merloni12,merloni24}. Each dataset differs significantly in its overall strategy and selection function, though they collectively suggest that the gas fractions are lower for groups and lower-mass clusters than pre-eROSITA data indicated. This preference for lower gas fractions in galaxy groups and clusters became apparent in the first release of the eROSITA All-Sky Survey \citep[eRASS:1, see][]{bulbul24,dev24}, though analysis of the eROSITA Final Equatorial Depth survey \citep[eFEDS,][]{liu22}, which has smaller coverage but 8$\times$ the depth, has demonstrated that eRASS:1 has very poor X-ray completeness for $M_{500}<10^{14}$~M$_\odot$, with the majority of groups remaining undetected \citep{popesso24a,marini24}. This was attributed to eROSITA's selection function being biased against haloes with low surface brightness and high core entropy, such as those in which AGN feedback has expelled a large fraction of the halo's baryons, and against cool-core clusters, which are often misclassified as point sources \citep[e.g.][]{balzer25}.

\citet{popesso26} attempt to mitigate this issue by using the deeper eFEDS sample for greater completeness. They stack eFEDS data for optically selected systems from GAMA \citep{robotham11,driver22}, for which optical completeness is good, and for which the total group optical luminosity can be used as a halo mass proxy \citep{marini25a}. Assuming gas temperatures obtained from the $M_{500}-T$ relation \citep{lovisari15} and a metallicity of $0.3Z_\odot$, where $Z_\odot$ is the solar metallicity, they obtain the gas fractions shown in \autoref{fig:flagship_xray_ksz}. The principal source of uncertainty in this approach lies in the accuracy of using total group luminosity as a proxy for halo mass, and in subsequently using this halo mass to infer the gas temperature. \citet{popesso25} used mock observations of simulated haloes to show that the choice of optical group finder can cause downstream differences in emissivity of 10-40\% from the true value, and differences in metallicity can induce further variations of 37\%. Optical group finders may also be contaminated by $>30\%$ \citep{seppi25}. As shown in \autoref{fig:flagship_xray_ksz}, for $M_{500}>10^{13}$~M$_\odot$, COLIBRE shows agreement with the eFEDS data at both resolution levels in all but the highest-mass bin, where the L400m7 relation agrees only tenuously and the L200m6 relation is too high. The gas fraction is poorly constrained below this mass, where the X-ray completeness is near-zero even at the depth of eFEDS \citep{marini24}.

\citet{siegel26b} aim to circumvent uncertainties in halo mass by binning X-ray detected haloes from eRASS:1 in halo mass, and obtaining new calibrated halo masses for the binned haloes using galaxy-galaxy lensing data. They attempt to mitigate selection effects by considering only more massive ($M_{500}>10^{13.3}$~M$_\odot$) haloes at low redshift ($z<0.2$). As seen in \autoref{fig:flagship_xray_ksz}, their results are in agreement with those of \citet{popesso26}, but with tighter constraints on the gas fraction that lie below the predicted relations from COLIBRE's fiducial model.

In \autoref{fig:flagship_xray_ksz} we also show results from analysis of eRASS:4 by \citet{zhang24}, who stacked X-ray profiles for samples derived from the Sloan Digital Sky Survey \citep[data release 7,][]{strauss02} and DESI Legacy Imaging Survey \citep[data release 9,][]{dey19} in bins of halo mass; the masses in question being derived from the calibrated group-finding algorithm of \citet{tinker21}. \citet{zhang26} convert these X-ray profiles to density profiles and then to gas fractions, assuming the gas follows a beta profile with a fixed slope and core radius, is isothermal with a fixed temperature, and has fixed metallicity (0.3~$Z_\odot$). These results are indicated by the data points in \autoref{fig:flagship_xray_ksz}. They also examine how the predicted gas fractions are sensitive to their methodology, by varying the shapes of the density, temperature and metallicity profiles, or varying the assumed halo mass; they show that the uncertainty introduced by these choices is much larger than the statistical uncertainty. The error bars in \autoref{fig:flagship_xray_ksz} for these results therefore denote the maximum and minimum gas fractions found across all variations of their analysis (except for the assumption of solar metallicity). The gas fractions in COLIBRE broadly agree with these constraints in the lower two mass bins, though the higher gas fractions in L200m6 are in better agreement at lower mass, and the lower gas fractions in L400m7 are preferred at higher mass. Their highest-mass constraint, however, suggests that COLIBRE's groups may be too gas-rich at $M_{500}\sim 10^{13.5}$~M$_\odot$.

Finally, for the galaxy cluster mass regime we show the value derived by \citet{popesso26} from the eROSITA stacking analysis of \citet{lyskova23}, who stacked 38 clusters from CHEX-MATE \citep{chexmate}, which is a sample of clusters at $z<0.2$ with masses $2\times 10^{14}<M_{500}/{\rm M}_\odot<9\times 10^{14}$ detected by {\it Planck} through the Sunyaev-Zel'dovich effect. In common with other eROSITA-derived measurements at lower mass, this value indicates lower gas fractions than the pre-eROSITA data from \citet{kugel23}, but with a smaller offset than is seen at $M_{500}\sim 10^{14}$~M$_\odot$. COLIBRE's cluster gas fractions at m7 resolution are in excellent agreement with this constraint, whilst at m6 resolution the few clusters present in the smaller L200m6 simulation are at the upper end of the uncertainty on the constraint.

\subsubsection{Baryonification models constrained by X-ray and kinetic Sunyaev-Zel'dovich effect measurements}
\label{sec:results:obs:ksz}

Study of the kinetic Sunyaev-Zel'dovich (kSZ) effect has emerged as a complementary method for assessing the properties of gas in and around dark matter haloes, benefitting from a direct proportionality to the electron number density (as opposed to the square of the density as in X-ray observations, which also depend sensitively on temperature and metallicity). While X-ray observations are biased to the most dense gas at the centres of haloes, and are best suited to studying the impact of feedback in the most massive haloes at low redshift, the kSZ effect probes the larger-scale gas distribution and is best suited to constraining feedback in galaxy groups at higher redshift \citep[e.g][]{luciesmith25}. With the advent of high-resolution CMB experiments such as the Atacama Cosmology Telescope \citep[ACT,][]{thornton16,naess20}, stacked kSZ measurements can now be used to produce gas profiles \citep{schaan21,hadzhiyska25b,riedguachalla25,roper25}, and the kSZ signal has indicated a stronger suppression of the matter power spectrum than is predicted by most cosmological, hydrodynamical simulations \citep{schneider22,bigwood24,bigwood25kSZ,mccarthy25,hadzhiyska25a,riedguachalla25,kovac25,siegel26a,siegel26b}.

Performing a completely robust and like-for-like comparison of the kSZ signal in COLIBRE with observational data is beyond the scope of this work \citep[see e.g.][for comparisons with the FLAMINGO simulations]{mccarthy25,siegel26b,bigwood25kSZ}. Instead, we make use of constraints on the $f_{\rm gas}^{500}-M_{500}$ relation provided by the baryonification framework, which uses kSZ and X-ray measurements to constrain the radial density profiles of gas, dark matter and stars \citep{schneiderteyssier15,girischneider21,schneider25}. We show the results of two applications of this framework by \citet[][light grey shading]{kovac25}, and \citet[][dark grey shading]{siegel26a} in \autoref{fig:flagship_xray_ksz}, where the shading indicates the 16$^{\rm th}$-84$^{\rm th}$ percentile confidence interval for the fit as a function of halo mass. Both studies supplemented kSZ data with X-ray data in order to constrain the suppression of the matter power spectrum, but differ in the data products used; \citet{kovac25} use eFEDS X-ray-derived gas masses from \citet{popesso26} and assume hydrostatic equilibrium to derive halo masses, whilst \citet{siegel26a} use eRASS:1 data from \citet{siegel26b}, including galaxy-galaxy lensing-calibrated halo masses.

These constraints are consistent with those derived from eROSITA data alone at $M_{500}\sim 10^{14}$~M$_\odot$, indicating that haloes of this mass have lower gas fractions than predicted by COLIBRE. At high mass, both X-ray+kSZ constraints are consistent with both the CHEX-MATE cluster stacks and COLIBRE at m7 resolution. At lower masses, however, the baryonification prediction suggests that the gas fraction is even lower than is predicted by eFEDS or eRASS:4 data alone, giving a larger discrepancy with COLIBRE's predictions. 

\subsubsection{Differences between observational constraints on the gas fraction}
\label{sec:results:obs:summary}

The degree to which the COLIBRE model produces realistic halo gas mass fractions is challenging to assess, as current observational constraints on the gas fraction (which are only available for galaxy groups and clusters) are not mutually consistent, for reasons that are not yet fully understood. Gas fractions inferred from stacked eROSITA X-ray profiles are lower than gas fractions inferred from Chandra and XMM-Newton X-ray data for samples of individually analyzed groups and clusters. Baryonification models employing both eROSITA and kSZ constraints appear to reinforce this result, though kSZ measurements are sensitive to gas at larger radii and higher redshift than X-ray measurements \citep{luciesmith25}. This inconsistency has been attributed to selection effects stemming from a bias towards more X-ray luminous systems in pre-eROSITA studies \citep[e.g][]{marini24} and led to suggestions that models with stronger AGN feedback should be favoured to match these newer constraints. 

However, \citet{eckert26} showed that the X-ray luminosities ($L_{\rm X}$) of a sample of groups observed with XMM-Newton, and the optically-selected eROSITA stacks of \citet{popesso25xray}, are both consistent with the fiducial FLAMINGO simulation, which was calibrated to match pre-eROSITA gas fractions. These luminosities are not compatible with the FLAMINGO model variant that employs the strongest AGN feedback, which is favoured by the {\it gas fractions} inferred from eROSITA data \citep[e.g.][]{siegel26a}. To eliminate the uncertainty introduced by the halo mass estimate, they examined the $L_{\rm X}-T$ scaling relation, finding an even stronger preference for the fiducial FLAMINGO model. \citet{eckert26} therefore attribute the discrepancy between pre- and post-eROSITA constraints on the gas fractions of groups to uncertainties in the conversion from X-ray surface brightness profiles to gas fractions. As discussed in \S\ref{sec:results:obs:xray}, this conversion is highly sensitive to the halo mass estimate and the assumed gas temperature and metallicity profiles; a further point of consideration is that many optically selected galaxy groups have not yet virialised within a common halo and have lower gas temperatures than assumed.

An alternative approach to this problem is to forward-model the X-ray signal from simulated haloes through realistic observational pipelines \citep[e.g.][]{oppenheimer20,shreeram25,grayson25}, facilitating like-for-like comparisons with observational constraints that avoid the need to derive gas fractions from X-ray data. In such an analysis, \citet{seppi26} constructed analogues of the XMM-Newton Group AGN project \citep[X-GAP,][]{eckert24} sample from the FLAMINGO simulations, forward modelling the full selection function and producing mock observations. Comparison of these simulated analogues with the X-GAP data revealed a preference for slightly stronger feedback than the fiducial FLAMINGO model, but ruled out the FLAMINGO model with the strongest AGN feedback at $>4\sigma$ significance. Therefore, whilst models employing particularly strong AGN feedback may appear to agree well with the gas fractions suggested by analysis of eROSITA data (as shown in \autoref{fig:flagship_xray_ksz}), they may produce poor agreement with the X-ray emission itself, once selection effects and observational systematics are accounted for.

These lines of evidence indicate that the gas fractions of galaxy groups (and less massive haloes) are not yet well constrained observationally, and as such, we refrain from assuming that any particular observational dataset shown in \autoref{fig:flagship_xray_ksz} is more reliable than another. Overall, COLIBRE shows an encouraging level of agreement with these constraints, which the model was not calibrated to match. We show in the following section that across COLIBRE's various available resolution levels and model variations, there is at least one simulation that provides a good match to any given dataset shown in \autoref{fig:flagship_xray_ksz}, and in \S\ref{sec:results:simcompare} we show that COLIBRE exhibits similar or better agreement with the data than a range of prominent galaxy formation models.

\subsection{Influence of the AGN feedback prescription}
\label{sec:results:hybrid}

\begin{figure*}
\includegraphics[width=\textwidth]{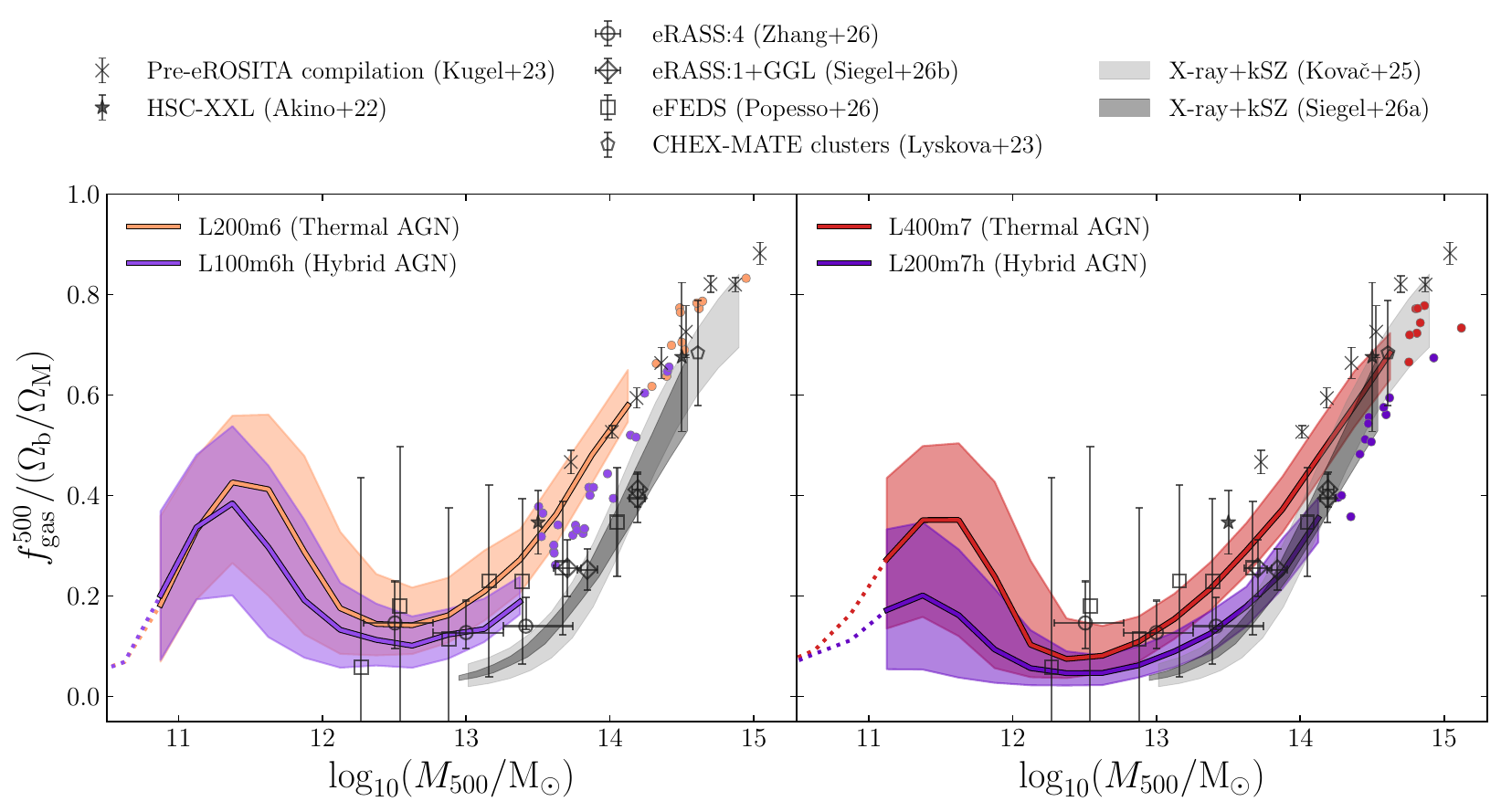}
\vspace{-4mm}
\caption{Influence of the AGN feedback prescription on the present-day $f_{\rm gas}^{500}-M_{500}$ relation in COLIBRE, at m6 (left panel) and m7 (right panel) resolution levels. We compare the fiducial L200m6 and L400m7 simulations, run with purely thermally-driven AGN feedback, to the largest available simulations at the same resolution that were run with hybrid thermal/jet AGN feedback, L100m6h and L200m7h. Simulation results and observational constraints are shown in the same fashion as in previous figures. The hybrid AGN feedback model yields lower gas fractions than the fiducial model at fixed mass resolution. The L200m7h simulation, which features COLIBRE's strongest feedback, produces group gas fractions in good agreement with constraints from eROSITA stacks and from baryonification models fitted to kSZ and X-ray data, but significantly lower gas fractions than indicated by pre-eROSITA X-ray data.}
\vspace{-4mm}
\label{fig:hybrid}
\end{figure*}

As detailed in \S\ref{sec:methods:bh:hybrid}, a second model for AGN feedback is available in COLIBRE, which distinguishes between several possible accretion and feedback states for SMBHs based on their Eddington ratios and spins. Unlike in the fiducial model, in all states feedback is injected through two channels: thermal feedback, using the same mechanism as in the fiducial model, and bipolar jets, implemented through a kinetic feedback prescription. The efficiencies and fractional contributions of each channel vary according to the accretion state, Eddington ratio, and BH spin. As is the case for the fiducial model, for thermal feedback the AGN heating temperature increases with the BH mass to ensure consistent sampling of feedback events, and similarly the jet velocity is scaled such that the kick energy scales linearly with the BH mass.

In \autoref{fig:hybrid} we compare the $f_{\rm gas}^{500}-M_{500}$ relations produced by the fiducial (thermal) and hybrid AGN feedback models in COLIBRE. In the left-hand panel we compare simulations at m6 resolution, L200m6 and L100m6h, and in the right-hand panel we compare simulations at m7 resolution, L400m7 and L200m7h. All data are shown in the same fashion as in previous figures.

At m6 resolution, the gas fractions are similar for both models below $M_{500}=10^{11.3}$~M$_\odot$; this is the mass regime in which BH masses are typically low, and hence AGN feedback is not yet significantly influencing the halo. At higher masses, the gas fractions are consistently lower for the hybrid model than for the fiducial model, albeit with significant overlap in the scatter. The difference appears to diminish towards the highest-mass haloes found in the L100m6h simulation ($M_{500}\sim10^{14}$~M$_\odot$), and the three most massive clusters appear to have gas fractions consistent with those produced by the fiducial model. The lower gas fractions for galaxy groups yield a slightly better agreement with those suggested by eROSITA stacks and baryonification models fitted to X-ray and kSZ data, but poorer agreement with pre-eROSITA measurements. It also appears that the onset of expulsive AGN feedback, and hence the turnover in the gas fraction, occurs at slightly lower mass in the hybrid model; we discuss possible reasons for this later in this section.

At m7 resolution, the hybrid model yields lower gas fractions than the fiducial model, and the difference is more significant than for m6. For $M_{500}\gtrsim 10^{13.5}$, the 16th-84th percentile scatter does not overlap, and the hybrid gas fractions are significantly lower; they remain lower even in $M_{500}=10^{14}$~M$_\odot$ clusters, where the difference in gas fractions is negligible at m6 resolution. The gas fractions produced by the hybrid model at m7 resolution show good agreement with constraints from eROSITA at the mass scales of galaxy groups, and with the CHEX-MATE cluster stacks, but are low in comparison to pre-eROSITA X-ray constraints.

The lower gas fractions in the hybrid AGN model may be the result of increased energy input, as the hybrid model injects $\approx 50\%$ more energy for $z<5$ than the fiducial model, decreasing to $\approx 25\%$ overall by $z=0$ \citep{husko26}. The directionality of the feedback is also important as it affects how much of the AGN energy couples to the CGM. Purely isotropic implementations of AGN feedback can produce jet-like, anisotropic behaviour, as outflows will naturally escape along the path of least resistance along a galaxy's minor axis; this is the case for simulations injecting energy thermally or kinetically \citep[e.g.][]{truong21,nica22,silich25}. However, the fraction of the outflow driven into the dense ISM is likely to be impeded and will not drive halo-scale outflows. The bipolar jets in the hybrid AGN model can invest all their energy into driving outflows along the path of least resistance if the BH spin is aligned with the minor axis of the galaxy, and they can influence the surrounding gas out to $\sim 10$ virial radii in massive galaxies and galaxy groups, dropping to 1-2 virial radii for galaxy clusters \citep{husko26}. Jets provide $\approx 50\%$ of the AGN energy in COLIBRE's hybrid model, dominated by thick-disc jets ($f_{\rm Edd}<0.01$) at $z<1$ and thin-disc jets ($0.01<f_{\rm Edd}<1$) at earlier times, with isotropic thermal feedback from the thin disc injecting most of the remaining energy. Injecting half of the AGN energy through jets appears to couple energy to the CGM more effectively, yielding more halo gas expulsion at a given mass. 

The differences between the fiducial and hybrid AGN feedback prescriptions appear to be amplified at m7 resolution for lower-mass haloes. This difference is attributable (at least in part) to differences in the SMBH seed mass, which is a calibrated parameter that varies between resolution levels and AGN feedback models (see \S\ref{sec:methods:bh}). Using a higher seed mass allows SMBHs to grow more effectively, attaining higher masses earlier, and in lower-mass haloes; they can therefore reach higher accretion rates and inject energy with higher $\Delta T_{\rm AGN}$ (equation \ref{eq:deltaTAGN}) and $v_{\rm jet}$ (equation \ref{eq:vjet}), driving stronger gas expulsion. The seed mass is higher at lower resolution, and when hybrid AGN feedback is used; we thus expect AGN feedback to be most effective in the L200m7h simulation at lower masses. This effect also offers an explanation for why the onset of AGN feedback, and hence the turnover of the $f_{\rm gas}^{500}-M_{500}$ relation, occurs at slightly lower mass when hybrid AGN feedback is used at m6 resolution. We note that differences due to the BH seed mass are likely confined to the lowest mass haloes, and that these simulations converge to nearly identical relations between BH mass and stellar mass for stellar masses above $10^{10}$~M$_\odot$. 

It is important to restate that the COLIBRE model variations using fiducial and hybrid AGN feedback have each been separately calibrated at each resolution, and both variations produce realistic galaxy populations. Both models yield similar galaxy stellar mass functions, size-mass relations, BH mass-stellar mass relations, star formation histories, and quenched fractions in good agreement with observational constraints. We note that the hybrid model was not designed with the intention of making AGN feedback stronger, but with the goal of implementing more realistic prescriptions for accretion discs and jets into a cosmological simulation.

\begin{figure*}
\includegraphics[width=\textwidth]{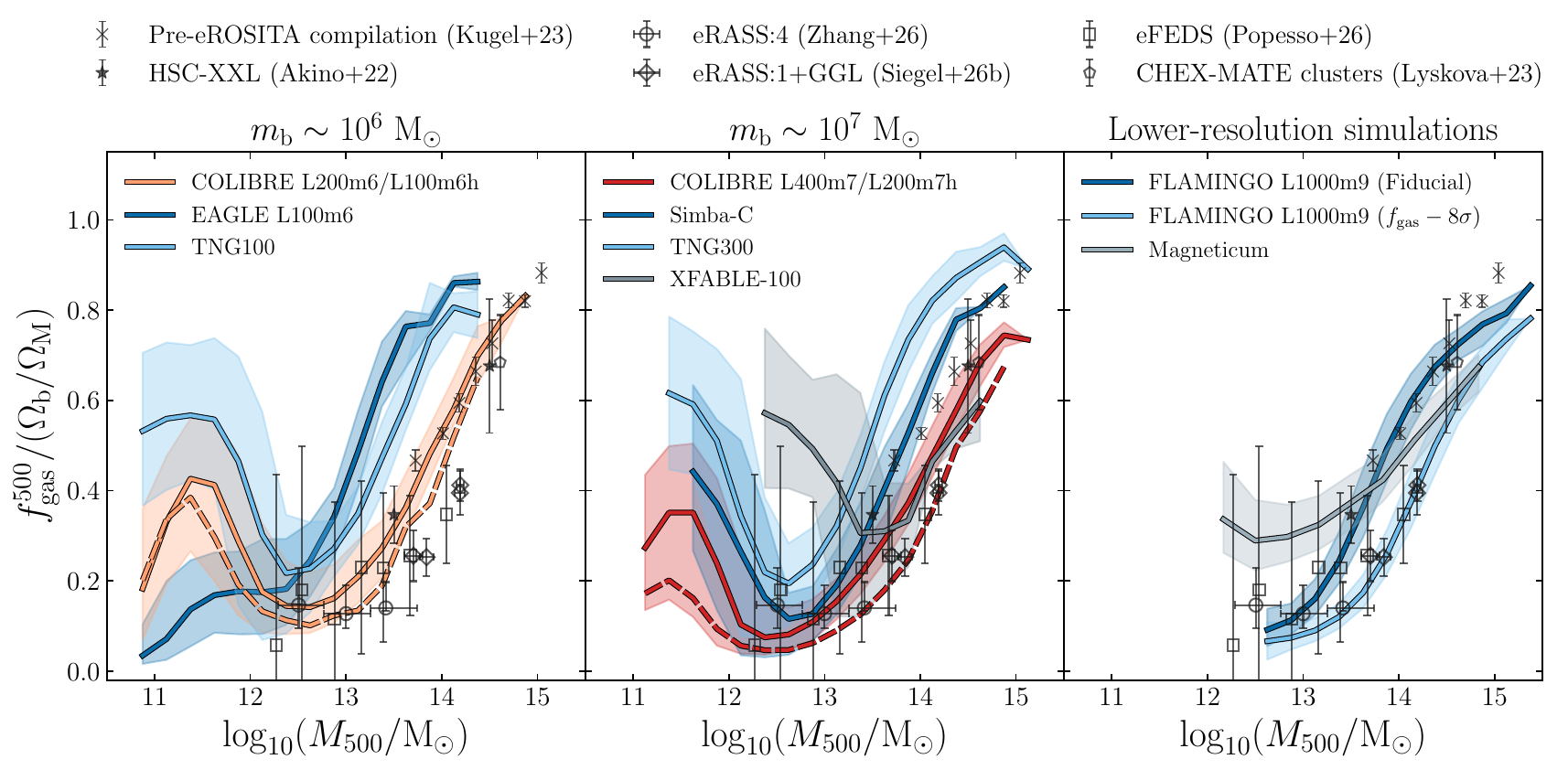}
\vspace{-4mm}
\caption{The present-day $f_{\rm gas}^{500}-M_{500}$ relation in COLIBRE, compared with the predictions of other contemporary simulations with similar mass resolution. The left-hand panel shows simulations with baryonic particle mass $m_{\rm b}\sim 10^6$~M$_\odot$: COLIBRE L200m6, EAGLE L100m6 (RefL100N1504 in their nomenclature), and Illustris-TNG100. The central panel shows simulations with $m_{\rm b}\sim 10^7$~M$_\odot$: COLIBRE L400m7, Simba-C, Illustris-TNG300, and XFABLE-100. The right-hand panel shows simulations with lower resolution: FLAMINGO L1000m9 (L1m9 in their nomenclature), an equivalent FLAMINGO simulation run with the $f_{\rm gas}-8\sigma$ model, which was calibrated to have lower gas fractions, and the Magneticum {\it Box2/hr} simulation. The largest available COLIBRE simulations with hybrid AGN feedback, L100m6h and L200m7h, are shown with dashed lines. Relations and observational data are shown as in previous figures, but for clarity we omit bins containing poorly-resolved low-mass haloes and continue showing the median even in poorly-sampled high-mass bins. Across the entire halo mass range, COLIBRE predicts lower $f_{\rm gas}$ than any other model, with the exception of EAGLE at lower masses ($M_{500}\lesssim 10^{12}$~M$_\odot$), XFABLE at higher masses ($M_{500}\gtrsim 10^{13.5}$~M$_\odot$), and FLAMINGO $f_{\rm gas}-8\sigma$. The L200m6 simulation produces similar group and cluster gas fractions to the fiducial FLAMINGO simulation, and both agree well with pre-eROSITA X-ray data. The L200m7h simulation produces similar group gas fractions to the FLAMINGO $f_{\rm gas}-8\sigma$ simulation, and a similar level of agreement with observational constraints from eROSITA.}
\vspace{-4mm}
\label{fig:simcompare}
\end{figure*}

The hybrid-AGN COLIBRE model calibrated for use at m7 resolution is able to produce a realistic galaxy population {\it and} the low halo gas fractions for galaxy groups that have been suggested by eROSITA X-ray stacks and baryonification models constrained by X-ray and kSZ data. The $f_{\rm gas}^{500}-M_{500}$ relation in this simulation is remarkably similar to the relation produced by the FLAMINGO $f_{\rm gas}-8\sigma$ model (see \S\ref{sec:results:simcompare} and \autoref{fig:simcompare}), which is also consistent with these constraints \citep[e.g.][]{siegel26a,siegel26b} and agrees with the observed kSZ signal in like-for-like comparisons \citep{mccarthy25,bigwood25kSZ}, but does not reproduce the observed thermodynamic profiles of clusters \citep{braspenning24}, or the X-ray--cosmic shear cross-correlation \citep{mcdonald26}, and is strongly disfavoured by like-for-like comparisons with XMM-Newton data \citep{seppi26}. Nonetheless, our results for the hybrid model at m7 resolution demonstrate that a ``stronger feedback'' scenario that yields lower group gas fractions is perfectly achievable within a successful galaxy formation model. 

\subsection{Comparison with EAGLE and other galaxy formation models}
\label{sec:results:simcompare}

We now consider how the new physics implemented in COLIBRE has shaped its prediction of the $f_{\rm gas}^{500}-M_{500}$ relation relative to its predecessor, EAGLE, and place the simulations in the context of other prominent galaxy formation models. In \autoref{fig:simcompare} we compare the $f_{\rm gas}^{500}-M_{500}$ relation in COLIBRE with the relations produced by other models in simulations of comparable resolution\footnote{Note that here we show {\it total} gas fractions for all simulations, and do not make any temperature/density cuts or exclude star-forming gas. Relations may therefore differ from those in published articles for which such cuts have been applied. Results were derived from particle data (EAGLE, IllustrisTNG), halo catalogues (COLIBRE, FLAMINGO, Simba-C, XFABLE), or provided directly (Magneticum).}. All curves and observational datasets are shown in the same fashion as in previous figures, though for clarity we omit bins containing poorly-resolved low-mass haloes and continue showing the median even in poorly-sampled high-mass bins. For COLIBRE, solid lines with shading represent the median and scatter for the fiducial model (L200m6 and L400m7), and dashed lines of the same colour but without shading represent the median for the hybrid model at the same resolution (L100m6h and L200m7h respectively).

\subsubsection{EAGLE}
\label{sec:results:simcompare:EAGLE}

In the left-hand panel, we compare the L200m6 and L100m6h COLIBRE simulations with others in which the baryonic particle mass $m_{\rm b}\sim 10^6$~M$_\odot$: the EAGLE RefL100N1504 simulation \citep[][here denoted as L100m6 in our nomenclature]{schaye15,crain15,mcalpine16}, and the Illustris TNG100 simulation \citep{pillepich18,nelson19dr}. EAGLE provides a natural reference point from which to explore the impact of COLIBRE's feedback models on the gas content of haloes, as the two models share much in terms of design philosophy, for example in the use of stochastic thermal feedback. The details of the feedback and cooling models in COLIBRE are significantly different to those in EAGLE, however, and COLIBRE directly models the cold ISM. \autoref{fig:simcompare} shows that COLIBRE consequently produces a significantly different $f_{\rm gas}^{500}-M_{500}$ relation to EAGLE.

For $M_{500}>10^{12}$~M$_\odot$, EAGLE's haloes are far more gas-rich than those in COLIBRE, particularly at the mass scale of galaxy groups. The overly-high gas fractions relative to pre-eROSITA constraints from X-ray observations are a well-documented shortcoming of the EAGLE model \citep[e.g.][]{schaye15,davies19,kelly21}, and the difference relative to new data from eROSITA is even greater. EAGLE's prescription for AGN feedback is similar to the fiducial model in COLIBRE, but employed a fixed heating temperature increment $\Delta T_{\rm AGN}=10^{8.5}$~K. The energy of individual feedback events ($\Delta E_{\rm AGN,thermal} \propto \Delta T_{\rm AGN}$, equation \ref{eq:deltaE_thermal}) strongly influences feedback's effectiveness; EAGLE's temperature increment is highly effective at expelling gas for $\sim L^\star$ galaxies and their haloes \citep[e.g.][]{davies19,davies20,wright24}, but becomes ineffective for halo gas with higher entropies in the deeper potentials of groups and clusters. 

AGN feedback in COLIBRE is far more effective in high-mass haloes, yielding a better match to observational constraints. A major contributor to this improvement is the fact that $\Delta T_{\rm AGN}$ scales linearly with the BH mass, $M_{\rm BH}$, and hence individual feedback events become more energetic as $M_{\rm BH}$ increases (see \S\ref{sec:methods:bh:fiducial}). Whilst this means that very massive BHs can inject more energetic and expulsive heating events than EAGLE, the reverse is true for lower-mass BHs. Since the gas fractions of haloes are influenced by the whole history of feedback from BH growth \citep[e.g.][]{davies22,luciesmith25}, we must therefore consider what the value of $\Delta T_{\rm AGN}$ was during the period when most baryons were expelled. For the mass range in which $f_{\rm gas}^{500}$ is significantly lower for COLIBRE than for EAGLE ($M_{500}\gtrsim 10^{13}$~M$_\odot$), we find that $\Delta T_{\rm AGN}$ exceeds the fixed value used in EAGLE when the halo is still gas-rich, and that the majority of baryon expulsion occurs with an $\approx 0.5-1$ dex higher $\Delta T_{\rm AGN}$, yielding more efficient expulsion. We show this in detail in Appendix \ref{sec:app:dTAGN}. In COLIBRE's hybrid AGN feedback prescription, $\Delta T$ for the thermally-driven wind component is scaled linearly with $M_{\rm BH}$ in the same fashion, and the kinetic energy injected in the form of jets is also proportional to $M_{\rm BH}$ (and hence the jet velocity $v_{\rm jet}\propto M_{\rm BH}^{1/2}$). Hybrid AGN feedback therefore also remains effective in high-mass haloes. In addition, other changes and improvements made to the modelling of BHs and AGN feedback may contribute to the lower gas fractions produced in COLIBRE. The new BH repositioning scheme may allow for more rapid accretion \citep{bahe22}, AGN feedback energy is now injected into the particle(s) nearest the BH rather than stochastically within the BH kernel, yielding improved efficiency \citep{chaikin22}, and super-Eddington accretion is allowed, though such events occur predominantly at high redshift \citep[see e.g.][]{chaikin_edd}.

The qualitatively different shape of the $f_{\rm gas}^{500}-M_{500}$ relation for $M_{500}\lesssim 10^{12}$~M$_\odot$ in COLIBRE relative to EAGLE is the result of major changes to star formation, CCSN feedback, and the ISM in the model, in conjunction with the use of a variable $\Delta T_{\rm AGN}$. The EAGLE model does not incorporate a cold interstellar gas phase, but features a pressurised ISM where any gas denser than a (metallicity-dependent) threshold can form stars at a rate governed by an imposed Kennicutt-Schmidt relation. To regulate this process with minimal radiative cooling losses, thermally-driven CCSN feedback with a high (fixed) heating temperature, $\Delta T_{\rm SN}=10^{7.5}$~K, was used. As a result, CCSN feedback is highly energetic and expulsive in EAGLE, and in the process of regulating star formation it causes low-mass haloes to be strongly depleted of gas. AGN feedback is also important for lower-mass haloes in EAGLE that host over-massive BHs \citep[e.g.][]{davies19,davies20}, and any such feedback is highly expulsive in lower-mass haloes due to EAGLE's fixed $\Delta T_{\rm AGN}$, resulting in the lowest gas fractions seen at these masses.

COLIBRE features a more sophisticated prescription for star formation, and ultimately, a more `gentle' CCSN feedback prescription. In COLIBRE a gas particle's eligibility for star formation depends on its density and both its thermal and turbulent velocity dispersions, requiring the gas to be gravitationally unstable at the resolution limit. Star formation can thus be regulated by driving turbulence in the ISM, which is achieved through a low-energy kinetic feedback prescription, and further prevention of star formation is provided by the modelling of HII regions and early stellar feedback. A more energetic and expulsive thermal feedback prescription is also present, but needs to expel less gas from the galaxy to regulate star formation than was the case for EAGLE's thermal feedback. Most importantly, in COLIBRE the heating temperature $\Delta T_{\rm SN}$ scales with the local ISM density to enable better sampling in low-mass galaxies and minimise radiative losses in very dense regions; when calibrated, at low gas densities this feedback mode is far less expulsive than in EAGLE, and hence $f_{\rm gas}^{500}$ is much higher for $M_{500}<10^{12}$~M$_\odot$.

Using the individually-calibrated variants of the COLIBRE model presented by \citet{chaikin_calibration}, we can test how several components of COLIBRE's feedback scheme individually impact the halo gas fraction. We refer the reader to Appendix \ref{sec:app:feedback} for details, and summarise our findings here. Models in which $\Delta T_{\rm SN}$ and $\Delta T_{\rm AGN}$ are fixed at high values always lead to low $f_{\rm gas}^{500}$ (similar to EAGLE); this is the case whether or not a fraction of the energy is injected kinetically, as the remaining thermal feedback is always highly expulsive. Allowing $\Delta T_{\rm SN}$ to vary for the thermal component of the feedback makes it less expulsive and yields higher $f_{\rm gas}^{500}$. In addition to these model features, allowing the fraction of SN energy that is coupled to the gas, $f_{\rm E}$, to vary with the gas pressure yields slightly lower $f_{\rm gas}^{500}$; the reasons for this are not trivial and are explored in Appendix \ref{sec:app:feedback}. \citet{chaikin_calibration} showed that only a model employing all these features can simultaneously match both the GSMF and SSMF; such a model produces higher gas fractions than EAGLE at lower halo masses.

The variable $\Delta T_{\rm AGN}$ used in COLIBRE also influences $f_{\rm gas}^{500}$ at lower halo masses. Scaling the energy of AGN feedback events with the BH mass ensures that for a fixed Eddington ratio, the feedback remains equally sampled across different BH masses; lower-mass BHs do not need to ``save up'' energy in a reservoir for long periods in order to be able to heat particles by a large, fixed $\Delta T_{\rm AGN}$, but can instead heat more frequently in less energetic events. We show in Appendix \ref{sec:app:feedback} that the fiducial model yields the highest gas fractions of any model variant at lower halo masses, because the AGN feedback injected by lower-mass BHs is better-sampled and less expulsive.

\subsubsection{Other simulations - m6 resolution}

The shape of the $f_{\rm gas}^{500}-M_{500}$ relation for IllustrisTNG-100, also shown in the left panel of \autoref{fig:simcompare}, bears a stronger resemblance to COLIBRE than to EAGLE, in that low-mass and high-mass haloes are gas-rich, and $\sim L^\star$ haloes are gas-poor. Baryon expulsion from haloes in IllustrisTNG is driven by its kinetic AGN feedback mode \citep{davies20,terrazas20,oren26}, which is generally more expulsive than the EAGLE prescription, as evidenced by lower $f_{\rm gas}^{500}$ for $M_{500}\gtrsim 10^{12.5}$~M$_\odot$, but less so than the COLIBRE prescription, and yields haloes that are too gas rich relative to observational constraints. The efficiency of AGN feedback in IllustrisTNG does depend on $M_{\rm BH}$ to some extent, as BHs transition from a low-efficiency thermal mode to the expulsive kinetic mode below a threshold Eddington ratio that scales with $M_{\rm BH}$. Once in the kinetic mode, however, the energy of feedback events scales only with the local gas mass and 1D dark matter velocity dispersion, and is insensitive to $M_{\rm BH}$.

Low-mass haloes ($M_{500}<10^{12}$~M$_\odot$) in IllustrisTNG are gas-rich, to a greater degree than in COLIBRE. As in EAGLE, IllustrisTNG galaxies lack a cold ISM, and rely on gas expulsion to regulate star formation in this mass regime, however outflows driven by stellar feedback typically recycle within the halo, rather than leave the halo altogether \citep{wright24}, keeping $f_{\rm gas}^{500}$ high. The IllustrisTNG model minimises cooling losses in the ISM by briefly decoupling particles kicked by feedback from hydrodynamical forces; feedback events can therefore be less energetic and less expulsive at the halo scale whilst still successfully regulating star formation.

\subsubsection{Other simulations - m7 resolution}

In the central panel of \autoref{fig:simcompare}, we compare COLIBRE L400m7 and L200m7h with other simulations in which the baryon mass resolution $m_{\rm b}\sim 10^7$~M$_\odot$: the Illustris TNG300 simulation \citep{pillepich18,nelson19dr}, the XFABLE-100 simulation \citep{bigwood25xfable}, and the largest simulation run with the Simba-C model \citep{hough23}, a new version of the Simba model \citep{dave19} with updated chemical enrichment and feedback physics.

As was the case at m6 resolution, the IllustrisTNG model predicts higher gas fractions than COLIBRE across the whole mass range. However, an interesting difference here is that in the IllustrisTNG model, the $f_{\rm gas}^{500}-M_{500}$ relation is consistent between resolutions, whereas in COLIBRE gas fractions decrease at a fixed mass as resolution decreases, for the reasons discussed in \S\ref{sec:results:convergence}. The gas content of haloes in IllustrisTNG is predominantly influenced by kinetic AGN feedback, for which the minimum energy per event is resolution-dependent as it scales with the mass of a fixed number of gas neighbours. This number of neighbours is adjusted for different resolutions \citep[see section 2.3 and appendix B of][]{weinberger17}, yielding similar $f_{\rm gas}^{500}-M_{500}$ relations for the TNG100 and TNG300 simulations.

The XFABLE simulation model was designed to feature stronger AGN feedback than the original FABLE model \citep{henden18}, with the aim of producing a strong enough suppression of the matter power spectrum to resolve the $S_8$ tension \citep{amon22,preston23} and improve agreement with joint constraints from weak lensing and kSZ measurements \citep{bigwood24}. The largest simulation from the suite, XFABLE-100, produces a similar $f_{\rm gas}^{500}-M_{500}$ relation to the COLIBRE L400m7 simulation for $M_{500}>10^{13.5}$~M$_\odot$, with slightly lower gas fractions at the highest masses that are in better agreement with the COLIBRE L200m7h simulation. Below $M_{500}\approx10^{13.5}$~M$_\odot$, XFABLE-100 haloes are significantly more gas-rich than in COLIBRE, likely because XFABLE's expulsive ``radio-mode'' AGN feedback is only active for BHs where $M_{\rm BH}>10^9$~M$_\odot$, corresponding to $M_{500}\gtrsim 10^{13}$~M$_\odot$.

The Simba-C model yields similar gas fractions to COLIBRE at $M_{500}\approx10^{13}$~M$_\odot$, but otherwise exhibits more gas-rich haloes across the mass range (though less gas-rich than TNG300). The $f_{\rm gas}^{500}-M_{500}$ relation in the original Simba model (not shown here) closely matches the Simba-C relation for $M_{500}>10^{13}$~M$_\odot$, and closely matches the COLIBRE L400m7 result for lower masses. 

\subsubsection{Other simulations at lower resolution}

At present, there are no COLIBRE simulations available at a lower resolution level than m7, however, for comparison, we show the $f_{\rm gas}^{500}-M_{500}$ relation produced by two variants of the FLAMINGO model \citep{schaye23,kugel23,helly26}, plus the flagship {\it Box2/hr} Magneticum simulation \citep{hirschmann14,dolag25} in the right-hand panel of \autoref{fig:simcompare}. We show the FLAMINGO L1m9 (i.e. 1 cGpc side length, hence L1000m9 in our nomenclature) simulations evolved with the fiducial FLAMINGO model, calibrated to gas fractions inferred from the pre-eROSITA X-ray data compiled by \citet[][also shown in our \autoref{fig:flagship_xray_ksz}]{kugel23}, and also the $f_{\rm gas}-8\sigma$ model, which was calibrated to produce gas fractions $8\sigma$ below those derived from this X-ray data. 

The fiducial FLAMINGO model exhibits a similar relation to COLIBRE L400m7 for $M_{500}<10^{13.5}$~M$_\odot$, but higher gas fractions above this mass. The AGN feedback prescriptions employed by the fiducial COLIBRE and FLAMINGO models are similar, but differ in that $\Delta T_{\rm AGN}$ is fixed in FLAMINGO (as it was in EAGLE), whereas in COLIBRE $\Delta T_{\rm AGN}$ can reach higher temperature increments. A like-for-like comparison of $\Delta T_{\rm AGN}$ at the same resolution, as done for EAGLE in Appendix \ref{sec:app:dTAGN}, is not possible here, however, as COLIBRE simulations have not been performed at the same resolution. The fiducial FLAMINGO model agrees well with the pre-eROSITA constraints to which it was calibrated, but poorly with constraints from some eROSITA analyses, as extensively explored in several studies \citep{bigwood24,bigwood25kSZ,kovac25,mccarthy25,siegel26a,siegel26b}. These studies have shown that the $f_{\rm gas}-8\sigma$ model provides a better match to constraints on the gas content of galaxy groups from eROSITA stacking analyses (as is also evident in \autoref{fig:simcompare}), and have shown that this model is favoured over the fiducial model in like-for-like comparisons with kSZ measurements (which we have not performed for COLIBRE). As previously discussed, however, the $f_{\rm gas}-8\sigma$ model is ruled out when forward-modelling the X-ray emission from the simulated haloes and performing a like-for-like comparison with XMM-Newton data, accounting for selection effects \citep{seppi26}. Notably, comparing the central and right-hand panels of \autoref{fig:simcompare} shows that COLIBRE's most expulsive variant, the L200m7h simulation with hybrid AGN feedback, yields a very similar $f_{\rm gas}^{500}-M_{500}$ relation to FLAMINGO $f_{\rm gas}-8\sigma$.

The gas fractions predicted by Magneticum are lower than those in the fiducial FLAMINGO model, and similar to those in COLIBRE L400m7, at $M_{500}>10^{14}$~M$_\odot$. At lower mass, Magneticum haloes are considerably more gas-rich than both COLIBRE and FLAMINGO. At group scales, these gas fractions are considerably higher than those shown by \citet{popesso26} for Magneticum; this difference arises because \citet{popesso26} only include hot gas ($T>10^6$~K) in their measurements. 

\subsection{Histories of gas expulsion \& re-accretion as a function of halo mass}
\label{sec:results:histories}

\begin{figure}
\includegraphics[width=\columnwidth]{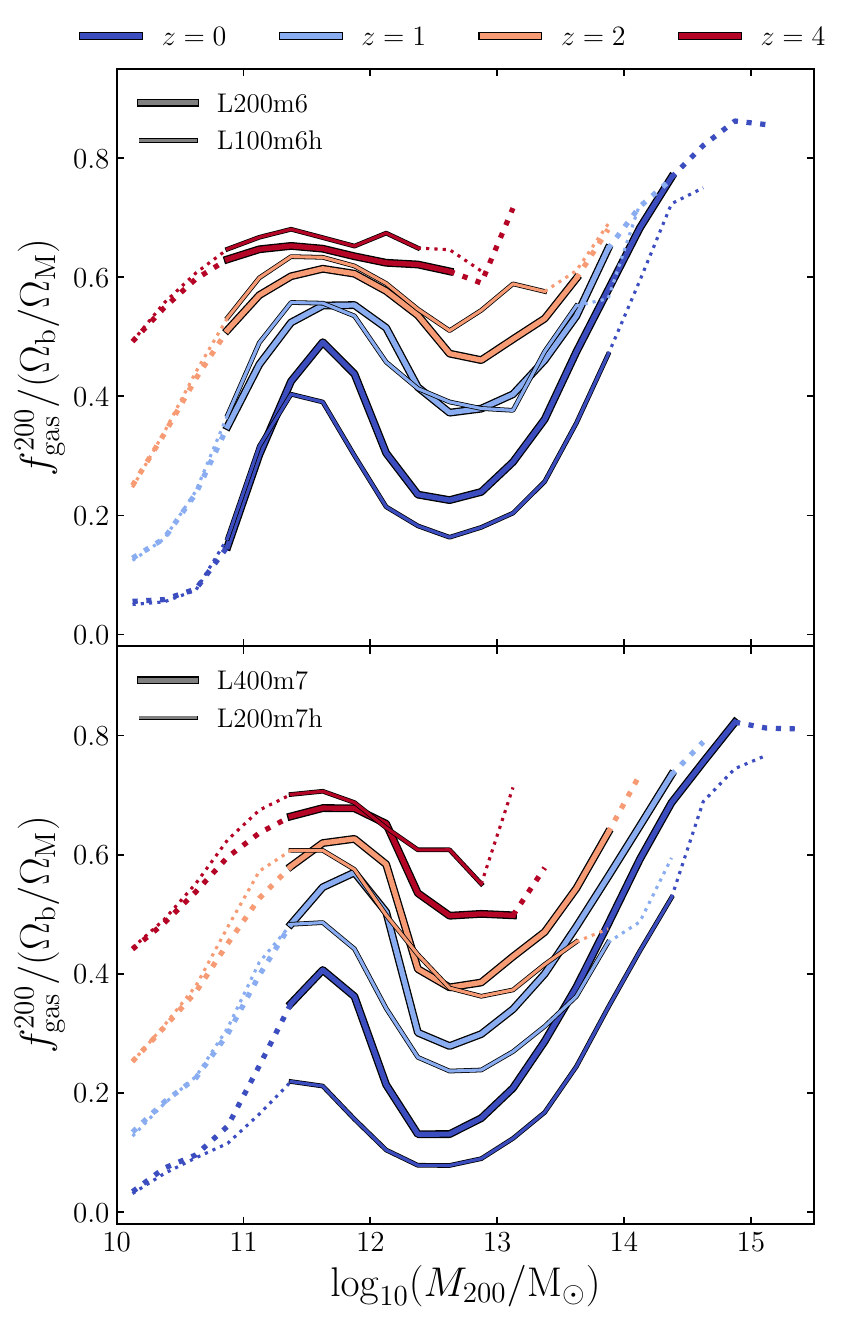}
\vspace{-4mm}
\caption{Evolution of the COLIBRE $f_{\rm gas}^{200}-M_{200}$ relation from $z=4$ to the present day, at m6 (upper panel) and m7 (lower panel) resolution, in the fiducial model (thick lines) and hybrid AGN model (thin lines). Dotted lines indicate low-mass bins containing poorly-resolved galaxies, and poorly-sampled high-mass bins. The shape of the relation is consistent between redshifts, except at $z=4$ where the relation remains relatively flat at high masses at m6 resolution, but exhibits a decrease towards higher mass at m7 resolution. For all but the most massive haloes the relation shifts to lower $f_{\rm gas}^{200}$ as time progresses, indicating that feedback is becoming more efficient.}
\vspace{-4mm}
\label{fig:fgas_relation_ev}
\end{figure}

We conclude our exploration of halo gas fractions in COLIBRE by examining how the gas content of haloes evolves through time in the simulations. In \autoref{fig:fgas_relation_ev}, we show the median $f_{\rm gas}^{200}-M_{200}$ relation at four epochs: $z=0$, 1, 2, and 4 for the fiducial (thick lines) and hybrid (thin lines) models. The upper and lower panels show simulations at m6 resolution (L200m6/L100m6h), and m7 resolution (L400m7/L200m7h) respectively. As in previous figures, bins at low mass containing poorly-resolved galaxies are shown with dotted lines, but for clarity we continue showing the median even in poorly-sampled high-mass bins, again using dotted lines.

The general shape of the $f_{\rm gas}^{200}-M_{200}$ relation is consistent at all epochs, and at both resolution levels, however its normalisation evolves significantly. At a given mass, $f_{\rm gas}^{200}$ is higher at higher redshift (for all but the most massive haloes) indicating that feedback-driven expulsion is less effective at early times. This is to be expected, as earlier-forming haloes of a given mass have higher binding energies and accretion rates, and because gas densities are higher at early times, leading to greater radiative losses; these effects together yield less effective expulsion for a fixed energy input\footnote{CCSN feedback may not necessarily be less expulsive in higher-density gas due to the scaling of $\Delta T_{\rm SN}$ with density (to the power $2/3$, equation \ref{eq:deltaTSN}).}. AGN feedback suppresses the gas fraction at the high mass end earlier at m7 resolution than at m6 resolution; at $z=4$ the relation is relatively flat at high mass in the L200m6 simulation, but exhibits a decline for $M_{200}>10^{12}$~M$_\odot$ in the L400m7 simulation.

The time evolution of the relation is more pronounced in the hybrid model (dashed lines) than in the fiducial model; the relation is similar at $z=2$ for the two model variants, but they diverge as time progresses, yielding the overall lower $z=0$ gas fractions in the hybrid model (also shown in \autoref{fig:hybrid}). This behaviour reflects the increasing prevalence of efficient thick-disc ($f_{\rm Edd}<0.01$) jet feedback in the hybrid model at $z<1$. This divergence is resolution-dependent; at m6 resolution, the models still produce similar results at $z=1$, whilst at m7 the gas fractions are already lower across the mass range by this epoch\footnote{Note that at both resolution levels, the simulations with hybrid AGN feedback are a factor of 8 times smaller and contain commensurately fewer haloes. Small differences seen at the high mass end may therefore be the result of poor sampling.}. The typical mass scale at which AGN feedback becomes efficient at expelling baryons beyond $r_{200}$ in most haloes, indicated by the location of the lower-mass peak in the relation, is consistent between resolution levels. This peak is shifted to slightly lower mass for the hybrid AGN model, indicating that hybrid AGN feedback becomes effective earlier in a halo's growth; as discussed in \S\ref{sec:results:hybrid}, this is likely due to the use of a higher SMBH seed mass, which facilitates earlier SMBH growth.

\begin{figure*}
\includegraphics[width=\textwidth]{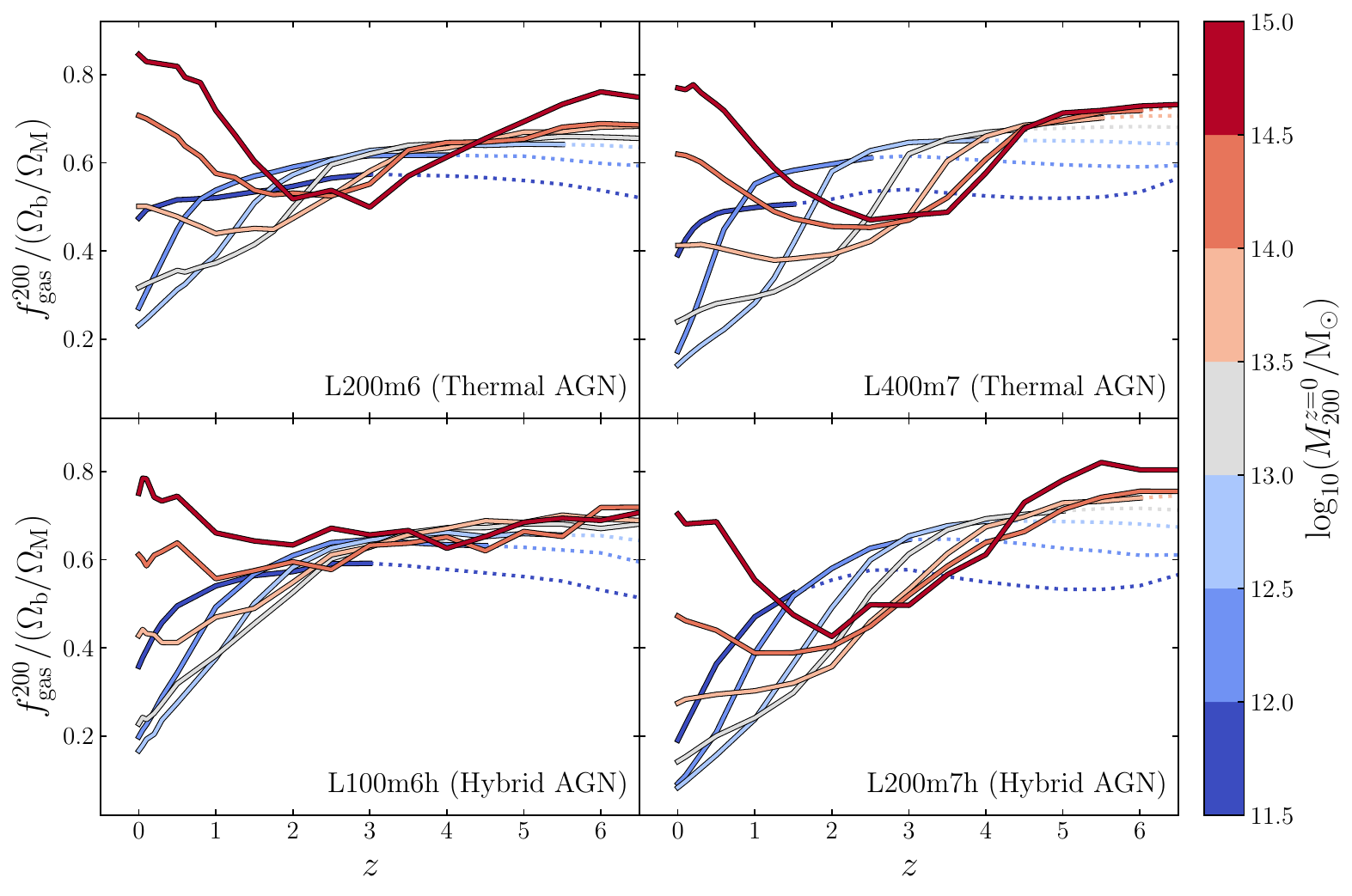}
\vspace{-4mm}
\caption{Evolution of $f_{\rm gas}^{200}$ with redshift for haloes binned by present-day halo mass, $M_{200}^{z=0}$, at both m6 and m7 resolutions, and with both thermal and hybrid AGN feedback. Solid lines show the median $f_{\rm gas}^{200}$, and dotted lines indicate where the median stellar mass is below 100 times the initial baryonic particle mass. As present-day mass increases, gas is expelled earlier, and for the two most massive bins, $f_{\rm gas}^{200}$ rises at late times as haloes are replenished by either fresh accretion or the re-accretion of previously-expelled gas. The high-redshift progenitors of present-day clusters were less strongly depleted of gas than present-day groups (and less massive haloes). The amount of gas expelled, and the times of expulsion and (re-)accretion are sensitive to resolution and AGN feedback prescription.}
\vspace{-4mm}
\label{fig:fgas_massbinned_ev}
\end{figure*}

We further explore how the $f_{\rm gas}^{200}-M_{200}$ relation is shaped at different epochs by showing the redshift evolution of $f_{\rm gas}^{200}$ in haloes of different present-day mass in \autoref{fig:fgas_massbinned_ev}. We divide haloes into 7 bins of $z=0$ halo mass, $M_{200}^{z=0}$, track each binned set of haloes back to their progenitors in every prior simulation snapshot back to $z\approx 7$ using the tracking information provided by HBT-HERONS, and show the median $f_{\rm gas}$ value in each bin as a function of $z$, with line colours reflecting the different $M_{200}^{z=0}$ bins. We indicate where the galaxies in these haloes are poorly-resolved with dotted lines, using the same criterion as in previous figures. Results are shown for the largest available simulations at m6 and m7 resolution levels, with fiducial and hybrid AGN feedback, in different panels of \autoref{fig:fgas_massbinned_ev}.

The evolution of $f_{\rm gas}^{200}$ varies significantly with present-day halo mass, and also with simulation resolution and AGN feedback model. Haloes begin gas rich, at an $f_{\rm gas}^{200}$ value primarily set by the efficiency of stellar feedback (see Appendix \ref{sec:app:feedback}). This initial value increases with $M_{200}^{z=0}$, since more massive present-day haloes form at earlier times, when physical densities are higher and hence feedback is less effective. In almost all cases\footnote{Present-day clusters ($M_{200}>10^{14}$~M$_\odot$) in the L100m6h simulation are an exception to this behaviour, as they do not experience a period of rapid depletion at high redshift. Very few haloes of this mass are present in a 100 Mpc volume, however, and this behaviour may not be representative of the typical behaviour of the hybrid AGN model at m6 resolution.}, upon reaching the mass scale at which SMBHs grow rapidly, AGN feedback becomes efficient at expelling gas beyond $r_{200}$ and $f_{\rm gas}^{200}$ drops significantly. \autoref{fig:fgas_relation_ev} shows that this mass scale does not change significantly with redshift, so the time at which $f_{\rm gas}^{200}$ begins to decrease is primarily set by the time at which this mass scale is reached; haloes with the highest present-day masses are thus depleted first, and those in the least massive bin are depleted last. The onset of expulsion is also slightly earlier at lower resolution, as was evident in \autoref{fig:fgas_relation_ev}.

The time of gas expulsion correlates with the fraction of gas expelled; at early times, clusters are only depleted to $f_{\rm gas}^{200}\approx 0.5 f_{\rm b}^{\rm cosmic}$, whereas the haloes of massive galaxies and low-mass groups ($M_{200}^{z=0}=10^{12.5-13.5}$~M$_\odot$) have $f_{\rm gas}^{200} \approx (0.1-0.3)f_{\rm b}^{\rm cosmic}$ at the present day and are still being depleted. The precise fractions are model and resolution dependent; lower resolution and/or the use of hybrid AGN feedback leads to stronger expulsion.

\autoref{fig:fgas_massbinned_ev} shows that once feedback ceases to be effective in the progenitors of present-day clusters, they begin replenishing their gas reservoirs, eventually attaining the high gas fractions seen in the present-day $f_{\rm gas}^{200}-M_{200}$ relation. This is achieved either through accretion of fresh material, or re-accretion of expelled material that did not fully escape the potential well. The likelihood of expelled material being re-accreted depends on the feedback strength and is hence model dependent; while clusters ($M_{200}>10^{14}$~M$_\odot$) always re-accrete, higher-mass groups ($M_{200}^{z=0}=10^{13.5-14}$~M$_\odot$) only do so in the fiducial model, and are still being depleted in the hybrid model. 

This process of re-accretion onto clusters offers an explanation for how models with stronger feedback can yield lower gas fractions for groups without over-depleting clusters. As discussed by \citet{luciesmith25}, AGN-driven depletion of the group-mass progenitors of present-day clusters is inefficient, and re-accretion effectively erases the earlier influence of feedback, making their properties appear consistent with models with `modest' AGN feedback strengths. Present-day groups, however, are gas-poor and have not begun re-accreting, and thus require stronger AGN feedback (such as that in FLAMINGO's $f_{\rm gas}-8\sigma$ model or COLIBRE's hybrid model) to match the lower $f_{\rm gas}^{500}$ constraints suggested by eROSITA stacks and baryonification analyses. The timing of gas expulsion and subsequent (re-)accretion has been shown to be crucial for setting the scatter in $f_{\rm gas}$ for group and cluster haloes in FLAMINGO \citep{costello25}; we will explore the scatter in COLIBRE in this mass range in a forthcoming study.

\section{Summary and Discussion}
\label{sec:summary}

In this study we have presented predictions for the relationship between the halo gas mass fraction, $f_{\rm gas}^{200}$, and halo mass, $M_{200}$, from the COLIBRE suite of cosmological hydrodynamical simulations, and explored the physical processes that shape it. We presented results for the largest-volume simulations performed at the three resolution levels in the COLIBRE suite that have reached $z=0$, L400m7, L200m6 and L025m5, which have side lengths of 400, 200, and 25 comoving Mpc and for which the baryon and DM particle masses are $\sim 10^7$~M$_\odot$ (m7), $\sim 10^6$~M$_\odot$ (m6) and $\sim 10^5$~M$_\odot$ (m5) respectively. We also showed results for the simulations incorporating COLIBRE's alternative `hybrid' thermal/jet implementation of AGN feedback, L200m7h and L100m6h. 

We began by showing how the baryon fractions, stellar fractions, and gas fractions of COLIBRE haloes vary with halo mass in each simulation. We then demonstrated how gas at different temperatures contributes to $f_{\rm gas}^{200}$, illustrating the multiphase nature of the gas in COLIBRE haloes, and examined how the resolution of the simulations influences the efficiency of feedback. We then compared the $f_{\rm gas}^{500}-M_{500}$ relation in each simulation with constraints inferred from X-ray observations and measurements of the kinetic Sunyaev-Zel'dovich effect, examined how the choice of AGN feedback prescription influences the relation, placed COLIBRE in the context of other prominent galaxy formation models, and explored how the gas content of haloes of different masses has evolved through cosmic time. Our key findings are:

\begin{itemize}

\item The $f_{\rm gas}^{200}-M_{200}$ relation in COLIBRE is non-monotonic, reflecting the changing physical processes that shape halo gas content across the mass range (\S\ref{sec:results:flagship}, \autoref{fig:flagship}). Low-mass haloes ($M_{200}\lesssim 10^{11}$~M$_\odot$) are gas-poor, as feedback from SNe efficiently expels gas beyond $r_{200}$ and suppresses inflow at the halo scale. Above this mass the gas fraction increases with halo mass, reaching a peak in the $M_{200}\sim 10^{11.5-12}$~M$_\odot$ range as SN feedback becomes less efficient due to the deepening potential well and the development of a shock-heated hot halo. At higher masses, expulsive AGN feedback drives a decline in $f_{\rm gas}^{200}$, which reaches a minimum at $M_{200}\sim 10^{12.5-13}$~M$_\odot$ as SMBHs begin self-regulating their own growth. Above this mass, AGN feedback becomes less efficient and gas fractions rise once more, due to fresh gas accretion and re-accretion of previously expelled material.

\item Gas in the haloes of galaxies is multiphase, with gas spanning temperatures from $T\sim 10$~K to $\gtrsim 10^7$~K contributing significantly to the total gas content of haloes with $M_{200}\lesssim 10^{12.5}$~M$_\odot$ (\S\ref{sec:results:temperature}, \autoref{fig:temperature}, see also Appendix \ref{sec:app:multiphase}). Cold gas ($T\lesssim 10^4$~K) constitutes up to 50\% of the halo gas mass at $M_{200}\approx 10^{11.4}$~M$_\odot$, and the cold ISM ($T < 10^{2.5}$~K) contributes up to 11\% at $M_{200}\approx 10^{11.6}$~M$_\odot$. Haloes with $M_{200} > 10^{12.5}$~M$_\odot$ are dominated by hot gas at approximately the virial temperature.

\item The $f_{\rm gas}^{200}-M_{200}$ relation is not fully converged with resolution (\S\ref{sec:results:convergence}, \autoref{fig:flagship}), as COLIBRE's feedback prescriptions were not calibrated to reproduce observed gas fractions. As resolution increases, feedback events are individually less energetic, and in the case of AGN feedback the densities of heated gas particles are higher, leading to greater radiative cooling losses. As a result, gas fractions increase with increasing resolution at fixed halo mass, though there is substantial overlap in the scatter between resolution levels across the whole mass range.

\item At m6 resolution, gas fractions produced by the fiducial COLIBRE model agree well with constraints derived from Chandra and XMM-Newton X-ray data compiled by \citet{kugel23} for groups and clusters (\S\ref{sec:results:obs}, \autoref{fig:flagship_xray_ksz}). X-ray measurements from eROSITA stacks of galaxy groups ($M_{500}\sim 10^{13-14}$~M$_\odot$) suggest lower gas fractions than are produced by the fiducial model at any resolution, and baryonification models jointly constrained by X-ray and kSZ effect measurements also indicate lower group gas fractions than the fiducial model produces. At higher masses, the haloes of galaxy clusters in COLIBRE are in good agreement with these observational constraints at both resolution levels. We caution that the available constraints on group-scale gas fractions are not mutually consistent, for reasons that are not yet fully understood. We therefore refrain from assuming that any particular dataset is more reliable than another.

\item COLIBRE's hybrid AGN feedback model produces consistently lower gas fractions than the fiducial model at fixed resolution (\S\ref{sec:results:hybrid}, \autoref{fig:hybrid}), as a consequence of both increased total AGN energy injection ($\approx 25\%$ more by $z=0$) and more effective coupling of energy to the CGM through jets. The hybrid model at m7 resolution produces a good match to constraints on group gas fractions from eROSITA and from baryonification models, yielding an $f_{\rm gas}^{500}-M_{500}$ relation remarkably similar to that of the FLAMINGO $f_{\rm gas}-8\sigma$ model (\S\ref{sec:results:simcompare}, \autoref{fig:simcompare}). Both of COLIBRE's AGN feedback variants have been independently calibrated to produce realistic galaxy populations, demonstrating that the stronger feedback strengths apparently required by these constraints are compatible with a successful galaxy formation model.

\item For haloes with $M_{500}\gtrsim 10^{12}$~M$_\odot$, COLIBRE produces lower gas fractions than its predecessor EAGLE and most other contemporary simulations at comparable resolution, yielding a better match with observational constraints (\S\ref{sec:results:simcompare}, \autoref{fig:simcompare}). This improvement can be attributed to COLIBRE's AGN feedback events becoming more energetic as the BH mass increases, achieved through either an increasing temperature increment, $\Delta T_{\rm AGN}$, for thermal feedback, or a higher kick velocity for jet feedback in the hybrid model. In the fiducial model, $\Delta T_{\rm AGN}$ is significantly higher than EAGLE's fixed $\Delta T_{\rm AGN}$ value at the time baryons are expelled from haloes, yielding lower present-day gas fractions (Appendix \ref{sec:app:dTAGN}). 

\item At lower halo masses ($M_{500}\lesssim 10^{12}$~M$_\odot$), gas fractions in COLIBRE are lower than in simulations such as IllustrisTNG and SIMBA-C, but are higher than those in EAGLE. This difference with respect to EAGLE is primarily due to the adoption of variable heating temperatures for both supernova and AGN feedback; this results in less energetic but more frequent feedback events in lower-mass haloes, and yields less gas expulsion overall (\S \ref{sec:results:simcompare:EAGLE}, Appendix \ref{sec:app:feedback}).

\item The present-day gas mass fractions of haloes are set by the cumulative history of feedback over the course of their growth (\S\ref{sec:results:histories}, figs. \ref{fig:fgas_relation_ev} \& \ref{fig:fgas_massbinned_ev}). Haloes are relatively gas-rich at early times, and are depleted rapidly once AGN feedback becomes efficient at expelling gas beyond $r_{200}$. More massive present-day haloes are depleted of gas earlier, and the progenitors of clusters subsequently replenish their gas reservoirs through fresh accretion or re-accretion of previously expelled material.

\end{itemize}

The lack of perfect convergence in the $f_{\rm gas}^{200}-M_{200}$ relation is not unexpected given that COLIBRE's subgrid parameters were calibrated to reproduce galaxy-scale properties \citep[for which the numerical convergence is generally good, see][]{schaye26} rather than halo gas fractions. This highlights a known \citep[e.g.][]{mitchell22b} and fundamental problem: a feedback model that is calibrated to successfully regulate star formation and BH growth does not uniquely determine the fraction of baryons that should be expelled from the halo. A wide range of $f_{\rm gas}^{200}-M_{200}$ relations are compatible with realistic galaxy populations \citep[e.g.][]{davies20,schaye23,crainvandevoort23}. Observational constraints on this relation, and on the interaction of galaxies with their gaseous haloes through accretion and feedback, are therefore urgently needed in order to break this degeneracy in galaxy formation models.

\section*{Acknowledgements}
We thank the reviewer for their helpful comments, which improved the clarity of this manuscript. JJD would like to thank Rob Crain, Emily Costello, Anna Durrant, Ryan Roberts, Ian McCarthy, Andrew Pontzen and Hiranya Peiris for helpful discussions about this study and the topics it covers. The authors would like to thank Leah Bigwood, Romeel Dav\'e, Ajay Dev, Michael Kova\v c, Ilaria Marini, Paola Popesso, and Jared Siegel for sharing their data with us. We thank the EAGLE, IllustrisTNG and FLAMINGO collaborations for making their simulation data publicly available. Data analysis for this study was carried out using the {\sc swiftsimio} \citep{borrow20} and {\sc swiftgalaxy} \citep{oman25} open source tools. 

JJD acknowledges support from STFC grant ST/Y002482/1. FH acknowledges funding from the Netherlands Organization for Scientific Research (NWO) through research programme Athena 184.034.002. RJW is supported by the Forrest Research Foundation through a Forrest Research Fellowship. EC acknowledges support from STFC consolidated grant ST/X001075/1. ABL acknowledges support from the Italian Ministry for Universities (MUR) program “Dipartimenti di Eccellenza 2023-2027” within the Centro Bicocca di Cosmologia Quantitativa (BiCoQ), and support from UNIMIB’s Fondo Di Ateneo Quota Competitiva (project 2024-ATEQC-0050). SP acknowledges support by the Austrian Science Fund (FWF) through grant-DOI: 10.55776/V982. 

This work made use of the Prospero high performance computing facility at Liverpool John Moores University, and the DiRAC@Durham facility managed by the Institute for Computational Cosmology on behalf of the STFC DiRAC HPC Facility (www.dirac.ac.uk). Equipment at the DiRAC@Durham facility was funded by BEIS capital funding via STFC capital grants ST/K00042X/1, ST/P002293/1, ST/R002371/1, and ST/S002502/1, Durham University and STFC operations grant ST/R000832/1. DiRAC is part of the National e-Infrastructure.

\vspace{-4mm}

\section*{Data Availability}

The data underlying this article will be shared on reasonable request to the corresponding author.

\vspace{-4mm}



\bibliographystyle{mnras}
\bibliography{bibliography}

@Preamble{ " \newcommand{\noop}[1]{} " }

@ARTICLE{schaye15,
       author = {{Schaye}, Joop and {Crain}, Robert A. and {Bower}, Richard G. and {Furlong}, Michelle and {Schaller}, Matthieu and {Theuns}, Tom and {Dalla Vecchia}, Claudio and {Frenk}, Carlos S. and {McCarthy}, I.~G. and {Helly}, John C. and et al.},
        title = "{The EAGLE project: simulating the evolution and assembly of galaxies and their environments}",
      journal = {\mnras},
         year = 2015,
        month = jan,
       volume = {446},
       number = {1},
        pages = {521-554},
          doi = {10.1093/mnras/stu2058},
archivePrefix = {arXiv},
       eprint = {1407.7040},
 primaryClass = {astro-ph.GA},
       adsurl = {https://ui.adsabs.harvard.edu/abs/2015MNRAS.446..521S}
}

@ARTICLE{schaye26,
       author = {{Schaye}, Joop and {Chaikin}, Evgenii and {Schaller}, Matthieu and {Ploeckinger}, Sylvia and {Hu{\v{s}}ko}, Filip and {McGibbon}, Robert J. and {Trayford}, James W. and {Ben{\'\i}tez-Llambay}, Alejandro and {Correa}, Camila and {Frenk}, Carlos S. and et al.},
        title = "{The COLIBRE project: cosmological hydrodynamical simulations of galaxy formation and evolution}",
      journal = {\mnras},
         year = 2026,
        month = may,
       volume = {548},
       number = {1},
          eid = {stag375},
        pages = {stag375},
          doi = {10.1093/mnras/stag375},
archivePrefix = {arXiv},
       eprint = {2508.21126},
 primaryClass = {astro-ph.GA},
       adsurl = {https://ui.adsabs.harvard.edu/abs/2026MNRAS.548ag375S}
}

@ARTICLE{chaikin_calibration,
       author = {{Chaikin}, Evgenii and {Schaye}, Joop and {Schaller}, Matthieu and {Ploeckinger}, Sylvia and {Bah{\'e}}, Yannick M. and {Ben{\'\i}tez-Llambay}, Alejandro and {Correa}, Camila and {Forouhar Moreno}, Victor J. and {Frenk}, Carlos S. and {Hu{\v{s}}ko}, Filip and et al.},
        title = "{COLIBRE: calibrating subgrid feedback in cosmological simulations that include a cold gas phase}",
      journal = {\mnras},
         year = 2026,
        month = may,
       volume = {548},
       number = {1},
          eid = {stag300},
        pages = {stag300},
          doi = {10.1093/mnras/stag300},
archivePrefix = {arXiv},
       eprint = {2509.04067},
 primaryClass = {astro-ph.GA},
       adsurl = {https://ui.adsabs.harvard.edu/abs/2026MNRAS.548ag300C}
}

@ARTICLE{chaikin_gsmf,
       author = {{Chaikin}, Evgenii and {Schaye}, Joop and {Schaller}, Matthieu and {Ploeckinger}, Sylvia and {Ben{\'\i}tez-Llambay}, Alejandro and {Frenk}, Carlos S. and {Hu{\v{s}}ko}, Filip and {McGibbon}, Robert J. and {Richings}, Alexander J. and {Trayford}, James W.},
        title = "{The evolution of the galaxy stellar mass function and star formation rates in the COLIBRE simulations from redshift 17 to 0}",
      journal = {\mnras},
         year = 2026,
        month = jun,
       volume = {548},
       number = {4},
          eid = {stag740},
        pages = {stag740},
          doi = {10.1093/mnras/stag740},
archivePrefix = {arXiv},
       eprint = {2509.07960},
 primaryClass = {astro-ph.GA},
       adsurl = {https://ui.adsabs.harvard.edu/abs/2026MNRAS.548ag740C}
}

@ARTICLE{schaller24,
       author = {{Schaller}, Matthieu and {Borrow}, Josh and {Draper}, Peter W. and {Ivkovic}, Mladen and {McAlpine}, Stuart and {Vandenbroucke}, Bert and {Bah{\'e}}, Yannick and {Chaikin}, Evgenii and {Chalk}, Aidan B.~G. and {Chan}, Tsang Keung and et al.},
        title = "{SWIFT: A modern highly-parallel gravity and smoothed particle hydrodynamics solver for astrophysical and cosmological applications}",
      journal = {\mnras},
         year = 2024,
        month = may,
       volume = {530},
       number = {2},
        pages = {2378-2419},
          doi = {10.1093/mnras/stae922},
archivePrefix = {arXiv},
       eprint = {2305.13380},
 primaryClass = {astro-ph.IM},
       adsurl = {https://ui.adsabs.harvard.edu/abs/2024MNRAS.530.2378S}
}

@ARTICLE{borrow22,
       author = {{Borrow}, Josh and {Schaller}, Matthieu and {Bower}, Richard G. and {Schaye}, Joop},
        title = "{SPHENIX: smoothed particle hydrodynamics for the next generation of galaxy formation simulations}",
      journal = {\mnras},
         year = 2022,
        month = apr,
       volume = {511},
       number = {2},
        pages = {2367-2389},
          doi = {10.1093/mnras/stab3166},
archivePrefix = {arXiv},
       eprint = {2012.03974},
 primaryClass = {astro-ph.GA},
       adsurl = {https://ui.adsabs.harvard.edu/abs/2022MNRAS.511.2367B}
}

@software{hahn20,
       author = {{Hahn}, Oliver and {Michaux}, Micha{\"e}l and {Rampf}, Cornelius and {Uhlemann}, Cora and {Angulo}, Raul E.},
        title = "{MUSIC2-monofonIC: 3LPT initial condition generator}",
 howpublished = {Astrophysics Source Code Library, record ascl:2008.024},
         year = 2020,
        month = aug,
          eid = {ascl:2008.024},
archivePrefix = {ascl},
       eprint = {2008.024},
       adsurl = {https://ui.adsabs.harvard.edu/abs/2020ascl.soft08024H}
}

@ARTICLE{michaux21,
       author = {{Michaux}, Micha{\"e}l and {Hahn}, Oliver and {Rampf}, Cornelius and {Angulo}, Raul E.},
        title = "{Accurate initial conditions for cosmological N-body simulations: minimizing truncation and discreteness errors}",
      journal = {\mnras},
         year = 2021,
        month = jan,
       volume = {500},
       number = {1},
        pages = {663-683},
          doi = {10.1093/mnras/staa3149},
archivePrefix = {arXiv},
       eprint = {2008.09588},
 primaryClass = {astro-ph.CO},
       adsurl = {https://ui.adsabs.harvard.edu/abs/2021MNRAS.500..663M}
}

@ARTICLE{ludlow19,
       author = {{Ludlow}, Aaron D. and {Schaye}, Joop and {Schaller}, Matthieu and {Richings}, Jack},
        title = "{Energy equipartition between stellar and dark matter particles in cosmological simulations results in spurious growth of galaxy sizes}",
      journal = {\mnras},
         year = 2019,
        month = sep,
       volume = {488},
       number = {1},
        pages = {L123-L128},
          doi = {10.1093/mnrasl/slz110},
archivePrefix = {arXiv},
       eprint = {1903.10110},
 primaryClass = {astro-ph.GA},
       adsurl = {https://ui.adsabs.harvard.edu/abs/2019MNRAS.488L.123L}
}

@ARTICLE{ludlow21,
       author = {{Ludlow}, Aaron D. and {Fall}, S. Michael and {Schaye}, Joop and {Obreschkow}, Danail},
        title = "{Spurious heating of stellar motions in simulated galactic discs by dark matter halo particles}",
      journal = {\mnras},
         year = 2021,
        month = dec,
       volume = {508},
       number = {4},
        pages = {5114-5137},
          doi = {10.1093/mnras/stab2770},
archivePrefix = {arXiv},
       eprint = {2105.03561},
 primaryClass = {astro-ph.GA},
       adsurl = {https://ui.adsabs.harvard.edu/abs/2021MNRAS.508.5114L}
}

@ARTICLE{ludlow23,
       author = {{Ludlow}, Aaron D. and {Fall}, S. Michael and {Wilkinson}, Matthew J. and {Schaye}, Joop and {Obreschkow}, Danail},
        title = "{Spurious heating of stellar motions by dark matter particles in cosmological simulations of galaxy formation}",
      journal = {\mnras},
         year = 2023,
        month = nov,
       volume = {525},
       number = {4},
        pages = {5614-5630},
          doi = {10.1093/mnras/stad2615},
archivePrefix = {arXiv},
       eprint = {2306.05753},
 primaryClass = {astro-ph.GA},
       adsurl = {https://ui.adsabs.harvard.edu/abs/2023MNRAS.525.5614L}
}

@ARTICLE{wilkinson23,
       author = {{Wilkinson}, Matthew J. and {Ludlow}, Aaron D. and {Lagos}, Claudia del P. and {Fall}, S. Michael and {Schaye}, Joop and {Obreschkow}, Danail},
        title = "{The impact of spurious collisional heating on the morphological evolution of simulated galactic discs}",
      journal = {\mnras},
         year = 2023,
        month = mar,
       volume = {519},
       number = {4},
        pages = {5942-5961},
          doi = {10.1093/mnras/stad055},
archivePrefix = {arXiv},
       eprint = {2208.07623},
 primaryClass = {astro-ph.GA},
       adsurl = {https://ui.adsabs.harvard.edu/abs/2023MNRAS.519.5942W}
}

@ARTICLE{forouhar25,
       author = {{Forouhar Moreno}, Victor J. and {Helly}, John and {McGibbon}, Robert and {Schaye}, Joop and {Schaller}, Matthieu and {Han}, Jiaxin and {Kugel}, Roi and {Bah{\'e}}, Yannick M.},
        title = "{Assessing subhalo finders in cosmological hydrodynamical simulations}",
      journal = {\mnras},
         year = 2025,
        month = sep,
          doi = {10.1093/mnras/staf1478},
archivePrefix = {arXiv},
       eprint = {2502.06932},
 primaryClass = {astro-ph.CO},
       adsurl = {https://ui.adsabs.harvard.edu/abs/2025MNRAS.tmp.1440F}
}

@ARTICLE{han18,
       author = {{Han}, Jiaxin and {Cole}, Shaun and {Frenk}, Carlos S. and {Benitez-Llambay}, Alejandro and {Helly}, John},
        title = "{HBT+: an improved code for finding subhaloes and building merger trees in cosmological simulations}",
      journal = {\mnras},
         year = 2018,
        month = feb,
       volume = {474},
       number = {1},
        pages = {604-617},
          doi = {10.1093/mnras/stx2792},
archivePrefix = {arXiv},
       eprint = {1708.03646},
 primaryClass = {astro-ph.CO},
       adsurl = {https://ui.adsabs.harvard.edu/abs/2018MNRAS.474..604H}
}

@ARTICLE{mcgibbon25,
       author = {{McGibbon}, Robert and {Helly}, John and {Schaye}, Joop and {Schaller}, Matthieu and {Vandenbroucke}, Bert},
        title = "{SOAP: A Python Package for Calculating the Properties of Galaxies and Halos Formed in Cosmological Simulations}",
      journal = {The Journal of Open Source Software},
         year = 2025,
        month = jul,
       volume = {10},
       number = {111},
          eid = {8252},
        pages = {8252},
          doi = {10.21105/joss.08252},
archivePrefix = {arXiv},
       eprint = {2507.22669},
 primaryClass = {astro-ph.IM},
       adsurl = {https://ui.adsabs.harvard.edu/abs/2025JOSS...10.8252M}
}

@ARTICLE{abbott22,
       author = {{Abbott}, T.~M.~C. and {Aguena}, M. and {Alarcon}, A. and {Allam}, S. and {Alves}, O. and {Amon}, A. and {Andrade-Oliveira}, F. and {Annis}, J. and {Avila}, S. and {Bacon}, D. and et al.},
        title = "{Dark Energy Survey Year 3 results: Cosmological constraints from galaxy clustering and weak lensing}",
      journal = {\prd},
         year = 2022,
        month = jan,
       volume = {105},
       number = {2},
          eid = {023520},
        pages = {023520},
          doi = {10.1103/PhysRevD.105.023520},
archivePrefix = {arXiv},
       eprint = {2105.13549},
 primaryClass = {astro-ph.CO},
       adsurl = {https://ui.adsabs.harvard.edu/abs/2022PhRvD.105b3520A}
}

@ARTICLE{kugel23,
       author = {{Kugel}, Roi and {Schaye}, Joop and {Schaller}, Matthieu and {Helly}, John C. and {Braspenning}, Joey and {Elbers}, Willem and {Frenk}, Carlos S. and {McCarthy}, Ian G. and {Kwan}, Juliana and {Salcido}, Jaime and et al.},
        title = "{FLAMINGO: calibrating large cosmological hydrodynamical simulations with machine learning}",
      journal = {\mnras},
         year = 2023,
        month = dec,
       volume = {526},
       number = {4},
        pages = {6103-6127},
          doi = {10.1093/mnras/stad2540},
archivePrefix = {arXiv},
       eprint = {2306.05492},
 primaryClass = {astro-ph.CO},
       adsurl = {https://ui.adsabs.harvard.edu/abs/2023MNRAS.526.6103K}
}

@ARTICLE{luciesmith25,
       author = {{Lucie-Smith}, Luisa and {Peiris}, Hiranya V. and {Pontzen}, Andrew and {Halder}, Anik and {Schaye}, Joop and {Schaller}, Matthieu and {Helly}, John and {McGibbon}, Robert J. and {Elbers}, Willem},
        title = "{Cosmological feedback from a halo assembly perspective}",
      journal = {\prd},
         year = 2025,
        month = sep,
       volume = {112},
       number = {6},
          eid = {063541},
        pages = {063541},
          doi = {10.1103/vh8n-9cr2},
archivePrefix = {arXiv},
       eprint = {2505.18258},
 primaryClass = {astro-ph.CO},
       adsurl = {https://ui.adsabs.harvard.edu/abs/2025PhRvD.112f3541L}
}

@ARTICLE{costello25,
       author = {{Costello}, Emily E. and {McCarthy}, Ian G. and {Salcido}, Jaime and {Helly}, John C. and {McGibbon}, Robert J. and {Schaller}, Matthieu and {Schaye}, Joop},
        title = "{FLAMINGO: tracing the co-evolution of hot gas and black holes in galaxy groups and clusters}",
      journal = {\mnras},
         year = 2026,
        month = jul,
       volume = {549},
       number = {4},
          eid = {stag945},
        pages = {stag945},
          doi = {10.1093/mnras/stag945},
archivePrefix = {arXiv},
       eprint = {2510.17980},
 primaryClass = {astro-ph.CO},
       adsurl = {https://scixplorer.org/abs/2026MNRAS.549ag945C}
}

@ARTICLE{bower17,
       author = {{Bower}, Richard G. and {Schaye}, Joop and {Frenk}, Carlos S. and {Theuns}, Tom and {Schaller}, Matthieu and {Crain}, Robert A. and {McAlpine}, Stuart},
        title = "{The dark nemesis of galaxy formation: why hot haloes trigger black hole growth and bring star formation to an end}",
      journal = {\mnras},
         year = 2017,
        month = feb,
       volume = {465},
       number = {1},
        pages = {32-44},
          doi = {10.1093/mnras/stw2735},
archivePrefix = {arXiv},
       eprint = {1607.07445},
 primaryClass = {astro-ph.GA},
       adsurl = {https://ui.adsabs.harvard.edu/abs/2017MNRAS.465...32B}
}

@ARTICLE{mcalpine18,
       author = {{McAlpine}, Stuart and {Bower}, Richard G. and {Rosario}, David J. and {Crain}, Robert A. and {Schaye}, Joop and {Theuns}, Tom},
        title = "{The rapid growth phase of supermassive black holes}",
      journal = {\mnras},
         year = 2018,
        month = dec,
       volume = {481},
       number = {3},
        pages = {3118-3128},
          doi = {10.1093/mnras/sty2489},
archivePrefix = {arXiv},
       eprint = {1805.08293},
 primaryClass = {astro-ph.GA},
       adsurl = {https://ui.adsabs.harvard.edu/abs/2018MNRAS.481.3118M}
}

@ARTICLE{boothschaye10,
       author = {{Booth}, C.~M. and {Schaye}, Joop},
        title = "{Dark matter haloes determine the masses of supermassive black holes}",
      journal = {\mnras},
         year = 2010,
        month = jun,
       volume = {405},
       number = {1},
        pages = {L1-L5},
          doi = {10.1111/j.1745-3933.2010.00832.x},
archivePrefix = {arXiv},
       eprint = {0911.0935},
 primaryClass = {astro-ph.CO},
       adsurl = {https://ui.adsabs.harvard.edu/abs/2010MNRAS.405L...1B}
}

@ARTICLE{terrazas20,
       author = {{Terrazas}, Bryan A. and {Bell}, Eric F. and {Pillepich}, Annalisa and {Nelson}, Dylan and {Somerville}, Rachel S. and {Genel}, Shy and {Weinberger}, Rainer and {Habouzit}, M{\'e}lanie and {Li}, Yuan and {Hernquist}, Lars and et al.},
        title = "{The relationship between black hole mass and galaxy properties: examining the black hole feedback model in IllustrisTNG}",
      journal = {\mnras},
         year = 2020,
        month = apr,
       volume = {493},
       number = {2},
        pages = {1888-1906},
          doi = {10.1093/mnras/staa374},
archivePrefix = {arXiv},
       eprint = {1906.02747},
 primaryClass = {astro-ph.GA},
       adsurl = {https://ui.adsabs.harvard.edu/abs/2020MNRAS.493.1888T}
}

@ARTICLE{davies19,
       author = {{Davies}, Jonathan J. and {Crain}, Robert A. and {McCarthy}, Ian G. and {Oppenheimer}, Benjamin D. and {Schaye}, Joop and {Schaller}, Matthieu and {McAlpine}, Stuart},
        title = "{The gas fractions of dark matter haloes hosting simulated {\ensuremath{\sim}}L$^{{\ensuremath{\star}}}$ galaxies are governed by the feedback history of their black holes}",
      journal = {\mnras},
         year = 2019,
        month = may,
       volume = {485},
       number = {3},
        pages = {3783-3793},
          doi = {10.1093/mnras/stz635},
archivePrefix = {arXiv},
       eprint = {1810.07696},
 primaryClass = {astro-ph.GA},
       adsurl = {https://ui.adsabs.harvard.edu/abs/2019MNRAS.485.3783D}
}

@ARTICLE{davies20,
       author = {{Davies}, Jonathan J. and {Crain}, Robert A. and {Oppenheimer}, Benjamin D. and {Schaye}, Joop},
        title = "{The quenching and morphological evolution of central galaxies is facilitated by the feedback-driven expulsion of circumgalactic gas}",
      journal = {\mnras},
         year = 2020,
        month = jan,
       volume = {491},
       number = {3},
        pages = {4462-4480},
          doi = {10.1093/mnras/stz3201},
archivePrefix = {arXiv},
       eprint = {1908.11380},
 primaryClass = {astro-ph.GA},
       adsurl = {https://ui.adsabs.harvard.edu/abs/2020MNRAS.491.4462D}
}

@ARTICLE{davies22,
       author = {{Davies}, Jonathan J. and {Pontzen}, Andrew and {Crain}, Robert A.},
        title = "{Galaxy mergers can initiate quenching by unlocking an AGN-driven transformation of the baryon cycle}",
      journal = {\mnras},
         year = 2022,
        month = sep,
       volume = {515},
       number = {1},
        pages = {1430-1443},
          doi = {10.1093/mnras/stac1742},
archivePrefix = {arXiv},
       eprint = {2203.08157},
 primaryClass = {astro-ph.GA},
       adsurl = {https://ui.adsabs.harvard.edu/abs/2022MNRAS.515.1430D}
}

@ARTICLE{davies24,
       author = {{Davies}, Jonathan J. and {Pontzen}, Andrew and {Crain}, Robert A.},
        title = "{Are the fates of supermassive black holes and galaxies determined by individual mergers, or by the properties of their host haloes?}",
      journal = {\mnras},
         year = 2024,
        month = jan,
       volume = {527},
       number = {3},
        pages = {4705-4716},
          doi = {10.1093/mnras/stad3456},
archivePrefix = {arXiv},
       eprint = {2301.04145},
 primaryClass = {astro-ph.GA},
       adsurl = {https://ui.adsabs.harvard.edu/abs/2024MNRAS.527.4705D}
}

@ARTICLE{crainvandevoort23,
       author = {{Crain}, Robert A. and {van de Voort}, Freeke},
        title = "{Hydrodynamical Simulations of the Galaxy Population: Enduring Successes and Outstanding Challenges}",
      journal = {\araa},
         year = 2023,
        month = aug,
       volume = {61},
        pages = {473-515},
          doi = {10.1146/annurev-astro-041923-043618},
archivePrefix = {arXiv},
       eprint = {2309.17075},
 primaryClass = {astro-ph.GA},
       adsurl = {https://ui.adsabs.harvard.edu/abs/2023ARA&A..61..473C}
}

@ARTICLE{schaye23,
       author = {{Schaye}, Joop and {Kugel}, Roi and {Schaller}, Matthieu and {Helly}, John C. and {Braspenning}, Joey and {Elbers}, Willem and {McCarthy}, Ian G. and {van Daalen}, Marcel P. and {Vandenbroucke}, Bert and {Frenk}, Carlos S. and et al.},
        title = "{The FLAMINGO project: cosmological hydrodynamical simulations for large-scale structure and galaxy cluster surveys}",
      journal = {\mnras},
         year = 2023,
        month = dec,
       volume = {526},
       number = {4},
        pages = {4978-5020},
          doi = {10.1093/mnras/stad2419},
archivePrefix = {arXiv},
       eprint = {2306.04024},
 primaryClass = {astro-ph.CO},
       adsurl = {https://ui.adsabs.harvard.edu/abs/2023MNRAS.526.4978S}
}

@ARTICLE{mccarthy17,
       author = {{McCarthy}, Ian G. and {Schaye}, Joop and {Bird}, Simeon and {Le Brun}, Amandine M.~C.},
        title = "{The BAHAMAS project: calibrated hydrodynamical simulations for large-scale structure cosmology}",
      journal = {\mnras},
         year = 2017,
        month = mar,
       volume = {465},
       number = {3},
        pages = {2936-2965},
          doi = {10.1093/mnras/stw2792},
archivePrefix = {arXiv},
       eprint = {1603.02702},
 primaryClass = {astro-ph.CO},
       adsurl = {https://ui.adsabs.harvard.edu/abs/2017MNRAS.465.2936M}
}

@ARTICLE{henden18,
       author = {{Henden}, Nicholas A. and {Puchwein}, Ewald and {Shen}, Sijing and {Sijacki}, Debora},
        title = "{The FABLE simulations: a feedback model for galaxies, groups, and clusters}",
      journal = {\mnras},
         year = 2018,
        month = oct,
       volume = {479},
       number = {4},
        pages = {5385-5412},
          doi = {10.1093/mnras/sty1780},
archivePrefix = {arXiv},
       eprint = {1804.05064},
 primaryClass = {astro-ph.GA},
       adsurl = {https://ui.adsabs.harvard.edu/abs/2018MNRAS.479.5385H}
}

@ARTICLE{bigwood25xfable,
       author = {{Bigwood}, Leah and {Bourne}, Martin A. and {Ir{\v{s}}i{\v{c}}}, Vid and {Amon}, Alexandra and {Sijacki}, Debora},
        title = "{The case for large-scale AGN feedback in galaxy formation simulations: insights from XFABLE}",
      journal = {\mnras},
         year = 2025,
        month = sep,
       volume = {542},
       number = {4},
        pages = {3206-3230},
          doi = {10.1093/mnras/staf1435},
archivePrefix = {arXiv},
       eprint = {2501.16983},
 primaryClass = {astro-ph.CO},
       adsurl = {https://ui.adsabs.harvard.edu/abs/2025MNRAS.542.3206H}
}

@ARTICLE{truong21,
       author = {{Truong}, Nhut and {Pillepich}, Annalisa and {Nelson}, Dylan and {Werner}, Norbert and {Hernquist}, Lars},
        title = "{Predictions for anisotropic X-ray signatures in the circumgalactic medium: imprints of supermassive black hole driven outflows}",
      journal = {\mnras},
         year = 2021,
        month = dec,
       volume = {508},
       number = {2},
        pages = {1563-1581},
          doi = {10.1093/mnras/stab2638},
archivePrefix = {arXiv},
       eprint = {2109.06884},
 primaryClass = {astro-ph.GA},
       adsurl = {https://ui.adsabs.harvard.edu/abs/2021MNRAS.508.1563T}
}

@ARTICLE{nica22,
       author = {{Nica}, Anna and {Oppenheimer}, Benjamin D. and {Crain}, Robert A. and {Bogd{\'a}n}, {\'A}kos and {Davies}, Jonathan J. and {Forman}, William R. and {Kraft}, Ralph P. and {ZuHone}, John A.},
        title = "{Predictions for the X-ray circumgalactic medium of edge-on discs and spheroids}",
      journal = {\mnras},
         year = 2022,
        month = dec,
       volume = {517},
       number = {2},
        pages = {1958-1969},
          doi = {10.1093/mnras/stac2020},
archivePrefix = {arXiv},
       eprint = {2112.14919},
 primaryClass = {astro-ph.GA},
       adsurl = {https://ui.adsabs.harvard.edu/abs/2022MNRAS.517.1958N}
}

@ARTICLE{silich25,
       author = {{Silich}, Emily M. and {ZuHone}, John and {Bellomi}, Elena and {Hummels}, Cameron and {Oppenheimer}, Benjamin and {Hopkins}, Philip F. and {Lochhaas}, Cassandra and {Ponnada}, Sam B. and {Vikhlinin}, Alexey},
        title = "{X-Ray Emission Signatures of Galactic Feedback in the Hot Circumgalactic Medium: Predictions from Cosmological Hydrodynamical Simulations}",
      journal = {\apj},
         year = 2025,
        month = nov,
       volume = {993},
       number = {1},
          eid = {125},
        pages = {125},
          doi = {10.3847/1538-4357/ae08a3},
archivePrefix = {arXiv},
       eprint = {2506.17440},
 primaryClass = {astro-ph.GA},
       adsurl = {https://ui.adsabs.harvard.edu/abs/2025ApJ...993..125S}
}

@ARTICLE{husko26,
       author = {{Hu{\v{s}}ko}, Filip and {Lacey}, Cedric G. and {Schaye}, Joop and {Schaller}, Matthieu and {Chaikin}, Evgenii and {Ploeckinger}, Sylvia and {Ben{\'\i}tez Llambay}, Alejandro and {Richings}, Alexander J. and {Trayford}, James W.},
        title = "{A hybrid active galactic nucleus feedback model with spinning black holes, winds and jets}",
      journal = {\mnras},
         year = 2026,
        month = apr,
       volume = {547},
       number = {2},
          eid = {stag324},
        pages = {stag324},
          doi = {10.1093/mnras/stag324},
archivePrefix = {arXiv},
       eprint = {2509.05179},
 primaryClass = {astro-ph.GA},
       adsurl = {https://ui.adsabs.harvard.edu/abs/2026MNRAS.547ag324H}
}

@ARTICLE{crain15,
       author = {{Crain}, Robert A. and {Schaye}, Joop and {Bower}, Richard G. and {Furlong}, Michelle and {Schaller}, Matthieu and {Theuns}, Tom and {Dalla Vecchia}, Claudio and {Frenk}, Carlos S. and {McCarthy}, Ian G. and {Helly}, John C. and et al.},
        title = "{The EAGLE simulations of galaxy formation: calibration of subgrid physics and model variations}",
      journal = {\mnras},
         year = 2015,
        month = jun,
       volume = {450},
       number = {2},
        pages = {1937-1961},
          doi = {10.1093/mnras/stv725},
archivePrefix = {arXiv},
       eprint = {1501.01311},
 primaryClass = {astro-ph.GA},
       adsurl = {https://ui.adsabs.harvard.edu/abs/2015MNRAS.450.1937C}
}

@ARTICLE{pillepich18,
       author = {{Pillepich}, Annalisa and {Springel}, Volker and {Nelson}, Dylan and {Genel}, Shy and {Naiman}, Jill and {Pakmor}, R{\"u}diger and {Hernquist}, Lars and {Torrey}, Paul and {Vogelsberger}, Mark and {Weinberger}, Rainer and et al.},
        title = "{Simulating galaxy formation with the IllustrisTNG model}",
      journal = {\mnras},
         year = 2018,
        month = jan,
       volume = {473},
       number = {3},
        pages = {4077-4106},
          doi = {10.1093/mnras/stx2656},
archivePrefix = {arXiv},
       eprint = {1703.02970},
 primaryClass = {astro-ph.GA},
       adsurl = {https://ui.adsabs.harvard.edu/abs/2018MNRAS.473.4077P}
}

@ARTICLE{kelly21,
       author = {{Kelly}, Ashley J. and {Jenkins}, Adrian and {Frenk}, Carlos S.},
        title = "{The origin of X-ray coronae around simulated disc galaxies}",
      journal = {\mnras},
         year = 2021,
        month = apr,
       volume = {502},
       number = {2},
        pages = {2934-2951},
          doi = {10.1093/mnras/stab255},
archivePrefix = {arXiv},
       eprint = {2005.12926},
 primaryClass = {astro-ph.GA},
       adsurl = {https://ui.adsabs.harvard.edu/abs/2021MNRAS.502.2934K}
}

@ARTICLE{wright24,
       author = {{Wright}, Ruby J. and {Somerville}, Rachel S. and {Lagos}, Claudia del P. and {Schaller}, Matthieu and {Dav{\'e}}, Romeel and {Angl{\'e}s-Alc{\'a}zar}, Daniel and {Genel}, Shy},
        title = "{The baryon cycle in modern cosmological hydrodynamical simulations}",
      journal = {\mnras},
         year = 2024,
        month = aug,
       volume = {532},
       number = {3},
        pages = {3417-3440},
          doi = {10.1093/mnras/stae1688},
archivePrefix = {arXiv},
       eprint = {2402.08408},
 primaryClass = {astro-ph.GA},
       adsurl = {https://ui.adsabs.harvard.edu/abs/2024MNRAS.532.3417W}
}

@ARTICLE{hough23,
       author = {{Hough}, Renier T. and {Rennehan}, Douglas and {Kobayashi}, Chiaki and {Loubser}, S. Ilani and {Dav{\'e}}, Romeel and {Babul}, Arif and {Cui}, Weiguang},
        title = "{SIMBA-C: an updated chemical enrichment model for galactic chemical evolution in the SIMBA simulation}",
      journal = {\mnras},
         year = 2023,
        month = oct,
       volume = {525},
       number = {1},
        pages = {1061-1076},
          doi = {10.1093/mnras/stad2394},
archivePrefix = {arXiv},
       eprint = {2308.03436},
 primaryClass = {astro-ph.GA},
       adsurl = {https://ui.adsabs.harvard.edu/abs/2023MNRAS.525.1061H}
}

@ARTICLE{dave19,
       author = {{Dav{\'e}}, Romeel and {Angl{\'e}s-Alc{\'a}zar}, Daniel and {Narayanan}, Desika and {Li}, Qi and {Rafieferantsoa}, Mika H. and {Appleby}, Sarah},
        title = "{SIMBA: Cosmological simulations with black hole growth and feedback}",
      journal = {\mnras},
         year = 2019,
        month = jun,
       volume = {486},
       number = {2},
        pages = {2827-2849},
          doi = {10.1093/mnras/stz937},
archivePrefix = {arXiv},
       eprint = {1901.10203},
 primaryClass = {astro-ph.GA},
       adsurl = {https://ui.adsabs.harvard.edu/abs/2019MNRAS.486.2827D}
}

@ARTICLE{hirschmann14,
       author = {{Hirschmann}, Michaela and {Dolag}, Klaus and {Saro}, Alexandro and {Bachmann}, Lisa and {Borgani}, Stefano and {Burkert}, Andreas},
        title = "{Cosmological simulations of black hole growth: AGN luminosities and downsizing}",
      journal = {\mnras},
         year = 2014,
        month = aug,
       volume = {442},
       number = {3},
        pages = {2304-2324},
          doi = {10.1093/mnras/stu1023},
archivePrefix = {arXiv},
       eprint = {1308.0333},
 primaryClass = {astro-ph.CO},
       adsurl = {https://ui.adsabs.harvard.edu/abs/2014MNRAS.442.2304H}
}

@ARTICLE{dolag25,
       author = {{Dolag}, Klaus and {Remus}, Rhea-Silvia and {Valenzuela}, Lucas M. and {Kimmig}, Lucas C. and {Seidel}, Benjamin and {Fortune}, Silvio and {Stoiber}, Johannes and {Ivleva}, Anna and {Hoffmann}, Tadziu and {Biffi}, Veronica and et al.},
        title = "{Encyclopedia Magneticum: Scaling Relations from Cosmic Dawn to Present Day}",
      journal = {arXiv e-prints},
         year = 2025,
        month = apr,
          eid = {arXiv:2504.01061},
        pages = {arXiv:2504.01061},
          doi = {10.48550/arXiv.2504.01061},
archivePrefix = {arXiv},
       eprint = {2504.01061},
 primaryClass = {astro-ph.CO},
       adsurl = {https://ui.adsabs.harvard.edu/abs/2025arXiv250401061D}
}

@ARTICLE{mccarthy25,
       author = {{McCarthy}, Ian G. and {Amon}, Alexandra and {Schaye}, Joop and {Schaan}, Emmanuel and {Angulo}, Raul E. and {Salcido}, Jaime and {Schaller}, Matthieu and {Bigwood}, Leah and {Elbers}, Willem and {Kugel}, Roi and et al.},
        title = "{FLAMINGO: combining kinetic SZ effect and galaxy{\textendash}galaxy lensing measurements to gauge the impact of feedback on large-scale structure}",
      journal = {\mnras},
         year = 2025,
        month = jun,
       volume = {540},
       number = {1},
        pages = {143-163},
          doi = {10.1093/mnras/staf731},
archivePrefix = {arXiv},
       eprint = {2410.19905},
 primaryClass = {astro-ph.CO},
       adsurl = {https://ui.adsabs.harvard.edu/abs/2025MNRAS.540..143M}
}

@ARTICLE{bigwood24,
       author = {{Bigwood}, L. and {Amon}, A. and {Schneider}, A. and {Salcido}, J. and {McCarthy}, I.~G. and {Preston}, C. and {Sanchez}, D. and {Sijacki}, D. and {Schaan}, E. and {Ferraro}, S. and et al.},
        title = "{Weak lensing combined with the kinetic Sunyaev-Zel'dovich effect: a study of baryonic feedback}",
      journal = {\mnras},
         year = 2024,
        month = oct,
       volume = {534},
       number = {1},
        pages = {655-682},
          doi = {10.1093/mnras/stae2100},
archivePrefix = {arXiv},
       eprint = {2404.06098},
 primaryClass = {astro-ph.CO},
       adsurl = {https://ui.adsabs.harvard.edu/abs/2024MNRAS.534..655B}
}

@ARTICLE{bigwood25kSZ,
       author = {{Bigwood}, Leah and {Yamamoto}, Masaya and {Siegel}, Jared and {Amon}, Alexandra and {McCarthy}, Ian G. and {Dave}, Romeel and {Salcido}, Jaime and {Schaller}, Matthieu and {Schaye}, Joop and {Yang}, Tianyi},
        title = "{The kinetic Sunyaev Zeldovich effect as a benchmark for AGN feedback models in hydrodynamical simulations: insights from DESI + ACT}",
      journal = {\mnras},
         year = 2026,
        month = aug,
       volume = {550},
       number = {4},
          eid = {stag1314},
        pages = {stag1314},
          doi = {10.1093/mnras/stag1314},
archivePrefix = {arXiv},
       eprint = {2510.15822},
 primaryClass = {astro-ph.CO},
       adsurl = {https://scixplorer.org/abs/2026MNRAS.550g1314B}
}

@ARTICLE{siegel26a,
       author = {{Siegel}, Jared and {Bigwood}, Leah and {Amon}, Alexandra and {McCullough}, Jamie and {Yamamoto}, Masaya and {McCarthy}, Ian G. and {Schaller}, Matthieu and {Schaye}, Joop and {Schneider}, Aurel},
        title = "{The suppression of the matter power spectrum: strong feedback from X-ray gas mass fractions, kSZ effect profiles, and galaxy-galaxy lensing}",
      journal = {\mnras},
         year = 2026,
        month = may,
          doi = {10.1093/mnras/stag993},
archivePrefix = {arXiv},
       eprint = {2512.02954},
 primaryClass = {astro-ph.CO},
       adsurl = {https://ui.adsabs.harvard.edu/abs/2026MNRAS.tmp..950S}
}

@ARTICLE{siegel26b,
       author = {{Siegel}, Jared C. and {Amon}, Alexandra and {McCarthy}, Ian G. and {Bigwood}, Leah and {Yamamoto}, Masaya and {Bulbul}, Esra and {Greene}, Jenny E. and {McCullough}, Jamie and {Schaller}, Matthieu and {Schaye}, Joop},
        title = "{Joint X-Ray, Kinetic Sunyaev─Zeldovich, and Weak Lensing Measurements: Toward a Consensus Picture of Efficient Gas Expulsion from Groups and Clusters}",
      journal = {\apj},
         year = 2026,
        month = jun,
       volume = {1003},
       number = {2},
          eid = {151},
        pages = {151},
          doi = {10.3847/1538-4357/ae5dc2},
archivePrefix = {arXiv},
       eprint = {2509.10455},
 primaryClass = {astro-ph.CO},
       adsurl = {https://ui.adsabs.harvard.edu/abs/2026ApJ..1003..151S}
}

@ARTICLE{kovac25,
       author = {{Kova{\v{c}}}, Michael and {Nicola}, Andrina and {Bucko}, Jozef and {Schneider}, Aurel and {Reischke}, Robert and {Giri}, Sambit K. and {Teyssier}, Romain and {Schaller}, Matthieu and {Schaye}, Joop},
        title = "{Baryonification II: constraining feedback with X-ray and kinematic Sunyaev-Zel'dovich observations}",
      journal = {\jcap},
         year = 2025,
        month = nov,
       volume = {2025},
       number = {11},
          eid = {046},
        pages = {046},
          doi = {10.1088/1475-7516/2025/11/046},
archivePrefix = {arXiv},
       eprint = {2507.07991},
 primaryClass = {astro-ph.CO},
       adsurl = {https://ui.adsabs.harvard.edu/abs/2025JCAP...11..046K}
}

@ARTICLE{popesso26,
       author = {{Popesso}, P. and {Biviano}, A. and {Marini}, I. and {Dolag}, K. and {Vladutescu-Zopp}, S. and {Csizi}, B. and {Biffi}, V. and {Lamer}, G. and {Robothan}, A. and {Bravo}, M. and et al.},
        title = "{The hot gas mass fraction in halos: From Milky Way-like groups to massive clusters}",
      journal = {\aap},
         year = 2026,
        month = mar,
       volume = {707},
          eid = {A362},
        pages = {A362},
          doi = {10.1051/0004-6361/202453256},
archivePrefix = {arXiv},
       eprint = {2411.16555},
 primaryClass = {astro-ph.GA},
       adsurl = {https://ui.adsabs.harvard.edu/abs/2026A&A...707A.362P}
}

@ARTICLE{popesso24a,
       author = {{Popesso}, P. and {Biviano}, A. and {Bulbul}, E. and {Merloni}, A. and {Comparat}, J. and {Clerc}, N. and {Igo}, Z. and {Liu}, A. and {Driver}, S. and {Salvato}, M. and et al.},
        title = "{The X-ray invisible Universe. A look into the haloes undetected by eROSITA}",
      journal = {\mnras},
         year = 2024,
        month = jan,
       volume = {527},
       number = {1},
        pages = {895-910},
          doi = {10.1093/mnras/stad3253},
archivePrefix = {arXiv},
       eprint = {2302.08405},
 primaryClass = {astro-ph.CO},
       adsurl = {https://ui.adsabs.harvard.edu/abs/2024MNRAS.527..895P}
}

@ARTICLE{popesso25xray,
       author = {{Popesso}, P. and {Marini}, I. and {Dolag}, K. and {Lamer}, G. and {Csizi}, B. and {Biffi}, V. and {Robothan}, A. and {Bravo}, M. and {Biviano}, A. and {Vladutescu-Zopp}, S. and et al.},
        title = "{Average X-ray properties of galaxy groups: From Milky Way-like halos to massive clusters}",
      journal = {\aap},
         year = 2025,
        month = dec,
       volume = {704},
          eid = {A278},
        pages = {A278},
          doi = {10.1051/0004-6361/202453255},
archivePrefix = {arXiv},
       eprint = {2411.17120},
 primaryClass = {astro-ph.GA},
       adsurl = {https://ui.adsabs.harvard.edu/abs/2025A&A...704A.278P}
}

@ARTICLE{zhang24,
       author = {{Zhang}, Yi and {Comparat}, Johan and {Ponti}, Gabriele and {Merloni}, Andrea and {Nandra}, Kirpal and {Haberl}, Frank and {Locatelli}, Nicola and {Zhang}, Xiaoyuan and {Sanders}, Jeremy and {Zheng}, Xueying and et al.},
        title = "{The hot circumgalactic medium in the eROSITA All-Sky Survey: I. X-ray surface brightness profiles}",
      journal = {\aap},
         year = 2024,
        month = oct,
       volume = {690},
          eid = {A267},
        pages = {A267},
          doi = {10.1051/0004-6361/202449412},
archivePrefix = {arXiv},
       eprint = {2401.17308},
 primaryClass = {astro-ph.GA},
       adsurl = {https://ui.adsabs.harvard.edu/abs/2024A&A...690A.267Z}
}

@ARTICLE{zhang26,
       author = {{Zhang}, Yi and {Shreeram}, Soumya and {Ponti}, Gabriele and {Comparat}, Johan and {Merloni}, Andrea and {Qu}, Zhijie and {Li}, Jiangtao and {Bregman}, Joel N. and {Fang}, Taotao},
        title = "{On the baryon budget in the X-ray-emitting circumgalactic medium of Milky Way-mass galaxies}",
      journal = {\aap},
         year = 2026,
        month = feb,
       volume = {706},
          eid = {A102},
        pages = {A102},
          doi = {10.1051/0004-6361/202556835},
archivePrefix = {arXiv},
       eprint = {2511.17313},
 primaryClass = {astro-ph.GA},
       adsurl = {https://ui.adsabs.harvard.edu/abs/2026A&A...706A.102Z}
}

@ARTICLE{lyskova23,
       author = {{Lyskova}, N. and {Churazov}, E. and {Khabibullin}, I.~I. and {Burenin}, R. and {Starobinsky}, A.~A. and {Sunyaev}, R.},
        title = "{X-ray surface brightness and gas density profiles of galaxy clusters up to 3 {\texttimes} R$_{500c}$ with SRG/eROSITA}",
      journal = {\mnras},
         year = 2023,
        month = oct,
       volume = {525},
       number = {1},
        pages = {898-907},
          doi = {10.1093/mnras/stad2305},
archivePrefix = {arXiv},
       eprint = {2305.07080},
 primaryClass = {astro-ph.CO},
       adsurl = {https://ui.adsabs.harvard.edu/abs/2023MNRAS.525..898L}
}

@ARTICLE{bulbul24,
       author = {{Bulbul}, E. and {Liu}, A. and {Kluge}, M. and {Zhang}, X. and {Sanders}, J.~S. and {Bahar}, Y.~E. and {Ghirardini}, V. and {Artis}, E. and {Seppi}, R. and {Garrel}, C. and et al.},
        title = "{The SRG/eROSITA All-Sky Survey. The first catalog of galaxy clusters and groups in the Western Galactic Hemisphere}",
      journal = {\aap},
         year = 2024,
        month = may,
       volume = {685},
          eid = {A106},
        pages = {A106},
          doi = {10.1051/0004-6361/202348264},
archivePrefix = {arXiv},
       eprint = {2402.08452},
 primaryClass = {astro-ph.CO},
       adsurl = {https://ui.adsabs.harvard.edu/abs/2024A&A...685A.106B}
}

@ARTICLE{dev24,
       author = {{Dev}, Ajay and {Driver}, Simon P. and {Meyer}, Martin and {Robotham}, Aaron and {Obreschkow}, Danail and {Popesso}, Paola and {Comparat}, Johan},
        title = "{The baryon census and the mass-density of stars, neutral gas, and hot gas as a function of halo mass}",
      journal = {\mnras},
         year = 2024,
        month = dec,
       volume = {535},
       number = {3},
        pages = {2357-2374},
          doi = {10.1093/mnras/stae2485},
archivePrefix = {arXiv},
       eprint = {2411.00456},
 primaryClass = {astro-ph.CO},
       adsurl = {https://ui.adsabs.harvard.edu/abs/2024MNRAS.535.2357D}
}

@ARTICLE{liu22,
       author = {{Liu}, A. and {Bulbul}, E. and {Ghirardini}, V. and {Liu}, T. and {Klein}, M. and {Clerc}, N. and {{\"O}zsoy}, Y. and {Ramos-Ceja}, M.~E. and {Pacaud}, F. and {Comparat}, J. and et al.},
        title = "{The eROSITA Final Equatorial-Depth Survey (eFEDS). Catalog of galaxy clusters and groups}",
      journal = {\aap},
         year = 2022,
        month = may,
       volume = {661},
          eid = {A2},
        pages = {A2},
          doi = {10.1051/0004-6361/202141120},
archivePrefix = {arXiv},
       eprint = {2106.14518},
 primaryClass = {astro-ph.CO},
       adsurl = {https://ui.adsabs.harvard.edu/abs/2022A&A...661A...2L}
}

@ARTICLE{marini24,
       author = {{Marini}, I. and {Popesso}, P. and {Lamer}, G. and {Dolag}, K. and {Biffi}, V. and {Vladutescu-Zopp}, S. and {Dev}, A. and {Toptun}, V. and {Bulbul}, E. and {Comparat}, J. and et al.},
        title = "{Detecting galaxy groups populating the local Universe in the eROSITA era}",
      journal = {\aap},
         year = 2024,
        month = sep,
       volume = {689},
          eid = {A7},
        pages = {A7},
          doi = {10.1051/0004-6361/202450442},
archivePrefix = {arXiv},
       eprint = {2404.12719},
 primaryClass = {astro-ph.CO},
       adsurl = {https://ui.adsabs.harvard.edu/abs/2024A&A...689A...7M}
}

@ARTICLE{marini25a,
       author = {{Marini}, I. and {Popesso}, P. and {Dolag}, K. and {Bravo}, M. and {Robotham}, A. and {Tempel}, E. and {Li}, Q. and {Yang}, X. and {Csizi}, B. and {Behroozi}, P. and et al.},
        title = "{Detecting clusters and groups of galaxies populating the local Universe in large optical spectroscopic surveys}",
      journal = {\aap},
         year = 2025,
        month = feb,
       volume = {694},
          eid = {A207},
        pages = {A207},
          doi = {10.1051/0004-6361/202452028},
archivePrefix = {arXiv},
       eprint = {2411.16455},
 primaryClass = {astro-ph.GA},
       adsurl = {https://ui.adsabs.harvard.edu/abs/2025A&A...694A.207M}
}

@ARTICLE{marini25b,
       author = {{Marini}, I. and {Popesso}, P. and {Dolag}, K. and {Biffi}, V. and {Vladutescu-Zopp}, S. and {Castro}, T. and {Toptun}, V. and {de Is{\'\i}dio}, N. and {Dev}, A. and {Mazengo}, D. and et al.},
        title = "{The impact of assembly history on the X-ray detectability of halos: From galaxy groups to galaxy clusters}",
      journal = {\aap},
         year = 2025,
        month = jun,
       volume = {698},
          eid = {A191},
        pages = {A191},
          doi = {10.1051/0004-6361/202554677},
archivePrefix = {arXiv},
       eprint = {2503.19121},
 primaryClass = {astro-ph.GA},
       adsurl = {https://ui.adsabs.harvard.edu/abs/2025A&A...698A.191M}
}

@ARTICLE{merloni12,
       author = {{Merloni}, A. and {Predehl}, P. and {Becker}, W. and {B{\"o}hringer}, H. and {Boller}, T. and {Brunner}, H. and {Brusa}, M. and {Dennerl}, K. and {Freyberg}, M. and {Friedrich}, P. and et al.},
        title = "{eROSITA Science Book: Mapping the Structure of the Energetic Universe}",
      journal = {arXiv e-prints},
         year = 2012,
        month = sep,
          eid = {arXiv:1209.3114},
        pages = {arXiv:1209.3114},
          doi = {10.48550/arXiv.1209.3114},
archivePrefix = {arXiv},
       eprint = {1209.3114},
 primaryClass = {astro-ph.HE},
       adsurl = {https://ui.adsabs.harvard.edu/abs/2012arXiv1209.3114M}
}

@ARTICLE{merloni24,
       author = {{Merloni}, A. and {Lamer}, G. and {Liu}, T. and {Ramos-Ceja}, M.~E. and {Brunner}, H. and {Bulbul}, E. and {Dennerl}, K. and {Doroshenko}, V. and {Freyberg}, M.~J. and {Friedrich}, S. and et al.},
        title = "{The SRG/eROSITA all-sky survey. First X-ray catalogues and data release of the western Galactic hemisphere}",
      journal = {\aap},
         year = 2024,
        month = feb,
       volume = {682},
          eid = {A34},
        pages = {A34},
          doi = {10.1051/0004-6361/202347165},
archivePrefix = {arXiv},
       eprint = {2401.17274},
 primaryClass = {astro-ph.HE},
       adsurl = {https://ui.adsabs.harvard.edu/abs/2024A&A...682A..34M}
}

@ARTICLE{robotham11,
       author = {{Robotham}, A.~S.~G. and {Norberg}, P. and {Driver}, S.~P. and {Baldry}, I.~K. and {Bamford}, S.~P. and {Hopkins}, A.~M. and {Liske}, J. and {Loveday}, J. and {Merson}, A. and {Peacock}, J.~A. and et al.},
        title = "{Galaxy and Mass Assembly (GAMA): the GAMA galaxy group catalogue (G$^{3}$Cv1)}",
      journal = {\mnras},
         year = 2011,
        month = oct,
       volume = {416},
       number = {4},
        pages = {2640-2668},
          doi = {10.1111/j.1365-2966.2011.19217.x},
archivePrefix = {arXiv},
       eprint = {1106.1994},
 primaryClass = {astro-ph.CO},
       adsurl = {https://ui.adsabs.harvard.edu/abs/2011MNRAS.416.2640R}
}

@ARTICLE{driver22,
       author = {{Driver}, Simon P. and {Bellstedt}, Sabine and {Robotham}, Aaron S.~G. and {Baldry}, Ivan K. and {Davies}, Luke J. and {Liske}, Jochen and {Obreschkow}, Danail and {Taylor}, Edward N. and {Wright}, Angus H. and {Alpaslan}, Mehmet and et al.},
        title = "{Galaxy And Mass Assembly (GAMA): Data Release 4 and the z < 0.1 total and z < 0.08 morphological galaxy stellar mass functions}",
      journal = {\mnras},
         year = 2022,
        month = jun,
       volume = {513},
       number = {1},
        pages = {439-467},
          doi = {10.1093/mnras/stac472},
archivePrefix = {arXiv},
       eprint = {2203.08539},
 primaryClass = {astro-ph.GA},
       adsurl = {https://ui.adsabs.harvard.edu/abs/2022MNRAS.513..439D}
}

@ARTICLE{lovisari15,
       author = {{Lovisari}, L. and {Reiprich}, T.~H. and {Schellenberger}, G.},
        title = "{Scaling properties of a complete X-ray selected galaxy group sample}",
      journal = {\aap},
         year = 2015,
        month = jan,
       volume = {573},
          eid = {A118},
        pages = {A118},
          doi = {10.1051/0004-6361/201423954},
archivePrefix = {arXiv},
       eprint = {1409.3845},
 primaryClass = {astro-ph.CO},
       adsurl = {https://ui.adsabs.harvard.edu/abs/2015A&A...573A.118L}
}

@ARTICLE{seppi25,
       author = {{Seppi}, R. and {Eckert}, D. and {Finoguenov}, A. and {Shreeram}, S. and {Tempel}, E. and {Gozaliasl}, G. and {Lorenz}, M. and {Wilms}, J. and {Mamon}, G.~A. and {Gastaldello}, F. and et al.},
        title = "{Modelling the selection of galaxy groups with end-to-end simulations}",
      journal = {\aap},
         year = 2025,
        month = jul,
       volume = {699},
          eid = {A206},
        pages = {A206},
          doi = {10.1051/0004-6361/202553977},
archivePrefix = {arXiv},
       eprint = {2506.04757},
 primaryClass = {astro-ph.CO},
       adsurl = {https://ui.adsabs.harvard.edu/abs/2025A&A...699A.206S}
}

@ARTICLE{popesso25,
       author = {{Popesso}, P. and {Marini}, I. and {Dolag}, K. and {Lamer}, G. and {Csizi}, B. and {Vladutescu-Zopp}, S. and {Biffi}, V. and {Robothan}, A. and {Bravo}, M. and {Tempel}, E. and et al.},
        title = "{The perils of stacking optically selected groups in eROSITA data: The Magneticum perspective}",
      journal = {\aap},
         year = 2025,
        month = dec,
       volume = {704},
          eid = {A277},
        pages = {A277},
          doi = {10.1051/0004-6361/202453253},
archivePrefix = {arXiv},
       eprint = {2411.16546},
 primaryClass = {astro-ph.GA},
       adsurl = {https://ui.adsabs.harvard.edu/abs/2025A&A...704A.277P}
}

@ARTICLE{strauss02,
       author = {{Strauss}, Michael A. and {Weinberg}, David H. and {Lupton}, Robert H. and {Narayanan}, Vijay K. and {Annis}, James and {Bernardi}, Mariangela and {Blanton}, Michael and {Burles}, Scott and {Connolly}, A.~J. and {Dalcanton}, Julianne and et al.},
        title = "{Spectroscopic Target Selection in the Sloan Digital Sky Survey: The Main Galaxy Sample}",
      journal = {\aj},
         year = 2002,
        month = sep,
       volume = {124},
       number = {3},
        pages = {1810-1824},
          doi = {10.1086/342343},
archivePrefix = {arXiv},
       eprint = {astro-ph/0206225},
 primaryClass = {astro-ph},
       adsurl = {https://ui.adsabs.harvard.edu/abs/2002AJ....124.1810S}
}

@ARTICLE{dey19,
       author = {{Dey}, Arjun and {Schlegel}, David J. and {Lang}, Dustin and {Blum}, Robert and {Burleigh}, Kaylan and {Fan}, Xiaohui and {Findlay}, Joseph R. and {Finkbeiner}, Doug and {Herrera}, David and {Juneau}, St{\'e}phanie and et al.},
        title = "{Overview of the DESI Legacy Imaging Surveys}",
      journal = {\aj},
         year = 2019,
        month = may,
       volume = {157},
       number = {5},
          eid = {168},
        pages = {168},
          doi = {10.3847/1538-3881/ab089d},
archivePrefix = {arXiv},
       eprint = {1804.08657},
 primaryClass = {astro-ph.IM},
       adsurl = {https://ui.adsabs.harvard.edu/abs/2019AJ....157..168D}
}

@ARTICLE{tinker21,
       author = {{Tinker}, Jeremy L.},
        title = "{A Self-Calibrating Halo-Based Group Finder: Application to SDSS}",
      journal = {\apj},
         year = 2021,
        month = dec,
       volume = {923},
       number = {2},
          eid = {154},
        pages = {154},
          doi = {10.3847/1538-4357/ac2aaa},
archivePrefix = {arXiv},
       eprint = {2010.02946},
 primaryClass = {astro-ph.CO},
       adsurl = {https://ui.adsabs.harvard.edu/abs/2021ApJ...923..154T}
}

@ARTICLE{chexmate,
       author = {{CHEX-MATE Collaboration} and {Arnaud}, M. and {Ettori}, S. and {Pratt}, G.~W. and {Rossetti}, M. and {Eckert}, D. and {Gastaldello}, F. and {Gavazzi}, R. and {Kay}, S.~T. and {Lovisari}, L. and et al.},
        title = "{The Cluster HEritage project with XMM-Newton: Mass Assembly and Thermodynamics at the Endpoint of structure formation. I. Programme overview}",
      journal = {\aap},
         year = 2021,
        month = jun,
       volume = {650},
          eid = {A104},
        pages = {A104},
          doi = {10.1051/0004-6361/202039632},
archivePrefix = {arXiv},
       eprint = {2010.11972},
 primaryClass = {astro-ph.CO},
       adsurl = {https://ui.adsabs.harvard.edu/abs/2021A&A...650A.104C}
}

@ARTICLE{thornton16,
       author = {{Thornton}, R.~J. and {Ade}, P.~A.~R. and {Aiola}, S. and {Angil{\`e}}, F.~E. and {Amiri}, M. and {Beall}, J.~A. and {Becker}, D.~T. and {Cho}, H.-M. and {Choi}, S.~K. and {Corlies}, P. and et al.},
        title = "{The Atacama Cosmology Telescope: The Polarization-sensitive ACTPol Instrument}",
      journal = {\apjs},
         year = 2016,
        month = dec,
       volume = {227},
       number = {2},
          eid = {21},
        pages = {21},
          doi = {10.3847/1538-4365/227/2/21},
archivePrefix = {arXiv},
       eprint = {1605.06569},
 primaryClass = {astro-ph.IM},
       adsurl = {https://ui.adsabs.harvard.edu/abs/2016ApJS..227...21T}
}

@ARTICLE{naess20,
       author = {{Naess}, Sigurd and {Aiola}, Simone and {Austermann}, Jason E. and {Battaglia}, Nick and {Beall}, James A. and {Becker}, Daniel T. and {Bond}, Richard J. and {Calabrese}, Erminia and {Choi}, Steve K. and {Cothard}, Nicholas F. and et al.},
        title = "{The Atacama Cosmology Telescope: arcminute-resolution maps of 18 000 square degrees of the microwave sky from ACT 2008-2018 data combined with Planck}",
      journal = {\jcap},
         year = 2020,
        month = dec,
       volume = {2020},
       number = {12},
          eid = {046},
        pages = {046},
          doi = {10.1088/1475-7516/2020/12/046},
archivePrefix = {arXiv},
       eprint = {2007.07290},
 primaryClass = {astro-ph.IM},
       adsurl = {https://ui.adsabs.harvard.edu/abs/2020JCAP...12..046N}
}

@ARTICLE{schaan21,
       author = {{Schaan}, Emmanuel and {Ferraro}, Simone and {Amodeo}, Stefania and {Battaglia}, Nicholas and {Aiola}, Simone and {Austermann}, Jason E. and {Beall}, James A. and {Bean}, Rachel and {Becker}, Daniel T. and {Bond}, Richard J. and et al.},
        title = "{Atacama Cosmology Telescope: Combined kinematic and thermal Sunyaev-Zel'dovich measurements from BOSS CMASS and LOWZ halos}",
      journal = {\prd},
         year = 2021,
        month = mar,
       volume = {103},
       number = {6},
          eid = {063513},
        pages = {063513},
          doi = {10.1103/PhysRevD.103.063513},
archivePrefix = {arXiv},
       eprint = {2009.05557},
 primaryClass = {astro-ph.CO},
       adsurl = {https://ui.adsabs.harvard.edu/abs/2021PhRvD.103f3513S}
}

@ARTICLE{hadzhiyska25a,
       author = {{Hadzhiyska}, B. and {Ferraro}, S. and {Ried Guachalla}, B. and {Schaan}, E. and {Aguilar}, J. and {Ahlen}, S. and {Battaglia}, N. and {Bond}, J.~R. and {Brooks}, D. and {Calabrese}, E. and et al.},
        title = "{Evidence for large baryonic feedback at low and intermediate redshifts from kinematic Sunyaev-Zel'dovich observations with ACT and DESI photometric galaxies}",
      journal = {\prd},
         year = 2025,
        month = oct,
       volume = {112},
       number = {8},
          eid = {083509},
        pages = {083509},
          doi = {10.1103/kclp-x5j1},
archivePrefix = {arXiv},
       eprint = {2407.07152},
 primaryClass = {astro-ph.CO},
       adsurl = {https://ui.adsabs.harvard.edu/abs/2025PhRvD.112h3509H}
}

@ARTICLE{hadzhiyska25b,
       author = {{Hadzhiyska}, Boryana and {Ferraro}, Simone and {Farren}, Gerrit S. and {Sailer}, Noah and {Zhou}, Rongpu},
        title = "{Missing baryons recovered: A measurement of the gas fraction in galaxies and groups with the kinematic Sunyaev-Zel'dovich effect and CMB lensing}",
      journal = {\prd},
         year = 2025,
        month = dec,
       volume = {112},
       number = {12},
          eid = {123507},
        pages = {123507},
          doi = {10.1103/mdhz-fgj8},
archivePrefix = {arXiv},
       eprint = {2507.14136},
 primaryClass = {astro-ph.CO},
       adsurl = {https://ui.adsabs.harvard.edu/abs/2025PhRvD.112l3507H}
}

@ARTICLE{riedguachalla25,
       author = {{Ried Guachalla}, Bernardita and {Schaan}, Emmanuel and {Hadzhiyska}, Boryana and {Ferraro}, Simone and {Aguilar}, Jessica N. and {Ahlen}, Steven and {Battaglia}, Nicholas and {Bianchi}, Davide and {Bond}, Richard and {Brooks}, David and et al.},
        title = "{Backlighting extended gas halos around luminous red galaxies: Kinematic Sunyaev-Zel'dovich effect from DESI Y1 and ACT data}",
      journal = {\prd},
         year = 2025,
        month = nov,
       volume = {112},
       number = {10},
          eid = {103512},
        pages = {103512},
          doi = {10.1103/lqbj-wcqj},
archivePrefix = {arXiv},
       eprint = {2503.19870},
 primaryClass = {astro-ph.GA},
       adsurl = {https://ui.adsabs.harvard.edu/abs/2025PhRvD.112j3512R}
}

@ARTICLE{roper25,
       author = {{Roper}, Finn A. and {Cai}, Yan-Chuan and {Peacock}, John A.},
        title = "{Mass dependence of halo baryon fractions from the kinetic Sunyaev-Zeldovich effect}",
      journal = {\mnras},
         year = 2026,
        month = sep,
       volume = {551},
       number = {3},
          eid = {stag1619},
        pages = {stag1619},
          doi = {10.1093/mnras/stag1619},
archivePrefix = {arXiv},
       eprint = {2510.12553},
 primaryClass = {astro-ph.CO},
       adsurl = {https://scixplorer.org/abs/2026MNRAS.551g1619R}
}

@ARTICLE{schneider22,
       author = {{Schneider}, Aurel and {Giri}, Sambit K. and {Amodeo}, Stefania and {Refregier}, Alexandre},
        title = "{Constraining baryonic feedback and cosmology with weak-lensing, X-ray, and kinematic Sunyaev-Zeldovich observations}",
      journal = {\mnras},
         year = 2022,
        month = aug,
       volume = {514},
       number = {3},
        pages = {3802-3814},
          doi = {10.1093/mnras/stac1493},
archivePrefix = {arXiv},
       eprint = {2110.02228},
 primaryClass = {astro-ph.CO},
       adsurl = {https://ui.adsabs.harvard.edu/abs/2022MNRAS.514.3802S}
}

@ARTICLE{schneiderteyssier15,
       author = {{Schneider}, Aurel and {Teyssier}, Romain},
        title = "{A new method to quantify the effects of baryons on the matter power spectrum}",
      journal = {\jcap},
         year = 2015,
        month = dec,
       volume = {2015},
       number = {12},
        pages = {049-049},
          doi = {10.1088/1475-7516/2015/12/049},
archivePrefix = {arXiv},
       eprint = {1510.06034},
 primaryClass = {astro-ph.CO},
       adsurl = {https://ui.adsabs.harvard.edu/abs/2015JCAP...12..049S}
}

@ARTICLE{girischneider21,
       author = {{Giri}, Sambit K. and {Schneider}, Aurel},
        title = "{Emulation of baryonic effects on the matter power spectrum and constraints from galaxy cluster data}",
      journal = {\jcap},
         year = 2021,
        month = dec,
       volume = {2021},
       number = {12},
          eid = {046},
        pages = {046},
          doi = {10.1088/1475-7516/2021/12/046},
archivePrefix = {arXiv},
       eprint = {2108.08863},
 primaryClass = {astro-ph.CO},
       adsurl = {https://ui.adsabs.harvard.edu/abs/2021JCAP...12..046G}
}

@ARTICLE{schneider25,
       author = {{Schneider}, Aurel and {Kova{\v{c}}}, Michael and {Bucko}, Jozef and {Nicola}, Andrina and {Reischke}, Robert and {Giri}, Sambit K. and {Teyssier}, Romain and {Tr{\"o}ster}, Tilman and {Refregier}, Alexandre and {Schaller}, Matthieu and et al.},
        title = "{Baryonification: an alternative to hydrodynamical simulations for cosmological studies}",
      journal = {\jcap},
         year = 2025,
        month = dec,
       volume = {2025},
       number = {12},
          eid = {043},
        pages = {043},
          doi = {10.1088/1475-7516/2025/12/043},
archivePrefix = {arXiv},
       eprint = {2507.07892},
 primaryClass = {astro-ph.CO},
       adsurl = {https://ui.adsabs.harvard.edu/abs/2025JCAP...12..043S}
}

@ARTICLE{chaikin_edd,
       author = {{Chaikin}, Evgenii and {Schaye}, Joop and {Hu{\v{s}}ko}, Filip and {Lacey}, Cedric G. and {Ploeckinger}, Sylvia and {Schaller}, Matthieu},
        title = "{The importance of super-Eddington black hole accretion for the emergence of massive quiescent galaxies at high redshift}",
      journal = {\mnras},
         year = 2026,
        month = aug,
       volume = {550},
       number = {2},
          eid = {stag1148},
        pages = {stag1148},
          doi = {10.1093/mnras/stag1148},
archivePrefix = {arXiv},
       eprint = {2601.15207},
 primaryClass = {astro-ph.GA},
       adsurl = {https://scixplorer.org/abs/2026MNRAS.550g1148C}
}

@ARTICLE{bahe22,
       author = {{Bah{\'e}}, Yannick M. and {Schaye}, Joop and {Schaller}, Matthieu and {Bower}, Richard G. and {Borrow}, Josh and {Chaikin}, Evgenii and {Kugel}, Roi and {Nobels}, Folkert and {Ploeckinger}, Sylvia},
        title = "{The importance of black hole repositioning for galaxy formation simulations}",
      journal = {\mnras},
         year = 2022,
        month = oct,
       volume = {516},
       number = {1},
        pages = {167-184},
          doi = {10.1093/mnras/stac1339},
archivePrefix = {arXiv},
       eprint = {2109.01489},
 primaryClass = {astro-ph.GA},
       adsurl = {https://ui.adsabs.harvard.edu/abs/2022MNRAS.516..167B}
}

@ARTICLE{chaikin22,
       author = {{Chaikin}, Evgenii and {Schaye}, Joop and {Schaller}, Matthieu and {Bah{\'e}}, Yannick M. and {Nobels}, Folkert S.~J. and {Ploeckinger}, Sylvia},
        title = "{The importance of the way in which supernova energy is distributed around young stellar populations in simulations of galaxies}",
      journal = {\mnras},
         year = 2022,
        month = jul,
       volume = {514},
       number = {1},
        pages = {249-264},
          doi = {10.1093/mnras/stac1132},
archivePrefix = {arXiv},
       eprint = {2203.07134},
 primaryClass = {astro-ph.GA},
       adsurl = {https://ui.adsabs.harvard.edu/abs/2022MNRAS.514..249C}
}

@ARTICLE{eckert26,
       author = {{Eckert}, D. and {Seppi}, R. and {Braspenning}, J. and {Finoguenov}, A. and {Gastaldello}, F. and {Lovisari}, L. and {O'Sullivan}, E. and {Ettori}, S. and {Oppenheimer}, B.~D. and {Bourne}, M.~A. and et al.},
        title = "{The impact of strong feedback on galaxy group scaling relations}",
      journal = {\aap},
         year = 2026,
        month = may,
       volume = {709},
          eid = {L4},
        pages = {L4},
          doi = {10.1051/0004-6361/202558334},
archivePrefix = {arXiv},
       eprint = {2512.04203},
 primaryClass = {astro-ph.CO},
       adsurl = {https://ui.adsabs.harvard.edu/abs/2026A&A...709L...4E}
}

@ARTICLE{chaikin23,
       author = {{Chaikin}, Evgenii and {Schaye}, Joop and {Schaller}, Matthieu and {Ben{\'\i}tez-Llambay}, Alejandro and {Nobels}, Folkert S.~J. and {Ploeckinger}, Sylvia},
        title = "{A thermal-kinetic subgrid model for supernova feedback in simulations of galaxy formation}",
      journal = {\mnras},
         year = 2023,
        month = aug,
       volume = {523},
       number = {3},
        pages = {3709-3731},
          doi = {10.1093/mnras/stad1626},
archivePrefix = {arXiv},
       eprint = {2211.04619},
 primaryClass = {astro-ph.GA},
       adsurl = {https://ui.adsabs.harvard.edu/abs/2023MNRAS.523.3709C}
}

@ARTICLE{correa26,
       author = {{Correa}, Camila A. and {Schaye}, Joop and {Schaller}, Matthieu and {Trayford}, James W. and {Chaikin}, Evgenii and {Ben{\'\i}tez-Llambay}, Alejandro and {Frenk}, Carlos S. and {Ploeckinger}, Sylvia and {Richings}, Alexander J.},
        title = "{A subgrid model for chemical enrichment in cosmological simulations}",
      journal = {\mnras},
         year = 2026,
        month = may,
       volume = {548},
       number = {3},
          eid = {stag645},
        pages = {stag645},
          doi = {10.1093/mnras/stag645},
archivePrefix = {arXiv},
       eprint = {2604.00980},
 primaryClass = {astro-ph.GA},
       adsurl = {https://ui.adsabs.harvard.edu/abs/2026MNRAS.548ag645C}
}

@ARTICLE{trayford26,
       author = {{Trayford}, James W. and {Schaye}, Joop and {Correa}, Camila and {Ploeckinger}, Sylvia and {Richings}, Alexander J. and {Chaikin}, Evgenii and {Schaller}, Matthieu and {Ben{\'\i}tez-Llambay}, Alejandro and {Frenk}, Carlos and {Hu{\v{s}}ko}, Filip},
        title = "{Modelling the evolution and influence of dust in cosmological simulations that include the cold phase of the interstellar medium}",
      journal = {\mnras},
         year = 2026,
        month = feb,
       volume = {545},
       number = {4},
          eid = {staf2040},
        pages = {staf2040},
          doi = {10.1093/mnras/staf2040},
archivePrefix = {arXiv},
       eprint = {2505.13056},
 primaryClass = {astro-ph.GA},
       adsurl = {https://ui.adsabs.harvard.edu/abs/2026MNRAS.545f2040T}
}

@ARTICLE{benitezllambay26,
       author = {{Ben{\'\i}tez-Llambay}, Alejandro and {Ploeckinger}, Sylvia and {Schaye}, Joop and {Richings}, Alexander J. and {Chaikin}, Evgenii and {Schaller}, Matthieu and {Trayford}, James W. and {Frenk}, Carlos S. and {Hu{\v{s}}ko}, Filip and {Correa}, Camila},
        title = "{Non-explosive pre-supernova feedback in the COLIBRE model of galaxy formation}",
      journal = {\mnras},
         year = 2026,
        month = mar,
       volume = {546},
       number = {4},
          eid = {stag268},
        pages = {stag268},
          doi = {10.1093/mnras/stag268},
archivePrefix = {arXiv},
       eprint = {2509.25309},
 primaryClass = {astro-ph.GA},
       adsurl = {https://ui.adsabs.harvard.edu/abs/2026MNRAS.546ag268B}
}

@ARTICLE{richings14a,
       author = {{Richings}, A.~J. and {Schaye}, J. and {Oppenheimer}, B.~D.},
        title = "{Non-equilibrium chemistry and cooling in the diffuse interstellar medium - I. Optically thin regime}",
      journal = {\mnras},
         year = 2014,
        month = jun,
       volume = {440},
       number = {4},
        pages = {3349-3369},
          doi = {10.1093/mnras/stu525},
archivePrefix = {arXiv},
       eprint = {1401.4719},
 primaryClass = {astro-ph.GA},
       adsurl = {https://ui.adsabs.harvard.edu/abs/2014MNRAS.440.3349R}
}

@ARTICLE{richings14b,
       author = {{Richings}, A.~J. and {Schaye}, J. and {Oppenheimer}, B.~D.},
        title = "{Non-equilibrium chemistry and cooling in the diffuse interstellar medium - II. Shielded gas}",
      journal = {\mnras},
         year = 2014,
        month = aug,
       volume = {442},
       number = {3},
        pages = {2780-2796},
          doi = {10.1093/mnras/stu1046},
archivePrefix = {arXiv},
       eprint = {1403.6155},
 primaryClass = {astro-ph.GA},
       adsurl = {https://ui.adsabs.harvard.edu/abs/2014MNRAS.442.2780R}
}

@ARTICLE{ploeckinger25,
       author = {{Ploeckinger}, Sylvia and {Richings}, Alexander J. and {Schaye}, Joop and {Trayford}, James W. and {Schaller}, Matthieu and {Chaikin}, Evgenii},
        title = "{HYBRID-CHIMES: a model for radiative cooling and the abundances of ions and molecules in simulations of galaxy formation}",
      journal = {\mnras},
         year = 2025,
        month = oct,
       volume = {543},
       number = {2},
        pages = {891-916},
          doi = {10.1093/mnras/staf1402},
archivePrefix = {arXiv},
       eprint = {2506.15773},
 primaryClass = {astro-ph.GA},
       adsurl = {https://ui.adsabs.harvard.edu/abs/2025MNRAS.543..891P}
}

@ARTICLE{dallavecchia12,
       author = {{Dalla Vecchia}, Claudio and {Schaye}, Joop},
        title = "{Simulating galactic outflows with thermal supernova feedback}",
      journal = {\mnras},
         year = 2012,
        month = oct,
       volume = {426},
       number = {1},
        pages = {140-158},
          doi = {10.1111/j.1365-2966.2012.21704.x},
archivePrefix = {arXiv},
       eprint = {1203.5667},
 primaryClass = {astro-ph.GA},
       adsurl = {https://ui.adsabs.harvard.edu/abs/2012MNRAS.426..140D}
}

@ARTICLE{nobels24,
       author = {{Nobels}, Folkert S.~J. and {Schaye}, Joop and {Schaller}, Matthieu and {Ploeckinger}, Sylvia and {Chaikin}, Evgenii and {Richings}, Alexander J.},
        title = "{Tests of subgrid models for star formation using simulations of isolated disc galaxies}",
      journal = {\mnras},
         year = 2024,
        month = aug,
       volume = {532},
       number = {3},
        pages = {3299-3321},
          doi = {10.1093/mnras/stae1390},
archivePrefix = {arXiv},
       eprint = {2309.13750},
 primaryClass = {astro-ph.GA},
       adsurl = {https://ui.adsabs.harvard.edu/abs/2024MNRAS.532.3299N}
}

@ARTICLE{oppenheimer20,
       author = {{Oppenheimer}, Benjamin D. and {Bogd{\'a}n}, {\'A}kos and {Crain}, Robert A. and {ZuHone}, John A. and {Forman}, William R. and {Schaye}, Joop and {Wijers}, Nastasha A. and {Davies}, Jonathan J. and {Jones}, Christine and {Kraft}, Ralph P. and {Ghirardini}, Vittorio},
        title = "{EAGLE and Illustris-TNG Predictions for Resolved eROSITA X-Ray Observations of the Circumgalactic Medium around Normal Galaxies}",
      journal = {\apjl},
         year = 2020,
        month = apr,
       volume = {893},
       number = {1},
          eid = {L24},
        pages = {L24},
          doi = {10.3847/2041-8213/ab846f},
archivePrefix = {arXiv},
       eprint = {2003.13889},
 primaryClass = {astro-ph.GA},
       adsurl = {https://ui.adsabs.harvard.edu/abs/2020ApJ...893L..24O}
}

@ARTICLE{shreeram25,
       author = {{Shreeram}, Soumya and {Comparat}, Johan and {Merloni}, Andrea and {Ponti}, Gabriele and {Popesso}, Paola and {Zhang}, Yi and {Nandra}, Kirpal and {Salvato}, Mara and {Marini}, Ilaria and {Buchner}, Johannes and {Locatelli}, Nicola and {Igo}, Zsofi},
        title = "{Retrieving the hot circumgalactic medium physics from the X-ray radial profile from eROSITA with an IlustrisTNG-based forward model}",
      journal = {\aap},
         year = 2025,
        month = nov,
       volume = {703},
          eid = {A137},
        pages = {A137},
          doi = {10.1051/0004-6361/202554508},
archivePrefix = {arXiv},
       eprint = {2504.03840},
 primaryClass = {astro-ph.CO},
       adsurl = {https://ui.adsabs.harvard.edu/abs/2025A&A...703A.137S}
}

@ARTICLE{grayson25,
       author = {{Grayson}, Skylar and {Scannapieco}, Evan and {Comparat}, Johan and {ZuHone}, John A. and {Zhang}, Yi and {Shreeram}, Soumya and {Br{\"u}ggen}, Marcus and {Bulbul}, Esra},
        title = "{The Hot Circumgalactic Medium in Stacked X-Rays: Observations versus Simulations}",
      journal = {\apj},
         year = 2025,
        month = nov,
       volume = {994},
       number = {1},
          eid = {89},
        pages = {89},
          doi = {10.3847/1538-4357/ae100f},
archivePrefix = {arXiv},
       eprint = {2506.09123},
 primaryClass = {astro-ph.GA},
       adsurl = {https://ui.adsabs.harvard.edu/abs/2025ApJ...994...89G}
}

@ARTICLE{liu25,
       author = {{Liu}, R. Henry and {Ferraro}, Simone and {Schaan}, Emmanuel and {Zhou}, Rongpu and {Aguilar}, Jessica Nicole and {Ahlen}, Steven and {Battaglia}, Nicholas and {Bianchi}, Davide and {Brooks}, David and {Claybaugh}, Todd and {Cole}, Shaun and {Coulton}, William R. and {de la Macorra}, Axel and {Dey}, Arjun and {Fanning}, Kevin and {Forero-Romero}, Jaime E. and {Gazta{\~n}aga}, Enrique and {Gong}, Yulin and {Gontcho}, Satya Gontcho A. and {Gruen}, Daniel and {Gutierrez}, Gaston and {Hadzhiyska}, Boryana and {Honscheid}, Klaus and {Howlett}, Cullan and {Kehoe}, Robert and {Kisner}, Theodore and {Kremin}, Anthony and {Kusiak}, Aleksandra and {Lambert}, Andrew and {Landriau}, Martin and {Le Guillou}, Laurent and {Levi}, Michael and {Lokken}, Martine and {Manera}, Marc and {Martini}, Paul and {Meisner}, Aaron and {Miquel}, Ramon and {Moodley}, Kavilan and {Newman}, Jeffrey A. and {Niz}, Gustavo and {Palanque-Delabrouille}, Nathalie and {Percival}, Will and {Prada}, Francisco and {P{\'e}rez-R{\`a}fols}, Ignasi and {Ried Guachalla}, Bernardita and {Rossi}, Graziano and {Sanchez}, Eusebio and {Schlegel}, David and {Schubnell}, Michael and {Seo}, Hee-Jong and {Sif{\'o}n}, Crist{\'o}bal and {Sprayberry}, David and {Tarl{\'e}}, Gregory and {Vavagiakis}, Eve M. and {Weaver}, Benjamin Alan and {Wollack}, Edward J. and {Zou}, Hu},
        title = "{Measurements of the thermal Sunyaev-Zel'dovich effect with ACT and DESI luminous red galaxies}",
      journal = {\prd},
         year = 2025,
        month = oct,
       volume = {112},
       number = {8},
          eid = {083561},
        pages = {083561},
          doi = {10.1103/jqn8-19gx},
archivePrefix = {arXiv},
       eprint = {2502.08850},
 primaryClass = {astro-ph.CO},
       adsurl = {https://ui.adsabs.harvard.edu/abs/2025PhRvD.112h3561L}
}

@article{gatti22,
  title = {Cross-correlation of Dark Energy Survey Year 3 lensing data with ACT and Planck thermal Sunyaev-Zel'dovich effect observations. I. Measurements, systematics tests, and feedback model constraints},
  author = {Gatti, M. and Pandey, S. and Baxter, E. and Hill, J. C. and Moser, E. and Raveri, M. and Fang, X. and DeRose, J. and Giannini, G. and Doux, C. and Huang, H. and Battaglia, N. and Alarcon, A. and Amon, A. and Becker, M. and Campos, A. and Chang, C. and Chen, R. and Choi, A. and Eckert, K. and Elvin-Poole, J. and Everett, S. and Ferte, A. and Harrison, I. and Maccrann, N. and Mccullough, J. and Myles, J. and Navarro Alsina, A. and Prat, J. and Rollins, R. P. and Sanchez, C. and Shin, T. and Troxel, M. and Tutusaus, I. and Yin, B. and Abbott, T. and Aguena, M. and Allam, S. and Andrade-Oliveira, F. and Annis, J. and Bernstein, G. and Bertin, E. and Bolliet, B. and Bond, J. R. and Brooks, D. and Burke, D. L. and Calabrese, E. and Carnero Rosell, A. and Carrasco Kind, M. and Carretero, J. and Cawthon, R. and Costanzi, M. and Crocce, M. and da Costa, L. N. and da Silva Pereira, M. E. and De Vicente, J. and Desai, S. and Diehl, H. T. and Dietrich, J. P. and Doel, P. and Dunkley, J. and Evrard, A. E. and Ferraro, S. and Ferrero, I. and Flaugher, B. and Fosalba, P. and Frieman, J. and Garc\'{\i}a-Bellido, J. and Gaztanaga, E. and Gerdes, D. W. and Giannantonio, T. and Gruen, D. and Gruendl, R. A. and Gschwend, J. and Gutierrez, G. and Herner, K. and Hincks, A. D. and Hinton, S. R. and Hollowood, D. L. and Honscheid, K. and Hughes, J. P. and Huterer, D. and Jain, B. and James, D. J. and Krause, E. and Kuehn, K. and Kuropatkin, N. and Lahav, O. and Lidman, C. and Lima, M. and Lokken, M. and Madhavacheril, M. S. and Maia, M. A. G. and Marshall, J. L. and Mcmahon, J. J. and Melchior, P. and Moodley, K. and Mohr, J. J. and Morgan, R. and Nati, F. and Niemack, M. D. and Page, L. and Palmese, A. and Paz-Chinch\'on, F. and Pieres, A. and Plazas Malag\'on, A. A. and Rodriguez-Monroy, M. and Romer, A. K. and Sanchez, E. and Scarpine, V. and Schaan, E. and Secco, L. F. and Serrano, S. and Sheldon, E. and Sherwin, B. D. and Sif\'on, C. and Smith, M. and Soares-Santos, M. and Spergel, D. and Suchyta, E. and Tarle, G. and Thomas, D. and To, C. and Tucker, D. L. and Varga, T. N. and Weller, J. and Wilkinson, R. D. and Wollack, E. J. and Xu, Z.},
  collaboration = {DES and ACT Collaboration},
  journal = {Phys. Rev. D},
  volume = {105},
  issue = {12},
  pages = {123525},
  numpages = {28},
  year = {2022},
  month = {Jun},
  publisher = {American Physical Society},
  doi = {10.1103/PhysRevD.105.123525},
  url = {https://link.aps.org/doi/10.1103/PhysRevD.105.123525}
}

@ARTICLE{troster22,
       author = {{Tr{\"o}ster}, Tilman and {Mead}, Alexander J. and {Heymans}, Catherine and {Yan}, Ziang and {Alonso}, David and {Asgari}, Marika and {Bilicki}, Maciej and {Dvornik}, Andrej and {Hildebrandt}, Hendrik and {Joachimi}, Benjamin and {Kannawadi}, Arun and {Kuijken}, Konrad and {Schneider}, Peter and {Shan}, Huan Yuan and {van Waerbeke}, Ludovic and {Wright}, Angus H.},
        title = "{Joint constraints on cosmology and the impact of baryon feedback: Combining KiDS-1000 lensing with the thermal Sunyaev-Zeldovich effect from Planck and ACT}",
      journal = {\aap},
         year = 2022,
        month = apr,
       volume = {660},
          eid = {A27},
        pages = {A27},
          doi = {10.1051/0004-6361/202142197},
archivePrefix = {arXiv},
       eprint = {2109.04458},
 primaryClass = {astro-ph.CO},
       adsurl = {https://ui.adsabs.harvard.edu/abs/2022A&A...660A..27T}
}

@ARTICLE{pandey25,
       author = {{Pandey}, S. and {Hill}, J.~C. and {Alarcon}, A. and {Alves}, O. and {Amon}, A. and {Anbajagane}, D. and {Andrade-Oliveira}, F. and {Battaglia}, N. and {Baxter}, E. and {Bechtol}, K. and {Becker}, M.~R. and {Bernstein}, G.~M. and {Blazek}, J. and {Bridle}, S.~L. and {Calabrese}, E. and {Camacho}, H. and {Campos}, A. and {Carnero Rosell}, A. and {Carrasco Kind}, M. and {Cawthon}, R. and {Chang}, C. and {Chen}, R. and {Chintalapati}, P. and {Choi}, A. and {Cordero}, J. and {Coulton}, W. and {Crocce}, M. and {Davis}, C. and {DeRose}, J. and {Devlin}, M. and {Diehl}, H.~T. and {Dodelson}, S. and {Doux}, C. and {Drlica-Wagner}, A. and {Eckert}, K. and {Eifler}, T.~F. and {Elvin-Poole}, J. and {Everett}, S. and {Fang}, X. and {Fert{\'e}}, A. and {Fosalba}, P. and {Friedrich}, O. and {Gatti}, M. and {Gaztanaga}, E. and {Giannini}, G. and {Gluscevic}, V. and {Gruen}, D. and {Gruendl}, R.~A. and {Ried Guachalla}, B. and {Harrison}, I. and {Hartley}, W.~G. and {Herner}, K. and {Huang}, H. and {Huff}, E.~M. and {Huterer}, D. and {Jain}, B. and {Jarvis}, M. and {Krause}, E. and {Kuropatkin}, N. and {Kusiak}, A. and {Leget}, P. and {Lemos}, P. and {Liddle}, A.~R. and {Lokken}, M. and {MacCrann}, N. and {McCullough}, J. and {Moodley}, K. and {Muir}, J. and {Myles}, J. and {Navarro-Alsina}, A. and {Omori}, Y. and {Park}, Y. and {Partridge}, B. and {Porredon}, A. and {Prat}, J. and {Raveri}, M. and {Refregier}, A. and {Rollins}, R.~P. and {Roodman}, A. and {Rosenfeld}, R. and {Ross}, A.~J. and {Rykoff}, E.~S. and {Samuroff}, S. and {Sanchez}, J. and {S{\'a}nchez}, C. and {Secco}, L.~F. and {Sevilla-Noarbe}, I. and {Shaikh}, S. and {Sheldon}, E. and {Shin}, T. and {Sif{\'o}n}, Crist{\'o}bal and {To}, C. and {Troja}, A. and {Troxel}, M.~A. and {Tutusaus}, I. and {Varga}, T.~N. and {Weaverdyck}, N. and {Wechsler}, R.~H. and {Wollack}, E.~J. and {Yanny}, B. and {Yin}, B. and {Zhang}, Y. and {Zuntz}, J. and {Allam}, S.~S. and {Bacon}, D. and {Bocquet}, S. and {Brooks}, D. and {Burke}, D.~L. and {Carretero}, J. and {Cawthon}, R. and {Costanzi}, M. and {da Costa}, L.~N. and {da Silva Pereira}, M.~E. and {Davis}, T.~M. and {Desai}, S. and {Frieman}, J. and {Garc{\'\i}a-Bellido}, J. and {Gutierrez}, G. and {Hinton}, S.~R. and {Hollowood}, D.~L. and {Honscheid}, K. and {James}, D.~J. and {Jeffrey}, N. and {Lee}, S. and {Marshall}, J.~L. and {Mena-Fern{\'a}ndez}, J. and {Miquel}, R. and {Mohr}, J.~J. and {Ogando}, R.~L.~C. and {Plazas Malag'on}, A.~A. and {Romer}, A.~K. and {Sanchez}, E. and {Santiago}, B. and {Smith}, M. and {Suchyta}, E. and {Swanson}, M.~E.~C. and {Thomas}, D. and {Vikram}, V. and {Walker}, A.~R. and {Weller}, J. and {Wiseman}, P.},
        title = "{Constraints on cosmology and baryonic feedback with joint analysis of Dark Energy Survey Year 3 lensing data and ACT DR6 thermal Sunyaev-Zel'dovich effect observations}",
      journal = {arXiv e-prints},
         year = 2025,
        month = jun,
          eid = {arXiv:2506.07432},
        pages = {arXiv:2506.07432},
          doi = {10.48550/arXiv.2506.07432},
archivePrefix = {arXiv},
       eprint = {2506.07432},
 primaryClass = {astro-ph.CO},
       adsurl = {https://ui.adsabs.harvard.edu/abs/2025arXiv250607432P}
}

@ARTICLE{dalal26,
       author = {{Dalal}, Nihar and {To}, Chun-Hao and {Hirata}, Chris and {Hyeon-Shin}, Tae and {Hilton}, Matt and {Pandey}, Shivam and {Richard Bond}, J.},
        title = "{Deciphering baryonic feedback from ACT tSZ galaxy clusters}",
      journal = {\jcap},
         year = 2026,
        month = mar,
       volume = {2026},
       number = {3},
          eid = {036},
        pages = {036},
          doi = {10.1088/1475-7516/2026/03/036},
archivePrefix = {arXiv},
       eprint = {2507.04476},
 primaryClass = {astro-ph.CO},
       adsurl = {https://ui.adsabs.harvard.edu/abs/2026JCAP...03..036D}
}

@ARTICLE{das25,
       author = {{Das}, Sanskriti and {Truong}, Nhut and {Chiang}, Yi-Kuan and {Mathur}, Smita},
        title = "{Thermal Sunyaev─Zel'dovich Effect in the Circumgalactic Medium. II. Dependence on Star Formation}",
      journal = {\apj},
         year = 2025,
        month = oct,
       volume = {991},
       number = {2},
          eid = {205},
        pages = {205},
          doi = {10.3847/1538-4357/adfdd6},
archivePrefix = {arXiv},
       eprint = {2508.09514},
 primaryClass = {astro-ph.GA},
       adsurl = {https://ui.adsabs.harvard.edu/abs/2025ApJ...991..205D}
}

@ARTICLE{sun09,
       author = {{Sun}, M. and {Voit}, G.~M. and {Donahue}, M. and {Jones}, C. and {Forman}, W. and {Vikhlinin}, A.},
        title = "{Chandra Studies of the X-Ray Gas Properties of Galaxy Groups}",
      journal = {\apj},
         year = 2009,
        month = mar,
       volume = {693},
       number = {2},
        pages = {1142-1172},
          doi = {10.1088/0004-637X/693/2/1142},
archivePrefix = {arXiv},
       eprint = {0805.2320},
 primaryClass = {astro-ph},
       adsurl = {https://ui.adsabs.harvard.edu/abs/2009ApJ...693.1142S}
}

@ARTICLE{eckert16,
       author = {{Eckert}, D. and {Ettori}, S. and {Coupon}, J. and {Gastaldello}, F. and {Pierre}, M. and {Melin}, J.-B. and {Le Brun}, A.~M.~C. and {McCarthy}, I.~G. and {Adami}, C. and {Chiappetti}, L. and et al.},
        title = "{The XXL Survey. XIII. Baryon content of the bright cluster sample}",
      journal = {\aap},
         year = 2016,
        month = jun,
       volume = {592},
          eid = {A12},
        pages = {A12},
          doi = {10.1051/0004-6361/201527293},
archivePrefix = {arXiv},
       eprint = {1512.03814},
 primaryClass = {astro-ph.CO},
       adsurl = {https://ui.adsabs.harvard.edu/abs/2016A&A...592A..12E}
}

@ARTICLE{akino22,
       author = {{Akino}, Daichi and {Eckert}, Dominique and {Okabe}, Nobuhiro and {Sereno}, Mauro and {Umetsu}, Keiichi and {Oguri}, Masamune and {Gastaldello}, Fabio and {Chiu}, I.-Non and {Ettori}, Stefano and {Evrard}, August E. and et al.},
        title = "{HSC-XXL: Baryon budget of the 136 XXL groups and clusters}",
      journal = {\pasj},
         year = 2022,
        month = feb,
       volume = {74},
       number = {1},
        pages = {175-208},
          doi = {10.1093/pasj/psab115},
archivePrefix = {arXiv},
       eprint = {2111.10080},
 primaryClass = {astro-ph.CO},
       adsurl = {https://ui.adsabs.harvard.edu/abs/2022PASJ...74..175A}
}

@ARTICLE{efstathiou25,
       author = {{Efstathiou}, George and {McCarthy}, Fiona},
        title = "{The power spectrum of the thermal Sunyaev─Zeldovich effect}",
      journal = {\mnras},
         year = 2025,
        month = jun,
       volume = {540},
       number = {1},
        pages = {1055-1068},
          doi = {10.1093/mnras/staf709},
archivePrefix = {arXiv},
       eprint = {2502.10232},
 primaryClass = {astro-ph.CO},
       adsurl = {https://ui.adsabs.harvard.edu/abs/2025MNRAS.540.1055E}
}

@ARTICLE{louis25,
       author = {{Louis}, Thibaut and {La Posta}, Adrien and {Atkins}, Zachary and {Jense}, Hidde T. and {Abril-Cabezas}, Irene and {Addison}, Graeme E. and {Ade}, Peter A.~R. and {Aiola}, Simone and {Alford}, Tommy and {Alonso}, David and et al.},
        title = "{The Atacama Cosmology Telescope: DR6 power spectra, likelihoods and {\ensuremath{\Lambda}}CDM parameters}",
      journal = {\jcap},
         year = 2025,
        month = nov,
       volume = {2025},
       number = {11},
          eid = {062},
        pages = {062},
          doi = {10.1088/1475-7516/2025/11/062},
archivePrefix = {arXiv},
       eprint = {2503.14452},
 primaryClass = {astro-ph.CO},
       adsurl = {https://ui.adsabs.harvard.edu/abs/2025JCAP...11..062L}
}

@ARTICLE{raghunathan26,
       author = {{Raghunathan}, S. and {Ade}, P.~A.~R. and {Anbajagane}, D. and {Anderson}, A.~J. and {Ansarinejad}, B. and {Archipley}, M. and {Austermann}, J.~E. and {Balkenhol}, L. and {Barron}, D.~R. and {Barry}, P.~S. and et al.},
        title = "{Measurement of the Full Shape of the Thermal Sunyaev─Zel'dovich Power Spectrum from the South Pole Telescope and Herschel─SPIRE Observations}",
      journal = {\apjl},
         year = 2026,
        month = may,
       volume = {1002},
       number = {1},
          eid = {L22},
        pages = {L22},
          doi = {10.3847/2041-8213/ae5c07},
archivePrefix = {arXiv},
       eprint = {2602.10107},
 primaryClass = {astro-ph.CO},
       adsurl = {https://scixplorer.org/abs/2026ApJ..1002L..22R}
}

@ARTICLE{roberts26,
       author = {{Roberts}, Ryan J. and {Davies}, Jonathan J. and {Crain}, Robert A.},
        title = "{Galaxy discs regulate the growth of supermassive black holes}",
      journal = {\mnras},
         year = 2026,
        month = apr,
          doi = {10.1093/mnras/stag629},
archivePrefix = {arXiv},
       eprint = {2604.06977},
 primaryClass = {astro-ph.GA},
       adsurl = {https://ui.adsabs.harvard.edu/abs/2026MNRAS.tmp..592R}
}

@ARTICLE{bondi52,
       author = {{Bondi}, H.},
        title = "{On spherically symmetrical accretion}",
      journal = {\mnras},
         year = 1952,
        month = jan,
       volume = {112},
        pages = {195},
          doi = {10.1093/mnras/112.2.195},
       adsurl = {https://ui.adsabs.harvard.edu/abs/1952MNRAS.112..195B}
}

@ARTICLE{hoylelyttleton39,
       author = {{Hoyle}, F. and {Lyttleton}, R.~A.},
        title = "{The effect of interstellar matter on climatic variation}",
      journal = {Proceedings of the Cambridge Philosophical Society},
         year = 1939,
        month = jan,
       volume = {35},
       number = {3},
        pages = {405},
          doi = {10.1017/S0305004100021150},
       adsurl = {https://ui.adsabs.harvard.edu/abs/1939PCPS...35..405H}
}

@ARTICLE{krumholz06,
       author = {{Krumholz}, Mark R. and {McKee}, Christopher F. and {Klein}, Richard I.},
        title = "{Bondi-Hoyle Accretion in a Turbulent Medium}",
      journal = {\apj},
         year = 2006,
        month = feb,
       volume = {638},
       number = {1},
        pages = {369-381},
          doi = {10.1086/498844},
archivePrefix = {arXiv},
       eprint = {astro-ph/0510410},
 primaryClass = {astro-ph},
       adsurl = {https://ui.adsabs.harvard.edu/abs/2006ApJ...638..369K}
}

@ARTICLE{rosasguevara15,
       author = {{Rosas-Guevara}, Y.~M. and {Bower}, R.~G. and {Schaye}, J. and {Furlong}, M. and {Frenk}, C.~S. and {Booth}, C.~M. and {Crain}, R.~A. and {Dalla Vecchia}, C. and {Schaller}, M. and {Theuns}, T.},
        title = "{The impact of angular momentum on black hole accretion rates in simulations of galaxy formation}",
      journal = {\mnras},
         year = 2015,
        month = nov,
       volume = {454},
       number = {1},
        pages = {1038-1057},
          doi = {10.1093/mnras/stv2056},
archivePrefix = {arXiv},
       eprint = {1312.0598},
 primaryClass = {astro-ph.CO},
       adsurl = {https://ui.adsabs.harvard.edu/abs/2015MNRAS.454.1038R}
}

@ARTICLE{boothschaye09,
       author = {{Booth}, C.~M. and {Schaye}, Joop},
        title = "{Cosmological simulations of the growth of supermassive black holes and feedback from active galactic nuclei: method and tests}",
      journal = {\mnras},
         year = 2009,
        month = sep,
       volume = {398},
       number = {1},
        pages = {53-74},
          doi = {10.1111/j.1365-2966.2009.15043.x},
archivePrefix = {arXiv},
       eprint = {0904.2572},
 primaryClass = {astro-ph.CO},
       adsurl = {https://ui.adsabs.harvard.edu/abs/2009MNRAS.398...53B}
}

@ARTICLE{narayan03,
       author = {{Narayan}, Ramesh and {Igumenshchev}, Igor V. and {Abramowicz}, Marek A.},
        title = "{Magnetically Arrested Disk: an Energetically Efficient Accretion Flow}",
      journal = {\pasj},
         year = 2003,
        month = dec,
       volume = {55},
        pages = {L69-L72},
          doi = {10.1093/pasj/55.6.L69},
archivePrefix = {arXiv},
       eprint = {astro-ph/0305029},
 primaryClass = {astro-ph},
       adsurl = {https://ui.adsabs.harvard.edu/abs/2003PASJ...55L..69N}
}

@ARTICLE{blandfordznajek77,
       author = {{Blandford}, R.~D. and {Znajek}, R.~L.},
        title = "{Electromagnetic extraction of energy from Kerr black holes.}",
      journal = {\mnras},
         year = 1977,
        month = may,
       volume = {179},
        pages = {433-456},
          doi = {10.1093/mnras/179.3.433},
       adsurl = {https://ui.adsabs.harvard.edu/abs/1977MNRAS.179..433B}
}

@ARTICLE{sanchez24,
       author = {{Sanchez}, N. Nicole and {Werk}, Jessica K. and {Christensen}, Charlotte and {Telford}, O. Grace and {Quinn}, Thomas R. and {Tremmel}, Michael and {Mead}, Jennifer and {Sharma}, Ray S. and {Brooks}, Alyson M.},
        title = "{The Scatter Matters: Circumgalactic Metal Content in the Context of the M─{\ensuremath{\sigma}} Relation}",
      journal = {\apj},
         year = 2024,
        month = jun,
       volume = {967},
       number = {2},
          eid = {100},
        pages = {100},
          doi = {10.3847/1538-4357/ad39eb},
archivePrefix = {arXiv},
       eprint = {2305.07672},
 primaryClass = {astro-ph.GA},
       adsurl = {https://ui.adsabs.harvard.edu/abs/2024ApJ...967..100S}
}

@ARTICLE{robson23,
       author = {{Robson}, Dylan and {Dav{\'e}}, Romeel},
        title = "{Redshift evolution of galaxy group X-ray properties in the SIMBA simulations}",
      journal = {\mnras},
         year = 2023,
        month = feb,
       volume = {518},
       number = {4},
        pages = {5826-5842},
          doi = {10.1093/mnras/stac2982},
archivePrefix = {arXiv},
       eprint = {2107.01206},
 primaryClass = {astro-ph.GA},
       adsurl = {https://ui.adsabs.harvard.edu/abs/2023MNRAS.518.5826R}
}

@ARTICLE{ni25,
       author = {{Ni}, Yueying and {Chen}, Nianyi and {Zhou}, Yihao and {Park}, Minjung and {Yang}, Yanhui and {Di Matteo}, Tiziana and {Bird}, Simeon and {Croft}, Rupert},
        title = "{The Astrid Simulation: Evolution of Black Holes and Galaxies to z = 0.5 and Different Evolution Pathways for Galaxy Quenching}",
      journal = {\apj},
         year = 2025,
        month = sep,
       volume = {990},
       number = {2},
          eid = {120},
        pages = {120},
          doi = {10.3847/1538-4357/adf3a7},
archivePrefix = {arXiv},
       eprint = {2409.10666},
 primaryClass = {astro-ph.GA},
       adsurl = {https://ui.adsabs.harvard.edu/abs/2025ApJ...990..120N}
}

@ARTICLE{appleby21,
       author = {{Appleby}, Sarah and {Dav{\'e}}, Romeel and {Sorini}, Daniele and {Storey-Fisher}, Kate and {Smith}, Britton},
        title = "{The low-redshift circumgalactic medium in SIMBA}",
      journal = {\mnras},
         year = 2021,
        month = oct,
       volume = {507},
       number = {2},
        pages = {2383-2404},
          doi = {10.1093/mnras/stab2310},
archivePrefix = {arXiv},
       eprint = {2102.10126},
 primaryClass = {astro-ph.GA},
       adsurl = {https://ui.adsabs.harvard.edu/abs/2021MNRAS.507.2383A}
}

@ARTICLE{voit24,
       author = {{Voit}, G. Mark and {Oppenheimer}, Benjamin D. and {Bell}, Eric F. and {Terrazas}, Bryan and {Donahue}, Megan},
        title = "{Black Hole Growth, Baryon Lifting, Star Formation, and IllustrisTNG}",
      journal = {\apj},
         year = 2024,
        month = jan,
       volume = {960},
       number = {1},
          eid = {28},
        pages = {28},
          doi = {10.3847/1538-4357/ad0039},
archivePrefix = {arXiv},
       eprint = {2309.14818},
 primaryClass = {astro-ph.GA},
       adsurl = {https://ui.adsabs.harvard.edu/abs/2024ApJ...960...28V}
}

@ARTICLE{chadayammuri22,
       author = {{Chadayammuri}, Urmila and {Bogd{\'a}n}, {\'A}kos and {Oppenheimer}, Benjamin D. and {Kraft}, Ralph P. and {Forman}, William R. and {Jones}, Christine},
        title = "{Testing Galaxy Feedback Models with Resolved X-Ray Profiles of the Hot Circumgalactic Medium}",
      journal = {\apjl},
         year = 2022,
        month = sep,
       volume = {936},
       number = {1},
          eid = {L15},
        pages = {L15},
          doi = {10.3847/2041-8213/ac8936},
archivePrefix = {arXiv},
       eprint = {2203.01356},
 primaryClass = {astro-ph.GA},
       adsurl = {https://ui.adsabs.harvard.edu/abs/2022ApJ...936L..15C}
}

@ARTICLE{zhang25sf,
       author = {{Zhang}, Yi and {Comparat}, Johan and {Ponti}, Gabriele and {Merloni}, Andrea and {Nandra}, Kirpal and {Haberl}, Frank and {Truong}, Nhut and {Pillepich}, Annalisa and {Popesso}, Paola and {Locatelli}, Nicola and et al.},
        title = "{The hot circumgalactic medium in the eROSITA All-Sky Survey: III. Star-forming and quiescent galaxies}",
      journal = {\aap},
         year = 2025,
        month = jan,
       volume = {693},
          eid = {A197},
        pages = {A197},
          doi = {10.1051/0004-6361/202452273},
archivePrefix = {arXiv},
       eprint = {2411.19945},
 primaryClass = {astro-ph.GA},
       adsurl = {https://ui.adsabs.harvard.edu/abs/2025A&A...693A.197Z}
}

@ARTICLE{zinger20,
       author = {{Zinger}, Elad and {Pillepich}, Annalisa and {Nelson}, Dylan and {Weinberger}, Rainer and {Pakmor}, R{\"u}diger and {Springel}, Volker and {Hernquist}, Lars and {Marinacci}, Federico and {Vogelsberger}, Mark},
        title = "{Ejective and preventative: the IllustrisTNG black hole feedback and its effects on the thermodynamics of the gas within and around galaxies}",
      journal = {\mnras},
         year = 2020,
        month = nov,
       volume = {499},
       number = {1},
        pages = {768-792},
          doi = {10.1093/mnras/staa2607},
archivePrefix = {arXiv},
       eprint = {2004.06132},
 primaryClass = {astro-ph.GA},
       adsurl = {https://ui.adsabs.harvard.edu/abs/2020MNRAS.499..768Z}
}

@ARTICLE{grayson26,
       author = {{Grayson}, Skylar and {Scannapieco}, Evan and {Dav{\'e}}, Romeel and {Babul}, Arif and {Hough}, Renier T.},
        title = "{Introducing RAFIKI: Refining Active Galactic Nuclei Feedback in Kinetic Implementations}",
      journal = {\apj},
         year = 2026,
        month = mar,
       volume = {999},
       number = {1},
          eid = {9},
        pages = {9},
          doi = {10.3847/1538-4357/ae3dac},
archivePrefix = {arXiv},
       eprint = {2510.19924},
 primaryClass = {astro-ph.GA},
       adsurl = {https://ui.adsabs.harvard.edu/abs/2026ApJ...999....9G}
}

@ARTICLE{sorini22,
       author = {{Sorini}, Daniele and {Dav{\'e}}, Romeel and {Cui}, Weiguang and {Appleby}, Sarah},
        title = "{How baryons affect haloes and large-scale structure: a unified picture from the SIMBA simulation}",
      journal = {\mnras},
         year = 2022,
        month = oct,
       volume = {516},
       number = {1},
        pages = {883-906},
          doi = {10.1093/mnras/stac2214},
archivePrefix = {arXiv},
       eprint = {2111.13708},
 primaryClass = {astro-ph.GA},
       adsurl = {https://ui.adsabs.harvard.edu/abs/2022MNRAS.516..883S}
}

@ARTICLE{ayromlou23,
       author = {{Ayromlou}, Mohammadreza and {Nelson}, Dylan and {Pillepich}, Annalisa},
        title = "{Feedback reshapes the baryon distribution within haloes, in halo outskirts, and beyond: the closure radius from dwarfs to massive clusters}",
      journal = {\mnras},
         year = 2023,
        month = oct,
       volume = {524},
       number = {4},
        pages = {5391-5410},
          doi = {10.1093/mnras/stad2046},
archivePrefix = {arXiv},
       eprint = {2211.07659},
 primaryClass = {astro-ph.GA},
       adsurl = {https://ui.adsabs.harvard.edu/abs/2023MNRAS.524.5391A}
}

@ARTICLE{husko24,
       author = {{Hu{\v{s}}ko}, Filip and {Lacey}, Cedric G. and {Schaye}, Joop and {Nobels}, Folkert S.~J. and {Schaller}, Matthieu},
        title = "{Winds versus jets: a comparison between black hole feedback modes in simulations of idealized galaxy groups and clusters}",
      journal = {\mnras},
         year = 2024,
        month = jan,
       volume = {527},
       number = {3},
        pages = {5988-6020},
          doi = {10.1093/mnras/stad3548},
archivePrefix = {arXiv},
       eprint = {2307.01409},
 primaryClass = {astro-ph.GA},
       adsurl = {https://ui.adsabs.harvard.edu/abs/2024MNRAS.527.5988H}
}

@ARTICLE{khrykin24,
       author = {{Khrykin}, Ilya S. and {Sorini}, Daniele and {Lee}, Khee-Gan and {Dav{\'e}}, Romeel},
        title = "{The cosmic baryon partition between the IGM and CGM in the SIMBA simulations}",
      journal = {\mnras},
         year = 2024,
        month = mar,
       volume = {529},
       number = {1},
        pages = {537-549},
          doi = {10.1093/mnras/stae525},
archivePrefix = {arXiv},
       eprint = {2310.01496},
 primaryClass = {astro-ph.GA},
       adsurl = {https://ui.adsabs.harvard.edu/abs/2024MNRAS.529..537K}
}

@ARTICLE{vandaalen20,
       author = {{van Daalen}, Marcel P. and {McCarthy}, Ian G. and {Schaye}, Joop},
        title = "{Exploring the effects of galaxy formation on matter clustering through a library of simulation power spectra}",
      journal = {\mnras},
         year = 2020,
        month = jan,
       volume = {491},
       number = {2},
        pages = {2424-2446},
          doi = {10.1093/mnras/stz3199},
archivePrefix = {arXiv},
       eprint = {1906.00968},
 primaryClass = {astro-ph.CO},
       adsurl = {https://ui.adsabs.harvard.edu/abs/2020MNRAS.491.2424V}
}

@ARTICLE{medlock25,
       author = {{Medlock}, Isabel and {Neufeld}, Chloe and {Nagai}, Daisuke and {Angl{\'e}s-Alc{\'a}zar}, Daniel and {Genel}, Shy and {Oppenheimer}, Benjamin D. and {Sims}, Xavier and {Singh}, Priyanka and {Villaescusa-Navarro}, Francisco},
        title = "{Quantifying Baryonic Feedback on the Warm─Hot Circumgalactic Medium in CAMELS Simulations}",
      journal = {\apj},
         year = 2025,
        month = feb,
       volume = {980},
       number = {1},
          eid = {61},
        pages = {61},
          doi = {10.3847/1538-4357/ada442},
archivePrefix = {arXiv},
       eprint = {2410.16361},
 primaryClass = {astro-ph.GA},
       adsurl = {https://ui.adsabs.harvard.edu/abs/2025ApJ...980...61M}
}

@article{nelson19,
    author = {Nelson, Dylan and Pillepich, Annalisa and Springel, Volker and Pakmor, Rüdiger and Weinberger, Rainer and Genel, Shy and Torrey, Paul and Vogelsberger, Mark and Marinacci, Federico and Hernquist, Lars},
    title = {First results from the TNG50 simulation: galactic outflows driven by supernovae and black hole feedback},
    journal = {Monthly Notices of the Royal Astronomical Society},
    volume = {490},
    number = {3},
    pages = {3234-3261},
    year = {2019},
    month = {12},
    issn = {0035-8711},
    doi = {10.1093/mnras/stz2306},
    url = {https://doi.org/10.1093/mnras/stz2306},
    eprint = {https://academic.oup.com/mnras/article-pdf/490/3/3234/30327914/stz2306.pdf},
}

@article{gebhardt24,
    author = {Gebhardt, Matthew and Anglés-Alcázar, Daniel and Borrow, Josh and Genel, Shy and Villaescusa-Navarro, Francisco and Ni, Yueying and Lovell, Christopher C and Nagai, Daisuke and Davé, Romeel and Marinacci, Federico and Vogelsberger, Mark and Hernquist, Lars},
    title = {Cosmological baryon spread and impact on matter clustering in CAMELS},
    journal = {Monthly Notices of the Royal Astronomical Society},
    volume = {529},
    number = {4},
    pages = {4896-4913},
    year = {2024},
    month = {04},
    issn = {0035-8711},
    doi = {10.1093/mnras/stae817},
    url = {https://doi.org/10.1093/mnras/stae817},
    eprint = {https://academic.oup.com/mnras/article-pdf/529/4/4896/57161320/stae817.pdf},
}

@article{sanchezalmeida14,
   title={Star formation sustained by gas accretion},
   volume={22},
   ISSN={1432-0754},
   url={http://dx.doi.org/10.1007/s00159-014-0071-1},
   DOI={10.1007/s00159-014-0071-1},
   number={1},
   journal={The Astronomy and Astrophysics Review},
   publisher={Springer Science and Business Media LLC},
   author={Sánchez Almeida, Jorge and Elmegreen, Bruce G. and Muñoz-Tuñón, Casiana and Elmegreen, Debra Meloy},
   year={2014},
   month=July }

@article{keres05,
    author = {Kereš, Dušan and Katz, Neal and Weinberg, David H. and Davé, Romeel},
    title = {How do galaxies get their gas?},
    journal = {Monthly Notices of the Royal Astronomical Society},
    volume = {363},
    number = {1},
    pages = {2-28},
    year = {2005},
    month = {10},
    issn = {0035-8711},
    doi = {10.1111/j.1365-2966.2005.09451.x},
    url = {https://doi.org/10.1111/j.1365-2966.2005.09451.x},
    eprint = {https://academic.oup.com/mnras/article-pdf/363/1/2/4126209/363-1-2.pdf},
}

@article{heckmanbest14,
   title={The Coevolution of Galaxies and Supermassive Black Holes: Insights from Surveys of the Contemporary Universe},
   volume={52},
   ISSN={1545-4282},
   url={http://dx.doi.org/10.1146/annurev-astro-081913-035722},
   DOI={10.1146/annurev-astro-081913-035722},
   number={1},
   journal={Annual Review of Astronomy and Astrophysics},
   publisher={Annual Reviews},
   author={Heckman, Timothy M. and Best, Philip N.},
   year={2014},
   month=Aug, pages={589–660} }

@ARTICLE{veilleux20,
       author = {{Veilleux}, Sylvain and {Maiolino}, Roberto and {Bolatto}, Alberto D. and {Aalto}, Susanne},
        title = "{Cool outflows in galaxies and their implications}",
      journal = {\aapr},
         year = 2020,
        month = apr,
       volume = {28},
       number = {1},
          eid = {2},
        pages = {2},
          doi = {10.1007/s00159-019-0121-9},
archivePrefix = {arXiv},
       eprint = {2002.07765},
 primaryClass = {astro-ph.GA},
       adsurl = {https://ui.adsabs.harvard.edu/abs/2020A&ARv..28....2V}
}

@ARTICLE{harrison18,
       author = {{Harrison}, C.~M. and {Costa}, T. and {Tadhunter}, C.~N. and {Fl{\"u}tsch}, A. and {Kakkad}, D. and {Perna}, M. and {Vietri}, G.},
        title = "{AGN outflows and feedback twenty years on}",
      journal = {Nature Astronomy},
         year = 2018,
        month = feb,
       volume = {2},
        pages = {198-205},
          doi = {10.1038/s41550-018-0403-6},
archivePrefix = {arXiv},
       eprint = {1802.10306},
 primaryClass = {astro-ph.GA},
       adsurl = {https://ui.adsabs.harvard.edu/abs/2018NatAs...2..198H}
}

@ARTICLE{tumlinson17,
       author = {{Tumlinson}, Jason and {Peeples}, Molly S. and {Werk}, Jessica K.},
        title = "{The Circumgalactic Medium}",
      journal = {\araa},
         year = 2017,
        month = aug,
       volume = {55},
       number = {1},
        pages = {389-432},
          doi = {10.1146/annurev-astro-091916-055240},
archivePrefix = {arXiv},
       eprint = {1709.09180},
 primaryClass = {astro-ph.GA},
       adsurl = {https://ui.adsabs.harvard.edu/abs/2017ARA&A..55..389T}
}

@ARTICLE{vandevoort11a,
       author = {{van de Voort}, Freeke and {Schaye}, Joop and {Booth}, C.~M. and {Haas}, Marcel R. and {Dalla Vecchia}, Claudio},
        title = "{The rates and modes of gas accretion on to galaxies and their gaseous haloes}",
      journal = {\mnras},
         year = 2011,
        month = jul,
       volume = {414},
       number = {3},
        pages = {2458-2478},
          doi = {10.1111/j.1365-2966.2011.18565.x},
archivePrefix = {arXiv},
       eprint = {1011.2491},
 primaryClass = {astro-ph.CO},
       adsurl = {https://ui.adsabs.harvard.edu/abs/2011MNRAS.414.2458V}
}

@ARTICLE{vandevoort11b,
       author = {{van de Voort}, Freeke and {Schaye}, Joop and {Booth}, C.~M. and {Dalla Vecchia}, Claudio},
        title = "{The drop in the cosmic star formation rate below redshift 2 is caused by a change in the mode of gas accretion and by active galactic nucleus feedback}",
      journal = {\mnras},
         year = 2011,
        month = aug,
       volume = {415},
       number = {3},
        pages = {2782-2789},
          doi = {10.1111/j.1365-2966.2011.18896.x},
archivePrefix = {arXiv},
       eprint = {1102.3912},
 primaryClass = {astro-ph.CO},
       adsurl = {https://ui.adsabs.harvard.edu/abs/2011MNRAS.415.2782V}
}

@ARTICLE{oppenheimer10,
       author = {{Oppenheimer}, Benjamin D. and {Dav{\'e}}, Romeel and {Kere{\v{s}}}, Du{\v{s}}an and {Fardal}, Mark and {Katz}, Neal and {Kollmeier}, Juna A. and {Weinberg}, David H.},
        title = "{Feedback and recycled wind accretion: assembling the z = 0 galaxy mass function}",
      journal = {\mnras},
         year = 2010,
        month = aug,
       volume = {406},
       number = {4},
        pages = {2325-2338},
          doi = {10.1111/j.1365-2966.2010.16872.x},
archivePrefix = {arXiv},
       eprint = {0912.0519},
 primaryClass = {astro-ph.CO},
       adsurl = {https://ui.adsabs.harvard.edu/abs/2010MNRAS.406.2325O}
}

@ARTICLE{fauchergiguere11,
       author = {{Faucher-Gigu{\`e}re}, Claude-Andr{\'e} and {Kere{\v{s}}}, Du{\v{s}}an and {Ma}, Chung-Pei},
        title = "{The baryonic assembly of dark matter haloes}",
      journal = {\mnras},
         year = 2011,
        month = nov,
       volume = {417},
       number = {4},
        pages = {2982-2999},
          doi = {10.1111/j.1365-2966.2011.19457.x},
archivePrefix = {arXiv},
       eprint = {1103.0001},
 primaryClass = {astro-ph.CO},
       adsurl = {https://ui.adsabs.harvard.edu/abs/2011MNRAS.417.2982F}
}

@ARTICLE{mitchell20a,
       author = {{Mitchell}, Peter D. and {Schaye}, Joop and {Bower}, Richard G. and {Crain}, Robert A.},
        title = "{Galactic outflow rates in the EAGLE simulations}",
      journal = {\mnras},
         year = 2020,
        month = may,
       volume = {494},
       number = {3},
        pages = {3971-3997},
          doi = {10.1093/mnras/staa938},
archivePrefix = {arXiv},
       eprint = {1910.09566},
 primaryClass = {astro-ph.GA},
       adsurl = {https://ui.adsabs.harvard.edu/abs/2020MNRAS.494.3971M}
}

@ARTICLE{mitchell20b,
       author = {{Mitchell}, Peter D. and {Schaye}, Joop and {Bower}, Richard G.},
        title = "{Galactic inflow and wind recycling rates in the EAGLE simulations}",
      journal = {\mnras},
         year = 2020,
        month = oct,
       volume = {497},
       number = {4},
        pages = {4495-4516},
          doi = {10.1093/mnras/staa2252},
archivePrefix = {arXiv},
       eprint = {2005.10262},
 primaryClass = {astro-ph.GA},
       adsurl = {https://ui.adsabs.harvard.edu/abs/2020MNRAS.497.4495M}
}

@ARTICLE{mitchell22a,
       author = {{Mitchell}, Peter D. and {Schaye}, Joop},
        title = "{Baryonic mass budgets for haloes in the EAGLE simulation, including ejected and prevented gas}",
      journal = {\mnras},
         year = 2022,
        month = apr,
       volume = {511},
       number = {2},
        pages = {2600-2609},
          doi = {10.1093/mnras/stab3686},
archivePrefix = {arXiv},
       eprint = {2112.08244},
 primaryClass = {astro-ph.GA},
       adsurl = {https://ui.adsabs.harvard.edu/abs/2022MNRAS.511.2600M}
}

@ARTICLE{mitchell22b,
       author = {{Mitchell}, Peter D. and {Schaye}, Joop},
        title = "{How gas flows shape the stellar-halo mass relation in the EAGLE simulation}",
      journal = {\mnras},
         year = 2022,
        month = apr,
       volume = {511},
       number = {2},
        pages = {2948-2967},
          doi = {10.1093/mnras/stab3339},
archivePrefix = {arXiv},
       eprint = {2103.10966},
 primaryClass = {astro-ph.GA},
       adsurl = {https://ui.adsabs.harvard.edu/abs/2022MNRAS.511.2948M}
}

@ARTICLE{mccarthy10,
       author = {{McCarthy}, I.~G. and {Schaye}, J. and {Ponman}, T.~J. and {Bower}, R.~G. and {Booth}, C.~M. and {Dalla Vecchia}, C. and {Crain}, R.~A. and {Springel}, V. and {Theuns}, T. and {Wiersma}, R.~P.~C.},
        title = "{The case for AGN feedback in galaxy groups}",
      journal = {\mnras},
         year = 2010,
        month = aug,
       volume = {406},
       number = {2},
        pages = {822-839},
          doi = {10.1111/j.1365-2966.2010.16750.x},
archivePrefix = {arXiv},
       eprint = {0911.2641},
 primaryClass = {astro-ph.CO},
       adsurl = {https://ui.adsabs.harvard.edu/abs/2010MNRAS.406..822M}
}

@ARTICLE{mccarthy14,
       author = {{McCarthy}, I.~G. and {Le Brun}, A.~M.~C. and {Schaye}, J. and {Holder}, G.~P.},
        title = "{The thermal Sunyaev-Zel'dovich effect power spectrum in light of Planck}",
      journal = {\mnras},
         year = 2014,
        month = jun,
       volume = {440},
       number = {4},
        pages = {3645-3657},
          doi = {10.1093/mnras/stu543},
archivePrefix = {arXiv},
       eprint = {1312.5341},
 primaryClass = {astro-ph.CO},
       adsurl = {https://ui.adsabs.harvard.edu/abs/2014MNRAS.440.3645M}
}

@ARTICLE{mccarthy23,
       author = {{McCarthy}, Ian G. and {Salcido}, Jaime and {Schaye}, Joop and {Kwan}, Juliana and {Elbers}, Willem and {Kugel}, Roi and {Schaller}, Matthieu and {Helly}, John C. and {Braspenning}, Joey and {Frenk}, Carlos S. and et al.},
        title = "{The FLAMINGO project: revisiting the S$_{8}$ tension and the role of baryonic physics}",
      journal = {\mnras},
         year = 2023,
        month = dec,
       volume = {526},
       number = {4},
        pages = {5494-5519},
          doi = {10.1093/mnras/stad3107},
archivePrefix = {arXiv},
       eprint = {2309.07959},
 primaryClass = {astro-ph.CO},
       adsurl = {https://ui.adsabs.harvard.edu/abs/2023MNRAS.526.5494M}
}

@ARTICLE{oppenheimer21,
       author = {{Oppenheimer}, Benjamin D. and {Babul}, Arif and {Bah{\'e}}, Yannick and {Butsky}, Iryna S. and {McCarthy}, Ian G.},
        title = "{Simulating Groups and the IntraGroup Medium: The Surprisingly Complex and Rich Middle Ground between Clusters and Galaxies}",
      journal = {Universe},
         year = 2021,
        month = jun,
       volume = {7},
       number = {7},
          eid = {209},
        pages = {209},
          doi = {10.3390/universe7070209},
archivePrefix = {arXiv},
       eprint = {2106.13257},
 primaryClass = {astro-ph.GA},
       adsurl = {https://ui.adsabs.harvard.edu/abs/2021Univ....7..209O}
}

@ARTICLE{lu26a,
       author = {{Lu}, Shengdong and {Frenk}, Carlos S. and {Lacey}, Cedric G. and {Gebek}, Andrea and {Schaye}, Joop and {Cole}, Shaun and {Bose}, Sownak and {Andreadis}, Nick and {Baes}, Maarten and {Ben{\'\i}tez-Llambay}, Alejandro and et al.},
        title = "{Galaxy luminosity functions from far-UV to submillimetre at z = 0 in the COLIBRE simulations}",
      journal = {\mnras},
         year = 2026,
        month = nov,
       volume = {552},
       number = {4},
          eid = {stag1723},
        pages = {stag1723},
          doi = {10.1093/mnras/stag1723},
archivePrefix = {arXiv},
       eprint = {2605.02022},
 primaryClass = {astro-ph.GA},
       adsurl = {https://scixplorer.org/abs/2026MNRAS.552g1723L}
}

@ARTICLE{ludlow26,
       author = {{Ludlow}, Aaron D. and {Proctor}, Katy L. and {Schaye}, Joop and {Hu{\v{s}}ko}, Filip and {Moreno}, Victor J. Forouhar and {Obreschkow}, Danail and {Chaikin}, Evgenii and {Schaller}, Matthieu and {Ploeckinger}, Sylvia and {Ben{\'\i}tez-Llambay}, Alejandro and et al.},
        title = "{The evolution of the sizes and angular momentum content of galaxies in the COLIBRE simulations}",
      journal = {\mnras},
         year = 2026,
        month = aug,
       volume = {550},
       number = {4},
          eid = {stag1319},
        pages = {stag1319},
          doi = {10.1093/mnras/stag1319},
archivePrefix = {arXiv},
       eprint = {2603.26200},
 primaryClass = {astro-ph.GA},
       adsurl = {https://scixplorer.org/abs/2026MNRAS.550g1319L}
}

@ARTICLE{lagos26,
       author = {{Lagos}, Claudia del P. and {Schaye}, Joop and {Schaller}, Matthieu and {Obreschkow}, Danail and {Bah{\'e}}, Yannick M. and {Ben{\'\i}tez-Llambay}, Alejandro and {Chaikin}, Evgenii and {Correa}, Camila and {Davis}, Timothy A. and {Frenk}, Carlos S. and et al.},
        title = "{Kennicutt─Schmidt relation of galaxies over 13 billion years in the COLIBRE hydrodynamical simulations}",
      journal = {\mnras},
         year = 2026,
        month = jun,
       volume = {549},
       number = {2},
          eid = {stag947},
        pages = {stag947},
          doi = {10.1093/mnras/stag947},
archivePrefix = {arXiv},
       eprint = {2512.11309},
 primaryClass = {astro-ph.GA},
       adsurl = {https://ui.adsabs.harvard.edu/abs/2026MNRAS.549ag947L}
}

@ARTICLE{dimatteo05,
       author = {{Di Matteo}, Tiziana and {Springel}, Volker and {Hernquist}, Lars},
        title = "{Energy input from quasars regulates the growth and activity of black holes and their host galaxies}",
      journal = {\nat},
         year = 2005,
        month = feb,
       volume = {433},
       number = {7026},
        pages = {604-607},
          doi = {10.1038/nature03335},
archivePrefix = {arXiv},
       eprint = {astro-ph/0502199},
 primaryClass = {astro-ph},
       adsurl = {https://ui.adsabs.harvard.edu/abs/2005Natur.433..604D}
}

@ARTICLE{pontzen26,
       author = {{Pontzen}, Andrew and {Peiris}, Hiranya V. and {Schaye}, Joop and {Schaller}, Matthieu},
        title = "{The limits of feedback from active galactic nuclei}",
      journal = {arXiv e-prints},
         year = 2026,
        month = may,
          eid = {arXiv:2605.16483},
        pages = {arXiv:2605.16483},
          doi = {10.48550/arXiv.2605.16483},
archivePrefix = {arXiv},
       eprint = {2605.16483},
 primaryClass = {astro-ph.GA},
       adsurl = {https://ui.adsabs.harvard.edu/abs/2026arXiv260516483P}
}

@ARTICLE{mcalpine16,
       author = {{McAlpine}, S. and {Helly}, J.~C. and {Schaller}, M. and {Trayford}, J.~W. and {Qu}, Y. and {Furlong}, M. and {Bower}, R.~G. and {Crain}, R.~A. and {Schaye}, J. and {Theuns}, T. and et al.},
        title = "{The EAGLE simulations of galaxy formation: Public release of halo and galaxy catalogues}",
      journal = {Astronomy and Computing},
         year = 2016,
        month = apr,
       volume = {15},
        pages = {72-89},
          doi = {10.1016/j.ascom.2016.02.004},
archivePrefix = {arXiv},
       eprint = {1510.01320},
 primaryClass = {astro-ph.GA},
       adsurl = {https://ui.adsabs.harvard.edu/abs/2016A&C....15...72M}
}

@ARTICLE{nelson19dr,
       author = {{Nelson}, Dylan and {Springel}, Volker and {Pillepich}, Annalisa and {Rodriguez-Gomez}, Vicente and {Torrey}, Paul and {Genel}, Shy and {Vogelsberger}, Mark and {Pakmor}, Ruediger and {Marinacci}, Federico and {Weinberger}, Rainer and et al.},
        title = "{The IllustrisTNG simulations: public data release}",
      journal = {Computational Astrophysics and Cosmology},
         year = 2019,
        month = may,
       volume = {6},
       number = {1},
          eid = {2},
        pages = {2},
          doi = {10.1186/s40668-019-0028-x},
archivePrefix = {arXiv},
       eprint = {1812.05609},
 primaryClass = {astro-ph.GA},
       adsurl = {https://ui.adsabs.harvard.edu/abs/2019ComAC...6....2N}
}

@ARTICLE{helly26,
       author = {{Helly}, John C. and {McGibbon}, Robert J. and {Schaye}, Joop and {Schaller}, Matthieu and {McDonald}, William and {Braspenning}, Joey and {Broxterman}, Jeger C. and {Costello}, Emily E. and {Elbers}, Willem and {Forouhar Moreno}, Victor J. and et al.},
        title = "{The FLAMINGO simulations data release}",
      journal = {Astronomy and Computing},
         year = 2026,
        month = oct,
       volume = {57},
          eid = {101159},
        pages = {101159},
          doi = {10.1016/j.ascom.2026.101159},
archivePrefix = {arXiv},
       eprint = {2604.24324},
 primaryClass = {astro-ph.CO},
       adsurl = {https://scixplorer.org/abs/2026A&C....5701159H}
}

@ARTICLE{borrow20,
       author = {{Borrow}, Josh and {Borrisov}, Alexei},
        title = "{swiftsimio: A Python library for reading SWIFT data}",
      journal = {The Journal of Open Source Software},
         year = 2020,
        month = aug,
       volume = {5},
       number = {52},
          eid = {2430},
        pages = {2430},
          doi = {10.21105/joss.02430},
       adsurl = {https://ui.adsabs.harvard.edu/abs/2020JOSS....5.2430B}
}

@ARTICLE{oman25,
    author = {{Oman}, Kyle A.},
    title = "{SWIFTGalaxy: a Python package to work with particle groups from SWIFT simulations}",
    journal = {The Journal of Open Source Software},
    year = 2025,
    month = oct,
    volume = {10},
    number = {114},
    eid = {9278},
    pages = {9278},
    doi = {10.21105/joss.09278},
    adsurl = {https://ui.adsabs.harvard.edu/abs/2025JOSS...10.9278O}
}

@ARTICLE{braspenning24,
       author = {{Braspenning}, Joey and {Schaye}, Joop and {Schaller}, Matthieu and {McCarthy}, Ian G. and {Kay}, Scott T. and {Helly}, John C. and {Kugel}, Roi and {Elbers}, Willem and {Frenk}, Carlos S. and {Kwan}, Juliana and {Salcido}, Jaime and {van Daalen}, Marcel P. and {Vandenbroucke}, Bert},
        title = "{The FLAMINGO project: galaxy clusters in comparison to X-ray observations}",
      journal = {\mnras},
         year = 2024,
        month = sep,
       volume = {533},
       number = {3},
        pages = {2656-2676},
          doi = {10.1093/mnras/stae1436},
archivePrefix = {arXiv},
       eprint = {2312.08277},
 primaryClass = {astro-ph.GA},
       adsurl = {https://ui.adsabs.harvard.edu/abs/2024MNRAS.533.2656B}
}

@ARTICLE{oren26,
       author = {{Oren}, Yossi and {Pandya}, Viraj and {Somerville}, Rachel S. and {Genel}, Shy and {Omoruyi}, Osase and {Sternberg}, Amiel},
        title = "{The Cosmic Baryon Cycle in IllustrisTNG: Flows of Mass, Energy, and Metals}",
      journal = {\apj},
         year = 2026,
        month = mar,
       volume = {999},
       number = {2},
          eid = {259},
        pages = {259},
          doi = {10.3847/1538-4357/ae41bc},
archivePrefix = {arXiv},
       eprint = {2510.23343},
 primaryClass = {astro-ph.GA},
       adsurl = {https://ui.adsabs.harvard.edu/abs/2026ApJ...999..259O}
}

@ARTICLE{weinberger17,
       author = {{Weinberger}, Rainer and {Springel}, Volker and {Hernquist}, Lars and {Pillepich}, Annalisa and {Marinacci}, Federico and {Pakmor}, R{\"u}diger and {Nelson}, Dylan and {Genel}, Shy and {Vogelsberger}, Mark and {Naiman}, Jill and et al.},
        title = "{Simulating galaxy formation with black hole driven thermal and kinetic feedback}",
      journal = {\mnras},
         year = 2017,
        month = mar,
       volume = {465},
       number = {3},
        pages = {3291-3308},
          doi = {10.1093/mnras/stw2944},
archivePrefix = {arXiv},
       eprint = {1607.03486},
 primaryClass = {astro-ph.GA},
       adsurl = {https://ui.adsabs.harvard.edu/abs/2017MNRAS.465.3291W}
}

@ARTICLE{seppi26,
       author = {{Seppi}, R. and {Eckert}, D. and {Schaye}, J. and {Braspenning}, J. and {Schaller}, M. and {Oppenheimer}, B.~D. and {O'Sullivan}, E. and {Gastaldello}, F. and {Lovisari}, L. and {Bourne}, M.~A. and et al.},
        title = "{Bound or blown: The fate of hot gas in galaxy groups}",
      journal = {\aap},
         year = 2026,
        month = jun,
       volume = {710},
          eid = {A153},
        pages = {A153},
          doi = {10.1051/0004-6361/202660011},
archivePrefix = {arXiv},
       eprint = {2604.24863},
 primaryClass = {astro-ph.CO},
       adsurl = {https://ui.adsabs.harvard.edu/abs/2026A&A...710A.153S}
}

@ARTICLE{eckert24,
       author = {{Eckert}, Dominique and {Gastaldello}, Fabio and {O'Sullivan}, Ewan and {Finoguenov}, Alexis and {Brienza}, Marisa and {X-GAP Collaboration}},
        title = "{Galaxy Groups as the Ultimate Probe of AGN Feedback}",
      journal = {Galaxies},
         year = 2024,
        month = may,
       volume = {12},
       number = {3},
          eid = {24},
        pages = {24},
          doi = {10.3390/galaxies12030024},
archivePrefix = {arXiv},
       eprint = {2403.17145},
 primaryClass = {astro-ph.GA},
       adsurl = {https://ui.adsabs.harvard.edu/abs/2024Galax..12...24E}
}

@ARTICLE{amon22,
       author = {{Amon}, Alexandra and {Efstathiou}, George},
        title = "{A non-linear solution to the S$_{8}$ tension?}",
      journal = {\mnras},
         year = 2022,
        month = nov,
       volume = {516},
       number = {4},
        pages = {5355-5366},
          doi = {10.1093/mnras/stac2429},
archivePrefix = {arXiv},
       eprint = {2206.11794},
 primaryClass = {astro-ph.CO},
       adsurl = {https://ui.adsabs.harvard.edu/abs/2022MNRAS.516.5355A}
}

@ARTICLE{preston23,
       author = {{Preston}, Calvin and {Amon}, Alexandra and {Efstathiou}, George},
        title = "{A non-linear solution to the S$_{8}$ tension - II. Analysis of DES Year 3 cosmic shear}",
      journal = {\mnras},
         year = 2023,
        month = nov,
       volume = {525},
       number = {4},
        pages = {5554-5564},
          doi = {10.1093/mnras/stad2573},
archivePrefix = {arXiv},
       eprint = {2305.09827},
 primaryClass = {astro-ph.CO},
       adsurl = {https://ui.adsabs.harvard.edu/abs/2023MNRAS.525.5554P}
}

@ARTICLE{sharda26,
       author = {{Sharda}, Piyush and {Schaye}, Joop and {McGibbon}, Robert J. and {Ben{\'\i}tez-Llambay}, Alejandro and {Chaikin}, Evgenii and {Frenk}, Carlos S. and {Hodge}, Jacqueline and {Hu{\v{s}}ko}, Filip and {Ploeckinger}, Sylvia and {Richings}, Alexander J. and {Schaller}, Matthieu},
        title = "{The evolution of the galaxy gas-phase mass-metallicity relation from $z=15$ to $z=0$ in the COLIBRE cosmological simulations}",
      journal = {arXiv e-prints},
         year = 2026,
        month = jun,
          eid = {arXiv:2606.25995},
        pages = {arXiv:2606.25995},
          doi = {10.48550/arXiv.2606.25995},
archivePrefix = {arXiv},
       eprint = {2606.25995},
 primaryClass = {astro-ph.GA},
       adsurl = {https://ui.adsabs.harvard.edu/abs/2026arXiv260625995S}
}

@ARTICLE{balzer25,
       author = {{Balzer}, F. and {Bulbul}, E. and {Kluge}, M. and {Liu}, A. and {Salvato}, M. and {Fabricius}, M. and {Seppi}, R. and {Artis}, E. and {Bahar}, Y.~E. and {Bender}, R. and {Clerc}, N. and {Comparat}, J. and {Ghirardini}, V. and {Grandis}, S. and {Krippendorf}, S. and {Lamer}, G. and {Malavasi}, N. and {Merloni}, A. and {Nandra}, K. and {Ramos-Ceja}, M.~E. and {Sanders}, J.~S. and {Zelmer}, S. and {Zhang}, X.},
        title = "{The first SRG/eROSITA All-Sky Survey: Characterization of clusters of galaxies misclassified in the eRASS1 point source catalog}",
      journal = {\aap},
         year = 2025,
        month = sep,
       volume = {701},
          eid = {A283},
        pages = {A283},
          doi = {10.1051/0004-6361/202553942},
archivePrefix = {arXiv},
       eprint = {2501.17238},
 primaryClass = {astro-ph.GA},
       adsurl = {https://ui.adsabs.harvard.edu/abs/2025A&A...701A.283B}
}

@ARTICLE{bucko26,
       author = {{Bucko}, Jozef and {Nicola}, Andrina and {Schneider}, Aurel and {Kova{\v{c}}}, Michael and {Giri}, Sambit K. and {Reischke}, Robert and {Bulbul}, Esra and {Clerc}, Nicolas and {Liu}, Ang and {Bartalucci}, Iacopo and et al.},
        title = "{Baryonification IV: Constraining baryonic feedback with X-ray gas fractions}",
      journal = {arXiv e-prints},
         year = 2026,
        month = sep,
          eid = {arXiv:2609.09144},
        pages = {arXiv:2609.09144},
          doi = {10.48550/arXiv.2609.09144},
archivePrefix = {arXiv},
       eprint = {2609.09144},
 primaryClass = {astro-ph.CO},
       adsurl = {https://ui.adsabs.harvard.edu/abs/2026arXiv260909144B}
}

@ARTICLE{khalil26,
       author = {{Khalil}, H. and {Finoguenov}, A. and {Eckert}, D. and {Seppi}, R. and {Tempel}, E. and {Lovisari}, L. and {Gastaldello}, F.},
        title = "{eROSITA cosmology with galaxy groups: Hot gas budget out to the virial radius}",
      journal = {\aap},
         year = 2026,
        month = oct,
       volume = {714},
          eid = {A194},
        pages = {A194},
          doi = {10.1051/0004-6361/202660299},
archivePrefix = {arXiv},
       eprint = {2608.17735},
 primaryClass = {astro-ph.CO},
       adsurl = {https://scixplorer.org/abs/2026A&A...714A.194K}
}

@ARTICLE{mcdonald26,
       author = {{McDonald}, William and {Schaye}, Joop and {Kuijken}, Konrad and {Helly}, John and {Braspenning}, Joey and {Schaller}, Matthieu},
        title = "{The FLAMINGO project: exploring the X-ray-cosmic-shear cross-correlation as a probe of large-scale structure}",
      journal = {\mnras},
         year = 2026,
        month = nov,
       volume = {552},
       number = {4},
          eid = {stag1746},
        pages = {stag1746},
          doi = {10.1093/mnras/stag1746},
archivePrefix = {arXiv},
       eprint = {2602.02484},
 primaryClass = {astro-ph.CO},
       adsurl = {https://ui.adsabs.harvard.edu/abs/2026MNRAS.552g1746M}
}

@ARTICLE{sharma26,
       author = {{Sharma}, Kritti and {Krause}, Elisabeth and {Ravi}, Vikram and {Connor}, Liam and {Anbajagane}, Dhayaa and {Rajendra Singh}, Pranjal},
        title = "{Signatures of suppressed matter clustering revealed by fast radio bursts}",
      journal = {Nature Astronomy},
         year = 2026,
        month = sep,
          doi = {10.1038/s41550-026-02957-9},
archivePrefix = {arXiv},
       eprint = {2604.17162},
 primaryClass = {astro-ph.CO},
       adsurl = {https://ui.adsabs.harvard.edu/abs/2026NatAs.tmp..196S}
}


\appendix

\section{Multiphase gas in galaxies and their haloes}
\label{sec:app:multiphase}

\begin{figure*}
\includegraphics[width=\textwidth]{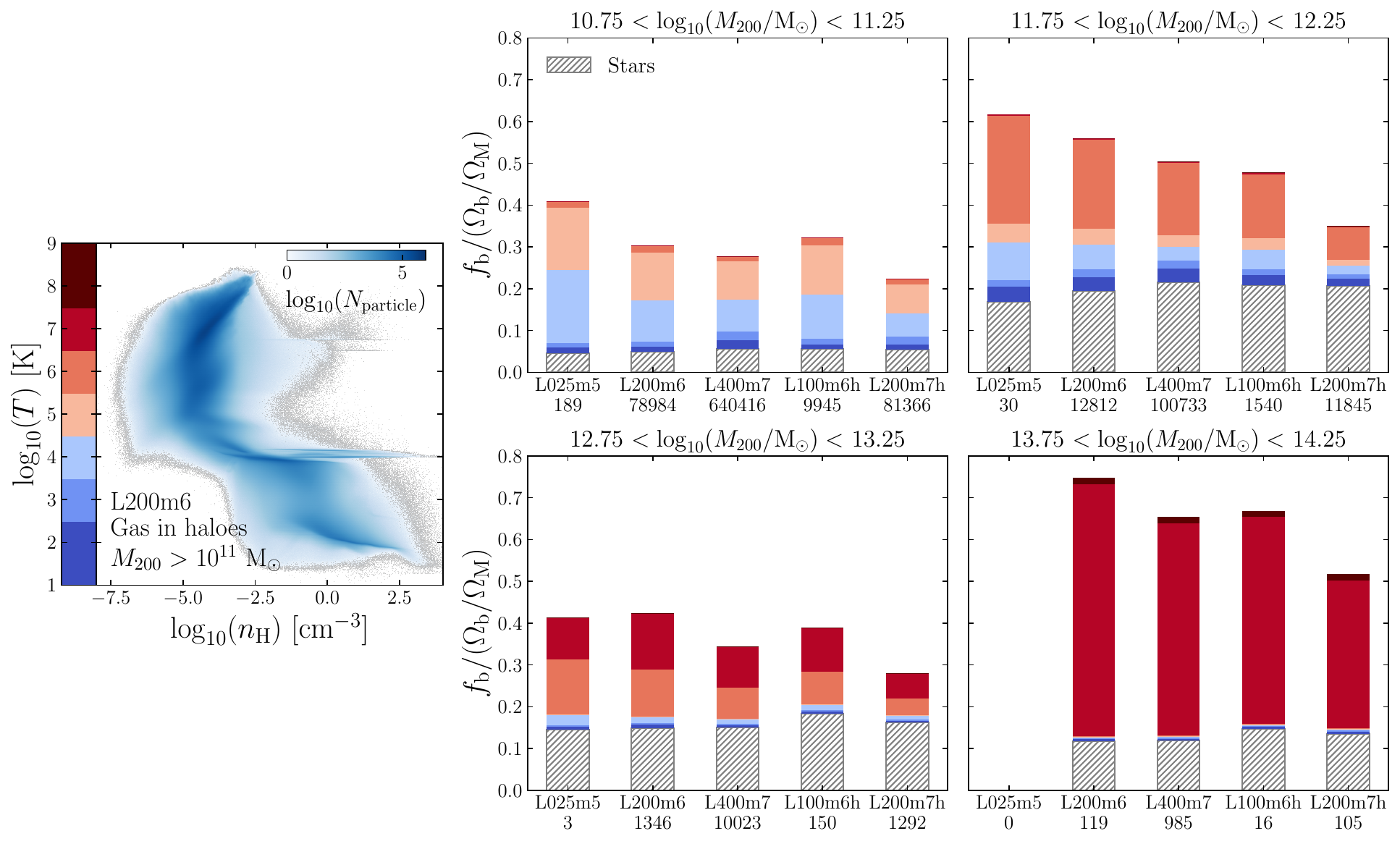}
\vspace{-4mm}
\caption{Left panel: the distribution of temperatures, $T$, and hydrogen number densities, $n_{\rm H}$, for gas particles bound to haloes with mass $M_{200}>10^{11}$~M$_\odot$ in the L200m6 simulation at $z=0$. The coloured regions highlighted along the temperature axis indicate the bins used in other panels. This panel is identical to the left panel of \autoref{fig:temperature} and is included here for convenience. Remaining panels: a census of the baryons within radius $r_{200}$ of haloes in four bins of $M_{200}$, for all simulations explored in this study. The stacked bars show the total gas mass within $r_{200}$ corresponding to each temperature bin, summed across all haloes in the mass bin, divided by the total halo mass to yield a baryon fraction, $f_{\rm b}$, and normalised to the cosmic average baryon fraction $\Omega_{\rm b}/\Omega_{\rm M}$. We show the contribution from stars with hatching. Each bar is labelled with the simulation to which it corresponds, and the number of haloes in the mass bin for that simulation. Comparison of the bars illustrates how changes in resolution or AGN feedback model affect the balance of gas temperatures at different halo masses.}
\vspace{-4mm}
\label{fig:temperature_balance_multi_sim}
\end{figure*}

In \S\ref{sec:results:temperature} we showed the contribution of halo gas in different temperature bins to the $f_{\rm b}^{200}-M_{200}$ relation for central haloes in the L200m6 simulation (\autoref{fig:temperature}). In \autoref{fig:temperature_balance_multi_sim} we show how this census of gas temperatures varies with simulation resolution and AGN feedback prescription. To aid interpretation of this figure, a copy of the density-temperature diagram from \autoref{fig:temperature} is shown at the left hand side, to illustrate the gas phases represented by each temperature bin, which are demarcated by colours along the temperature axis. We present results for all the COLIBRE simulations used in this study: the L025m5, L200m6, L400m7 simulations with thermal AGN feedback, and the L100m6h and L200m7h simulations with hybrid AGN feedback. We compare the halo baryon census between these simulations in 0.5 dex-wide bins of halo mass, as shown in the header of each panel. These bins, from left-to-right and top-to-bottom, approximately correspond to the haloes of dwarf galaxies, Milky Way-mass galaxies, groups, and clusters. To produce the stacked bars, we find the total gas mass within $r_{200}$ corresponding to each temperature bin, sum this across all haloes in the mass bin, divide by the total halo mass to yield a baryon fraction, $f_{\rm b}$, and normalise to the cosmic average baryon fraction $\Omega_{\rm b}/\Omega_{\rm M}$. We show the contribution from stars with hatching. Each bar is labelled with the simulation to which it corresponds, and the number of haloes in the mass bin for that simulation. This number varies significantly between simulation volume and mass bin; notably, the relatively small-volume L025m5 simulation contains no haloes for the most massive bin and hence no bar is shown.

In the haloes of dwarf galaxies (upper centre panel), the contribution from stars is consistent across simulations, but the halo gas content varies significantly. As resolution decreases, the overall gas fraction decreases, a larger fraction of the gas is in the cold ISM phase ($T< 10^{3.5}$~K), and a smaller fraction is in cool CGM gas at $T\sim 10^{4}$~K and warmer phases. The L200m6 and L100m6h simulations have very similar gas fractions in this mass range (see also \autoref{fig:hybrid}), and a very similar temperature distribution. Whilst the hybrid AGN model produces lower gas fractions for the L200m7h simulation relative to the L400m7 simulation, the relative distributions of gas temperatures are near-identical for these simulations. We note that most galaxies in this mass bin are poorly resolved (according to our definition) at m7 resolution, and galaxies in haloes with $M_{200}\lesssim10^{11}$~M$_\odot$ are poorly resolved at m6 resolution.

In the haloes of Milky Way-mass galaxies (upper right panel), the stellar fractions are more sensitive to resolution, as was noted in \S\ref{sec:results:convergence}, increasing slightly as resolution decreases. For the L025m5, L200m6 and L400m7 simulations, the stellar fractions are 15\%, 19\% and 22\% of the cosmic average respectively; the overall baryon fractions change more significantly, at 62\%, 56\% and 50\% respectively. The fraction of the total mass in the cold ISM phase ($T< 10^{3.5}$~K) is consistent with resolution, though this represents an increasing fraction of the halo gas content as resolution decreases. Cool gas at $T\sim 10^{4}$~K represents a decreasing fraction of both the total mass and the gas mass as resolution decreases. The relative contributions of the hotter phases to the gas content are consistent with resolution, despite the overall baryon fraction changing. Likewise, the temperature distribution is very similar for simulations run with the fiducial model and their counterparts run with the hybrid AGN model, despite their different overall baryon fractions.

In galaxy groups and clusters (lower centre and right panels respectively), the stellar fractions do not change significantly with resolution, though they are slightly higher in simulations with hybrid AGN feedback. In these mass bins, the halo gas is predominantly hot; groups contain an approximately even balance of gas in the $T=10^{5.5-6.5}$~K and $T=10^{6.5-7.5}$~K bins, while clusters are almost completely dominated by gas at $T=10^{6.5-7.5}$~K, with a small fraction of hotter gas. For groups and clusters, the relative temperature distribution is nearly identical between m6 and m7 resolution and for both AGN feedback models. Groups in the L025m5 simulation contain larger proportions of cool ($T\sim 10^{4}$~K) gas and hot ($T=10^{5.5-6.5}$~K) gas, though there are only three groups in the volume, so this difference may stem from poorer sampling.

\section{Using different temperature cuts in comparisons with X-ray observations}
\label{sec:app:xraytemp}

\begin{figure}
\includegraphics[width=\columnwidth]{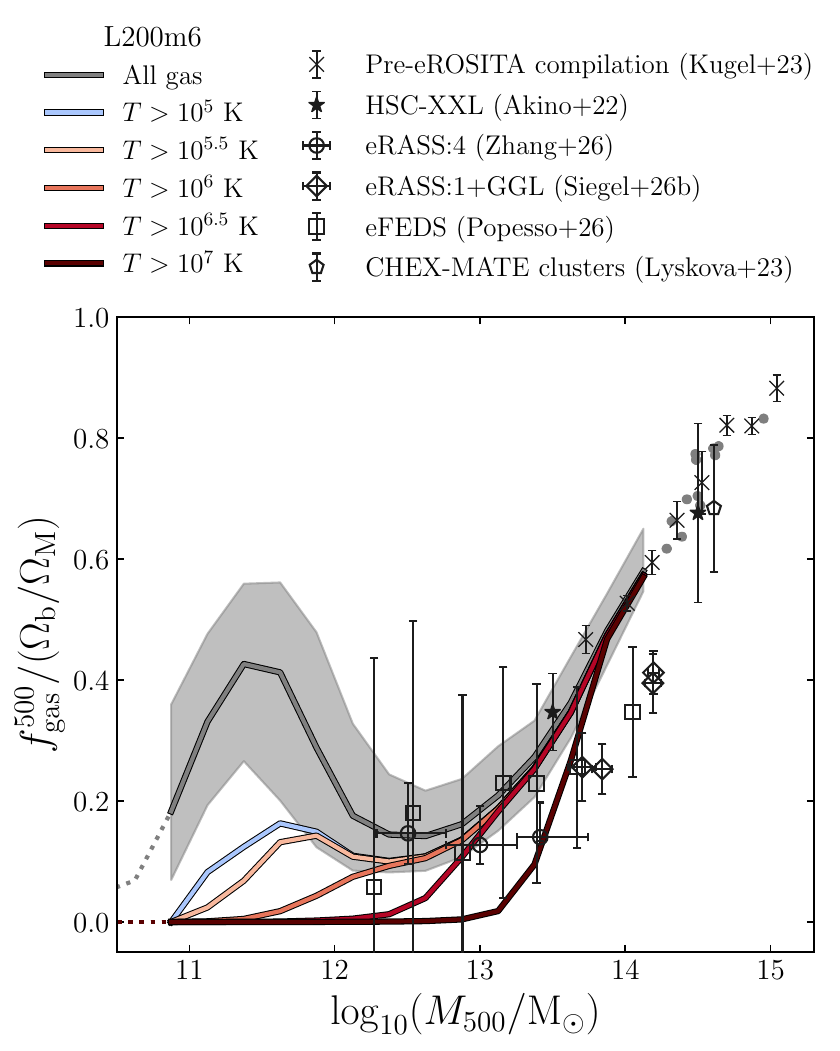}
\vspace{-4mm}
\caption{The present-day $f_{\rm gas}^{500}-M_{500}$ relations for the L200m6 simulation, obtained for different cuts in gas temperature, $T$, compared with constraints from X-ray observations. The grey line, shaded region, and scatter show the relation when including all gas within $r_{500}$, as in \autoref{fig:flagship_xray_ksz}. Coloured lines show the median relations found when only including gas at temperatures above five different thresholds in the calculation of $f_{\rm gas}^{500}$. The line corresponding to a threshold $T>10^6$~K is equivalent to the dashed line representing hot gas in \autoref{fig:flagship_xray_ksz}.}
\vspace{-4mm}
\label{fig:test_temp_cuts}
\end{figure}

When comparing the $f_{\rm gas}^{500}-M_{500}$ relations in the COLIBRE simulations to constraints derived from observations in \S\ref{sec:results:obs}, we showed two relations for each simulation: one including all gas within $r_{500}$, and another including only hot gas with temperature $T>10^6$~K. In this Appendix, we show how the $f_{\rm gas}^{500}-M_{500}$ relation depends on this choice of temperature cut.

In \autoref{fig:test_temp_cuts} we show the $f_{\rm gas}^{500}-M_{500}$ relation for all gas within $r_{500}$, along with constraints from X-ray observations, in the same fashion as in \autoref{fig:flagship_xray_ksz}. With coloured lines, we also show the median $f_{\rm gas}^{500}-M_{500}$ relations produced when adopting five different gas temperature cuts in the calculation of $f_{\rm gas}^{500}$. The temperature cuts are spaced 0.5 dex apart, centred on our fiducial choice of $T>10^6$~K. As in \autoref{fig:flagship_xray_ksz}, we do not show results including temperature cuts for poorly-sampled high mass bins, as for all cuts the difference relative to the fiducial relation is negligible.

In the mass range for which X-ray observations are available ($M_{500}\gtrsim 10^{12.5}$~M$_\odot$), temperature cuts between $10^5$~K and $10^6$~K yield near-identical results and make no difference to our comparison with the X-ray data. Adopting a cut of $10^{6.5}$~K yields a significantly lower gas mass fraction at $M_{500}\approx 10^{12.5}$~M$_\odot$, but produces similar results for $M_{500}\gtrsim 10^{13}$~M$_\odot$. Such a cut is likely too severe for our purposes, as eROSITA is sensitive down to energies of $\approx 0.2$~keV, corresponding to $T\approx 10^{6.37}$~K. Adopting an even higher temperature cut of $T>10^7$~K yields significantly lower $f_{\rm gas}^{500}$ for galaxy groups; again, such a cut is not appropriate for our comparison.

\section{The influence of feedback modelling choices on halo gas fractions}
\label{sec:app:feedback}

\begin{figure*}
\includegraphics[width=\textwidth]{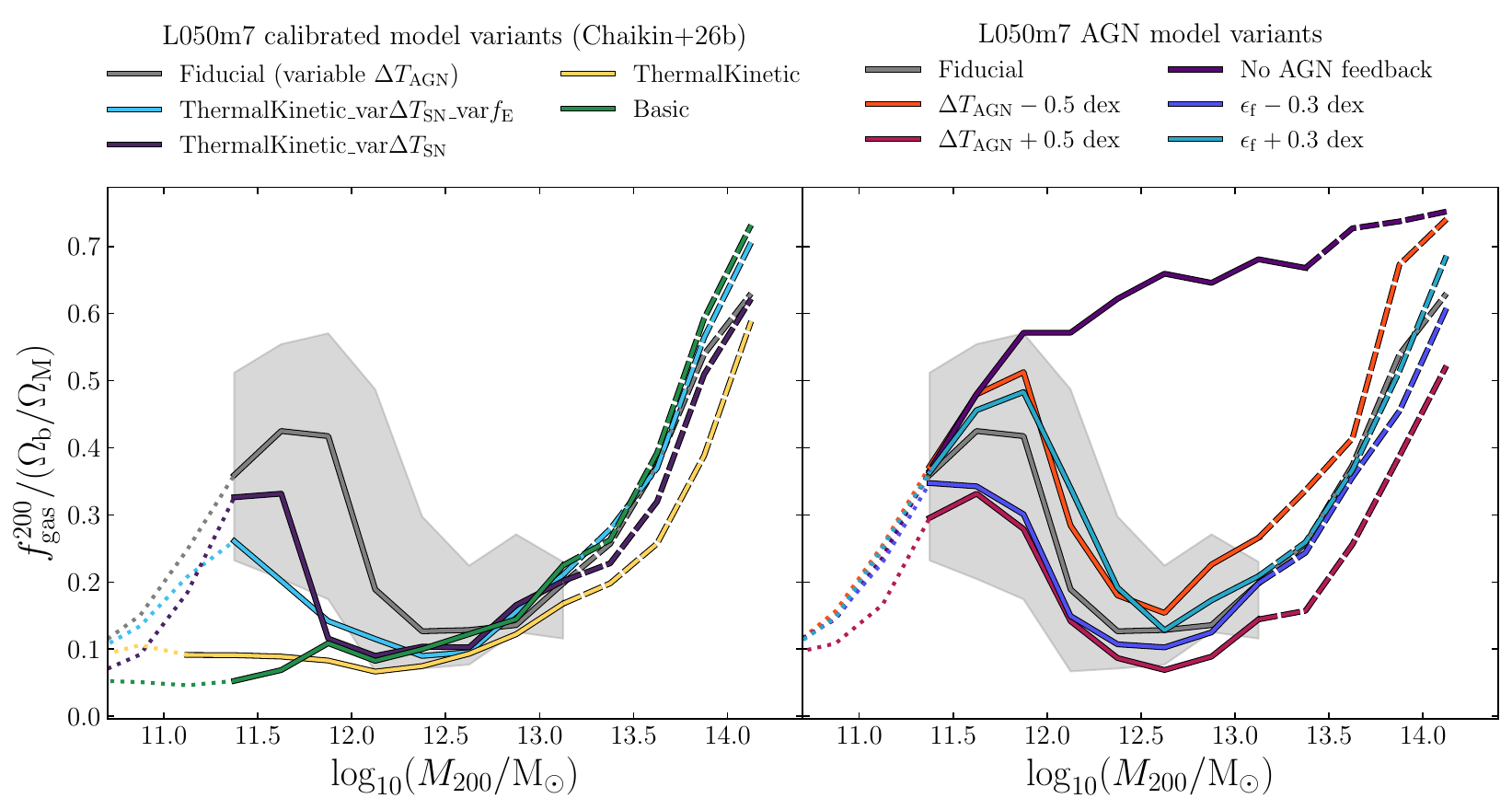}
\vspace{-4mm}
\caption{The present-day $f_{\rm gas}^{200}-M_{200}$ relation in L050m7 simulations performed with modified variants of the COLIBRE model. Results are shown in the same fashion as in previous figures, though for clarity we only show the 16$^{\rm th}$-84$^{\rm th}$ percentile scatter for the fiducial simulation, and the median relations for poorly-sampled bins at high mass (containing fewer than 5 haloes) are shown with dashed lines. {\it Left panel:} individually-calibrated model variants in which features of COLIBRE's feedback prescriptions are successively deactivated (see main text for details), as presented by \citet{chaikin_calibration}. Line colours match those in \citet{chaikin_calibration}, to facilitate comparison. {\it Right panel:} non-calibrated model variants in which the parameters $\Delta T_{\rm AGN}$ and $\epsilon_{\rm f}$ of the fiducial AGN feedback model are adjusted, or where AGN feedback is disabled altogether.}
\vspace{-4mm}
\label{fig:calibration}
\end{figure*}

In this Appendix we examine how the $f_{\rm gas}^{200}-M_{200}$ relation in COLIBRE is influenced by key components of the model's subgrid prescriptions for feedback from SNe and AGN. We will particularly focus on the model components that are new to COLIBRE relative to its predecessor, EAGLE, and how they have produced the differences seen with respect to EAGLE in \autoref{fig:simcompare}, and a much improved match to available observational constraints.

To test the importance of these modelling features, we explore the individually-calibrated variants of the COLIBRE model from \citet{chaikin_calibration}, which were considered during the calibration of the subgrid physics. These model variants were each used to run simulations in periodic volumes of side length 50 comoving Mpc, at m7 resolution, and hence are termed L050m7 in our nomenclature. The $f_{\rm gas}^{200}-M_{200}$ relations produced in these simulations are shown in the left-hand panel of \autoref{fig:calibration}. An L050m7 simulation performed with COLIBRE's fiducial model (as detailed in \S\ref{sec:methods:sf} and \S\ref{sec:methods:bh}) is shown in grey, in the same fashion as figures from the main text, though we show poorly-sampled high-mass bins with dashed lines to facilitate comparison between the median relations for different simulations\footnote{Note that the simulations considered here are 512 times smaller in volume than the flagship L400m7 simulation performed at this resolution, and so the mass range that is both well-sampled with haloes and contains well-resolved galaxies is far narrower. For this reason, we are less conservative here and show bins populated with 5 or fewer galaxies with dashed lines, rather than 10 as adopted in other figures.}. For clarity, we only show the 16$^{\rm th}$-84$^{\rm th}$ percentile scatter for the fiducial simulation. The other lines show the median relations for COLIBRE model variants in which features of the feedback model are successively deactivated. These are, in the nomenclature used by \citet{chaikin_calibration}:
\begin{itemize}
    \item ThermalKinetic\_var$\Delta T_{\rm SN}$\_var$f_{\rm E}$: the full COLIBRE model, but with a fixed AGN heating temperature, $\Delta T_{\rm AGN}=10^9$~K, rather than the BH mass-dependent variable increment given by equation \ref{eq:deltaTAGN}. This fixed value of $\Delta T_{\rm AGN}$ was used throughout the emulator-based calibration of the COLIBRE model. Note that this value is 0.5 dex higher than the value used in the EAGLE (RefL100N1504) simulation.

    \item ThermalKinetic\_var$\Delta T_{\rm SN}$: as the previous model, but without the pressure-dependent variable energy injected per supernova, $f_{\rm E}$, as given by equation \ref{eq:fE}.

    \item ThermalKinetic: as the ThermalKinetic\_var$\Delta T_{\rm SN}$ model, but with a fixed SN heating temperature, $\Delta T_{\rm SN}=10^{7.5}$~K (the same value as used in EAGLE), rather than the density-dependent variable increment given by equation \ref{eq:deltaTSN}.

    \item Basic: as the ThermalKinetic model, but with no kinetic stellar feedback.
\end{itemize}
We note that each of these models was individually calibrated to produce the best possible match to the observed present-day galaxy stellar mass function and size-mass relation, and so the calibrated subgrid parameters vary between the models. The removal of individual model features listed above are therefore not strictly the only changes made. \citet{chaikin_calibration} showed that only a model incorporating all of COLIBRE's stellar feedback features (i.e. the fiducial or ThermalKinetic\_var$\Delta T_{\rm SN}$\_var$f_{\rm E}$ models) can be calibrated to match both these observables simultaneously.

The feedback prescriptions in these model variants have the following effects on the $f_{\rm gas}^{200}-M_{200}$ relation:

\begin{itemize}
    \item The Basic model produces a similar $f_{\rm gas}^{200}-M_{200}$ relation to EAGLE, in that the haloes of galaxies are gas poor, with the gas fraction climbing to higher values only for group and cluster haloes. The purely thermal feedback from SNe is bursty and highly energetic in this model due to the use of a fixed $\Delta T_{\rm SN}$ and, as in EAGLE, is highly efficient at depleting lower-mass haloes of gas and preventing further accretion onto them.

    \item In the ThermalKinetic model, 10\% of the CCSN energy is injected as less expulsive kinetic feedback, however the gas fractions of lower-mass haloes are still low, and similar to those in the Basic model, as the remaining thermal feedback is highly expulsive. The gas fractions of higher-mass haloes are lower in the ThermalKinetic model; this is likely because this model features a higher calibrated BH seed mass and slightly weaker CCSN feedback, causing an earlier onset of AGN feedback in the progenitors of these haloes.

    \item Allowing $\Delta T_{\rm SN}$ to vary in the ThermalKinetic\_var$\Delta T_{\rm SN}$ model produces much higher gas fractions at low mass, as SN feedback becomes better sampled, less energetic, and less expulsive.

    \item Enabling a pressure-dependent energy injected per supernova in the ThermalKinetic\_var$\Delta T_{\rm SN}$\_var$f_{\rm E}$ model produces slightly lower gas fractions in lower-mass galaxies. This may be the result of several effects; pressures are typically higher at early times, leading to stronger SN feedback at higher redshift, whilst lower values of $f_{\rm E}$ at later times could permit higher ISM densities, increasing the value of $\Delta T_{\rm SN}$. This model also favours a higher BH seed mass in the emulator-based calibration than the previous models, yielding higher BH masses \citep[see fig. 9 of][]{chaikin_calibration} and stronger AGN feedback.

    \item Finally, allowing $\Delta T_{\rm AGN}$ to vary in the fiducial model yields the highest gas fractions in lower-mass haloes. At these masses $\Delta T_{\rm AGN}$ is typically lower than the fixed value of $10^9$~K used in the other model variants, and hence AGN feedback is better sampled and less expulsive.
\end{itemize}

In the right-hand panel of \autoref{fig:calibration} we show how the choice of AGN feedback parameters in the fiducial model influences the $f_{\rm gas}^{200}-M_{200}$ relation. Note that unlike the model variants shown in the left panel, these variants were not individually calibrated; in each case the specified change is the only one made to the model. Along with the fiducial relation, we show simulations performed without any AGN feedback, and for which the coupling efficiency $\epsilon_{\rm f}$ is adjusted by $\pm 0.3$ dex. We also show simulations for which the relationship between the variable $\Delta T_{\rm AGN}$ and the BH mass is adjusted by $\pm 0.5$ dex; this is done by adjusting both the prefactor of equation \ref{eq:deltaTAGN} and the maximal temperature increment $\Delta T_{\rm AGN,max}$. These adjustments to the model have the following effects:

\begin{itemize}
    \item When AGN feedback is absent, the gas fractions are drastically different for haloes with $M_{200}\gtrsim10^{12}$~M$_\odot$. The fall in the gas fraction (above the typical mass scale where AGN feedback becomes effective) is absent and the haloes of massive galaxies and groups remain gas rich. The gas fractions come back into agreement with the fiducial model at the mass scale of clusters, however in the fiducial model these haloes have previously been depleted by AGN feedback and then replenished (\autoref{fig:fgas_massbinned_ev}) and in the absence of AGN the evolution of their gas fractions will be wholly different.

    \item Increasing $\Delta T_{\rm AGN}$ reduces $f_{\rm gas}^{200}$ across the entire mass range, and vice-versa. As discussed throughout this study, the parameter $\Delta T_{\rm AGN}$ sets how energetic feedback events are, how readily they can overcome radiative losses, and hence how expulsive AGN feedback is on halo scales. This is true even for AGN in lower-mass haloes, as allowing $\Delta T_{\rm AGN}$ to vary with BH mass improves the sampling of AGN feedback and makes it more active in this mass range. 

    \item The effects of adjusting $\epsilon_{\rm f}$ may appear counter-intuitive, and are less straightforward to interpret. The self-regulatory nature of BH growth and AGN feedback means that BHs typically produce the same outflow rate on scales local to the BH regardless of the choice of $\epsilon_{\rm f}$; this parameter affects the mass at which BHs are able to achieve this self-regulation \citep[][]{boothschaye09,boothschaye10}. The value of $\epsilon_{\rm f}$ is therefore typically chosen to achieve a satisfactory match to the BH mass-stellar mass relation, which COLIBRE achieves \citep[][fig. 15]{schaye26}. However, unlike in EAGLE, in the COLIBRE model this parameter does influence the efficiency of feedback on halo scales, and hence the $f_{\rm gas}^{200}-M_{200}$ relation, because it affects the BH mass and hence the value of $\Delta T_{\rm AGN}$. In the haloes of galaxies, increasing $\epsilon_{\rm f}$ means that BHs can locally self-regulate at lower masses, and hence at lower $\Delta T_{\rm AGN}$, making feedback less expulsive on halo scales and yielding higher gas fractions. The reverse happens upon reducing $\epsilon_{\rm f}$. At higher masses, this effect is not present, likely because BH growth is primarily driven by BH-BH mergers rather than self-regulating accretion.
\end{itemize}

\section{Evolution of the AGN heating temperature}
\label{sec:app:dTAGN}

\begin{figure*}
\includegraphics[width=\textwidth]{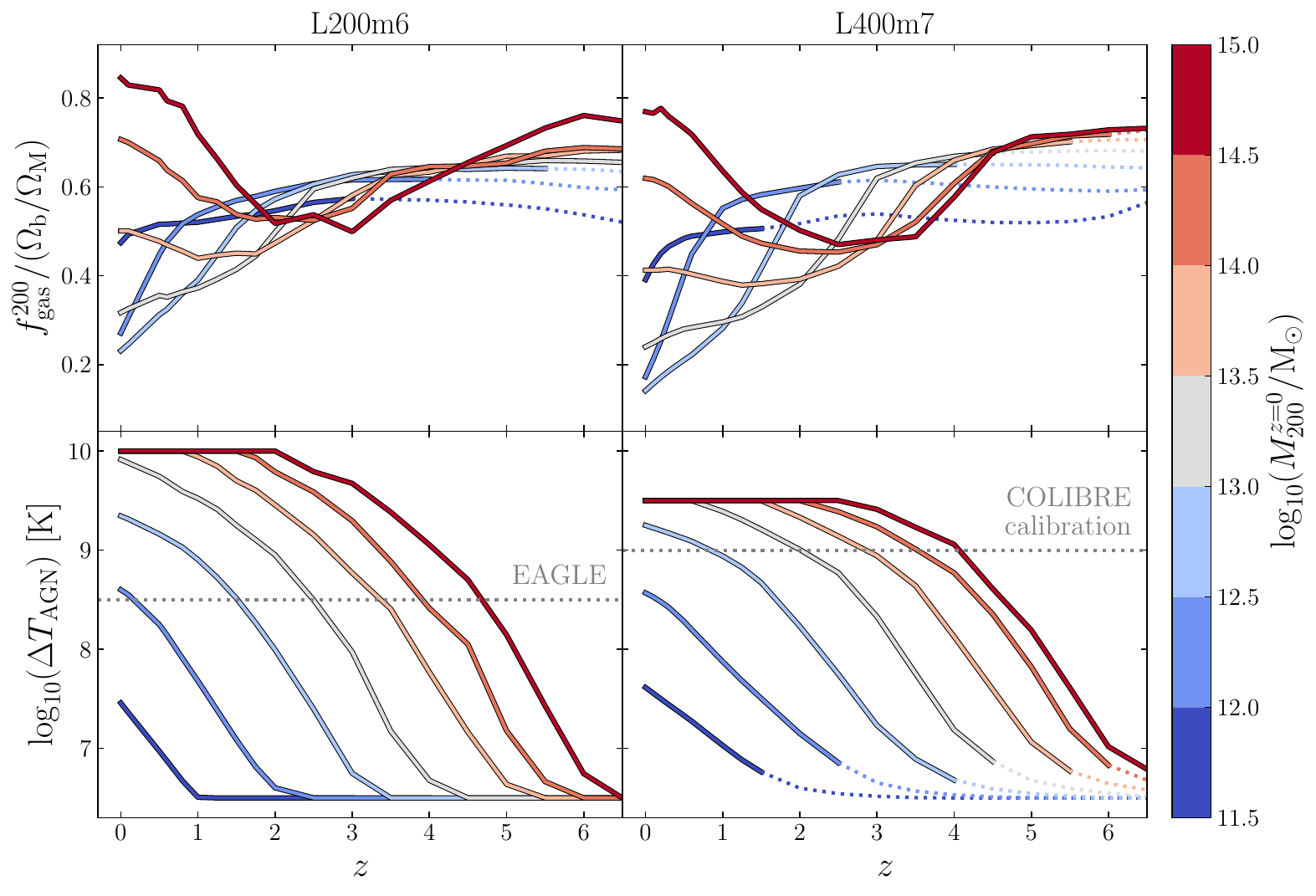}
\vspace{-4mm}
\caption{Evolution of $f_{\rm gas}^{200}$ (upper panels) and $\Delta T_{\rm AGN}$ (bottom panels) with redshift for haloes binned by present-day halo mass, $M_{200}^{z=0}$, in the L200m6 (left column) and L400m7 (right column) simulations. Lines are shown in the same fashion as in \autoref{fig:fgas_massbinned_ev}, from which the upper panels are replicated. Horizontal dotted lines in the bottom panel show the fixed value of $\Delta T_{\rm AGN}=10^{8.5}$~K that was used in the EAGLE RefL100N1504 simulation, and the fixed value of $\Delta T_{\rm AGN}=10^{9}$~K that was used during the calibration of COLIBRE. In haloes with $M_{200}^{z=0}\gtrsim 10^{12.5}$~M$_\odot$, the majority of AGN-driven gas expulsion from haloes occurs with a higher $\Delta T_{\rm AGN}$ in the L200m6 simulation than in EAGLE. In the L400m7 simulation, most expulsion happens at similar $\Delta T_{\rm AGN}$ to the fixed value used during calibration.}
\vspace{-4mm}
\label{fig:dTAGN}
\end{figure*}

In \S\ref{sec:results:simcompare} we showed that the gas fractions of group and cluster haloes in COLIBRE are significantly lower than in EAGLE (and other contemporary simulations), providing a better match with observational constraints. The normalisation of the $f_{\rm gas}^{200}-M_{200}$ relation is strongly influenced by changes in $\Delta T_{\rm AGN}$, as shown in the right-hand panel of \autoref{fig:calibration}. We therefore attribute this improvement to the fact that most AGN-driven gas expulsion in COLIBRE occurs with a higher $\Delta T_{\rm AGN}$ than the fixed value of $10^{8.5}$~K used in EAGLE.

We demonstrate this in \autoref{fig:dTAGN} by showing the evolution of $\Delta T_{\rm AGN}$ (bottom panels) alongside the evolution of $f_{\rm gas}^{200}$ for haloes in different bins of present-day halo mass, $M_{200}^{z=0}$, in the L200m6 (left column) and L400m7 (right column) simulations. The upper panels show the evolution of the median $f_{\rm gas}^{200}$ in each present-day mass bin, and are identical to those in \autoref{fig:fgas_massbinned_ev}; we include them here for comparison with the evolution of the median value of $\Delta T_{\rm AGN}$, which we convert from the median central SMBH mass\footnote{We select the most massive BH particle bound to the halo to be the central SMBH.} using equation \ref{eq:deltaTAGN}. For the L200m6 simulation, we show the fixed value of $\Delta T_{\rm AGN}=10^{8.5}$~K that was used in the EAGLE RefL100N1504 simulation (which has the same mass resolution) with a horizontal dotted line. For the L400m7 simulation, we show (again with a horizontal dotted line) the fixed value of $\Delta T_{\rm AGN}=10^{9}$~K that was used during the calibration of COLIBRE, and is used for all simulations in the left-hand panel of \autoref{fig:calibration} except for the fiducial case.

The value of $\Delta T_{\rm AGN}$ is directly proportional to the central SMBH mass, which is strongly correlated with both halo mass and stellar mass. We therefore see in \autoref{fig:dTAGN} that the final $\Delta T_{\rm AGN}$ is higher as $M_{200}^{z=0}$ increases. More massive present-day haloes also grow their mass earlier, and hence we see that in higher $M_{200}^{z=0}$ bins, $\Delta T_{\rm AGN}$ increases at earlier times as SMBHs co-evolve with their haloes. In the L200m6 simulation, SMBHs in haloes where $M_{200}^{z=0}>10^{13.5}$~M$_\odot$ have grown massive enough that $\Delta T_{\rm AGN}$ is limited to $\Delta T_{\rm AGN, max}=10^{10}$~K, whilst in the L400m7 the lower value of $\Delta T_{\rm AGN, max}=10^{9.5}$~K is also attained by haloes with $M_{200}^{z=0}>10^{13}$~M$_\odot$.

Comparing the evolution of $f_{\rm gas}^{200}$ and $\Delta T_{\rm AGN}$ reveals what the AGN heating temperature was at the time gas was expelled from the halo. We first consider the L200m6 simulations to investigate differences relative to EAGLE. The gas fractions of all haloes begin at a high value that is set primarily by stellar feedback, before a period of significant depletion begins. For the progenitors of very massive galaxies, groups and clusters at the present day ($M_{200}^{z=0}>10^{12.5}$~M$_\odot$), the $\Delta T_{\rm AGN}$ value used in EAGLE is exceeded early in this period, and the majority of AGN-driven expulsion occurs with a heating temperature up to 1.5 dex higher than in EAGLE, leading to lower present-day gas fractions. In the progenitors of the haloes of present-day Milky Way-mass galaxies ($M_{200}^{z=0}=10^{12-12.5}$~M$_\odot$), expulsion occurs with a similar $\Delta T_{\rm AGN}$ to EAGLE, perhaps explaining the similar gas fractions for both models in this mass regime, though differences in the modelling of stellar feedback are likely also relevant here. As in the main text, we note here that other improvements to BH modelling, such as BH repositioning and super-Eddington accretion, may also contribute to the increased efficiency of AGN feedback in COLIBRE.

The results for the L400m7 simulation allow us to investigate why allowing $\Delta T_{\rm AGN}$ to vary from the fixed value of $10^9$~K assumed in COLIBRE's calibration does not appear to drive a significant change to the high-mass end of the $f_{\rm gas}^{200}-M_{200}$ relation in \autoref{fig:calibration}. The fixed-$\Delta T_{\rm AGN}$ value is typically exceeded mid-way through the process of expulsion in group and cluster haloes, with most expulsion happening at a $\Delta T_{\rm AGN}$ within $\pm0.5$ dex of the fixed value. Earlier expulsion occurs with a lower temperature increment, and later expulsion with a higher increment, and the net effect appears to yield similar present-day gas fractions as for the fixed-$\Delta T_{\rm AGN}$ case. These results also demonstrate why using a variable $\Delta T_{\rm AGN}$ leads to higher gas fractions in lower-mass haloes ($M_{200}^{z=0}=10^{11.5-12}$~M$_\odot$); the present-day $\Delta T_{\rm AGN}$ is $\approx 1.5$ dex lower than the fixed calibration value in these haloes, making any AGN activity far less expulsive.


\bsp	
\label{lastpage}
\end{document}